\documentclass{article}

\usepackage{arxiv}

\usepackage[utf8]{inputenc} 
\usepackage[T1]{fontenc}    
\usepackage{hyperref}       
\usepackage{url}            
\usepackage{booktabs}       
\usepackage{amsfonts}       
\usepackage{nicefrac}       
\usepackage{microtype}      
\usepackage{lipsum}
\usepackage{graphicx}
\graphicspath{ {./images/} }
\usepackage{xcolor}

\title{Psychoacoustic study of simple-tone dyads: \\ frequency ratio and pitch}

\author{Stefania Kaklamani and Constantinos Simserides \\
Department of Physics, National and Kapodistrian University of Athens, \\ Panepistimiopolis, Zografos, GR-15784, Athens, Greece \\
\texttt{csimseri@phys.uoa.gr} \\
}

\begin{document}
\maketitle
\begin{abstract}
This study investigates how listeners perceive consonance and dissonance in dyads composed of simple (sine) tones, focusing on the effects of frequency ratio ($R$) and mean frequency ($F$). Seventy adult participants - categorized by musical training, gender, and age group - rated randomly ordered dyads using binary preference responses (``like'' or ``dislike''). Dyads represented standard Western intervals but were constructed with sine tones rather than musical notes, preserving interval ratios while varying absolute pitch. Statistical analyses reveal a consistent decrease in preference with increasing mean frequency, regardless of interval class or participant group. Octaves, fifths, fourths, and sixths showed a nearly linear decline in preference with increasing $F$. Major seconds were among the least preferred. Musicians rated octaves and certain consonant intervals more positively than non-musicians, while gender and age groups exhibited different sensitivity to high frequencies. The findings suggest that both interval structure and pitch range shape the perception of consonance in simple-tone dyads, with possible psychoacoustic explanations involving frequency sensitivity and auditory fatigue at higher frequencies.
\end{abstract}



\section{Introduction}
\label{sec:introduction}
The perception of consonance and dissonance is a long-standing topic in music psychology and psychoacoustics. Helmholtz's  proposed that dissonance arises from interference (``roughness'') caused by beating between closely spaced frequencies, which stimulate overlapping regions of the inner ear basilar membrane~\cite{Helmholtz:1863,Helmholtz:1895,Helmholtz:1954}. Later work challenged this purely physiological account. For instance, diotic presentations of dyads (both tones presented to both ears) are rated as more consonant than dichotic presentations~\cite{Bones:2014}, even though the latter should reduce interference effects, suggesting cognitive and cultural factors play significant roles~\cite{LahdelmaEerola:2020}.
Specifically, Bones et al. conclude that ``When  both notes of a consonant dyad were presented to both ears, the dyad was perceived as being more consonant than when the two notes were presented to separate ears.''
Many studies focused on the perception differences of music and musical tones between musicians and non-musicians and on the impact that musical training and musical expertise can have on the way people receive and process this kind of information. In an experiment by McLachlan et al.~\cite{McLachlan:2013}, it was found that musical training improves pitch matching and accuracy and that recognition mechanisms are integral to pitch processing. The results showed that the dissonance of a stimulus is not linked to its physical properties and their ratio, but, rather, to familiarity to that pitch combination \cite{Guernsey:1928,LahdelmaEerola:2020}. This supports the cognitive incongruence model, which suggests that the dissonance grade of a musical sound has to do with a mismatch between pitch information arriving at the auditory cortex and recognition mechanisms as well as periodicity processing in the brainstem and the mid brain. Consequently, the dissonance rate of a sound must be reduced after we hear it many times and as we become more familiar with the relationship of frequencies that constitute the sound~\cite{McDermott:2010,Bones:2014}. The connection between consonance and simple frequency rational ratios perhaps shows a familiarity with rational numbers due to cultural reasons.
Recent research showed connection between the preference of a musical sound when its spectrum is  similar to the spectra of a human voice \cite{Bowling:2017,BowlingPurves:2015:PNAS}.
Bowling et al.~\cite{Bowling:2017} managed to categorize every possible chromatic dyad (two note chords) played on a piano, in consonance rank, beginning with the least consonant and reaching an octave. This work was expanded in three note chords and four note chords. 
Terhardt~\cite{Terhardt:1984} conducted experiments in which subjects listened to dyads composed of simple and composite tones, including harmonics. These dyads consisted of two tones, one fixed at La4 and the other varying gradually, increasing the interval between them up to an octave (La5). Subjects rated the dyads for ``consonance'' and ``roughness". A key question is whether the specific choice of musical notes to construct dyads influences these results. Our study contributes to this discussion by focusing exclusively on simple (sine) tones, maintaining frequency ratios but excluding standard musical notes.

Gender differences have been studied in the past, e.g. in Refs.~\cite{Pearson:1995, McFadden:1998}.
An experiment studied preferences of men and women relative to male or female voices~\cite{Re:2012} in a frequency regime (143-285 Hz for female voices, 86-152 Hz for male voices) which is very small relative to the frequency regime of our experiment (200-5000 Hz). 
Other studies \cite{HarrisonPearce:2020, McDermott:2016, Marjieh:2024} advanced our understanding of consonance by disentangling psychoacoustic and cultural factors using rich harmonic or timbral stimuli. In contrast, the present study isolates pitch and frequency ratio using simple sine dyads to explore how mean frequency and interval class affect preference, with minimal influence from learned musical timbre or harmonic complexity.
Aging effects have also been studied in the past, e.g. in Refs.~\cite{BonesPlack:2015, Lentz:2022, Tufts:2005,KoernerZhang:2018}.
Morrell et al.~\cite{Morrell:1996} studied age-specific reference ranges for hearing level and change in the hearing level for men and women at 500, 1000, 2000, and 4000 Hz, using data from the Baltimore Longitudinal Study of Aging. These persons were screened for otological disorders and noise-induced hearing loss. Their results provide reference for detecting when a person deviates from the normal pattern. Pearson et al. \cite{Pearson:1995} also used data from the Baltimore Longitudinal Study of Aging.
It appears that females as a group have greater hearing sensitivity, greater susceptibility to noise exposure at high frequencies, shorter latencies in their auditory brain‐stem responses, more spontaneous otoacoustic emissions (SOAEs), and stronger click‐evoked otoacoustic emissions than males as a group ~\cite{McFadden:1998}.

In this study, we explore how listeners evaluate the consonance of dyads constructed from sine tones, varying both the frequency ratio and the mean frequency. Unlike studies that use continuous or Likert-type scales, we used a binary preference response ("like" vs. "dislike") to simplify the task and encourage intuitive judgments. We analyze how preference is modulated by musical training, gender, and age group. Our stimuli cover a wide frequency range (200–5000 Hz), allowing us to assess pitch range effects across listener groups.
Our study aims to clarify how interval class and pitch height jointly influence perceived consonance when stripped of timbral and harmonic context. We seek to determine whether preferences for particular intervals persist across pitch ranges and listener demographics, or whether they diminish in the absence of learned harmonic cues.

The participants were 70 Greek adults with diverse musical background. While most psychoacoustic studies implicitly assume cultural homogeneity, it is worth noting that Greek popular and traditional music often incorporates interval structures that differ from the Western classical canon, which may influence listener expectations. We divided subjects into groups: musicians vs. non-musicians, men vs. women, and age groups. 

To construct a simple-tone dyad (that could be called for brevity \textit{ditonia}) we used simple (sine) tones that do not correspond to notes but preserve the frequency rational ratios of western musical intervals.

Section~\ref{sec:methods} is devoted to methods: In Subsec.~\ref{subsec:ditonies} (Stimuli) we define the physical properties of simple-tone dyads. In Subsec.~\ref{subsec:Procedure} (Procedure) we explain experimental details and software. 
In Subsec.~\ref{subsec:Participants} (Participants) we describe the persons involved as subjects. 
In Subsec.~\ref{subsec:ResponsePreferenceStatistics} (Data Analysis) we explain response, preference and  statistics used to analyze experimental data. In Sec.~\ref{sec:ResultsDiscussion} we present and discuss our results: In Subsec.~\ref{subsec:TotalSet} we study the total set of subjects as a whole. Then, we study musicians vs. non-musicians (Subsec.~\ref{subsec:MusiciansNonmusicians}), 
men vs. women (Subsec.~\ref{subsec:MenWomen}), and 
age groups (Subsec.~\ref{subsec:35and50}). In Sec.~\ref{sec:conclusion} we state conclusions.

\begin{figure*}[htb]
\centering
\vspace{-0.4cm}
\includegraphics[width=0.24\textwidth]{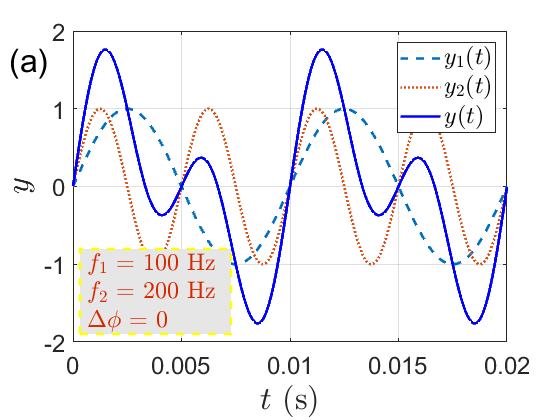}
\includegraphics[width=0.24\textwidth]{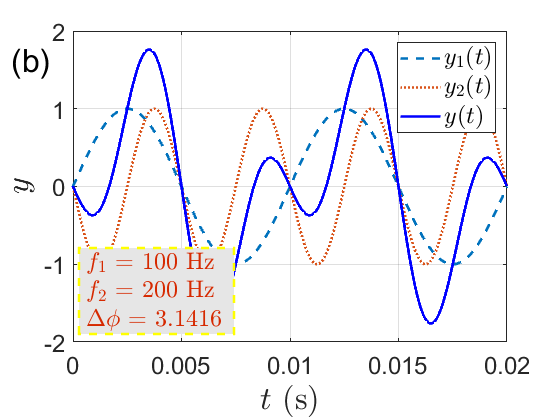}
\includegraphics[width=0.24\textwidth]{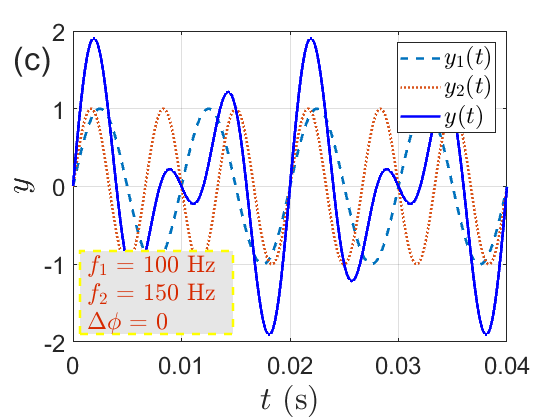}
\includegraphics[width=0.24\textwidth]{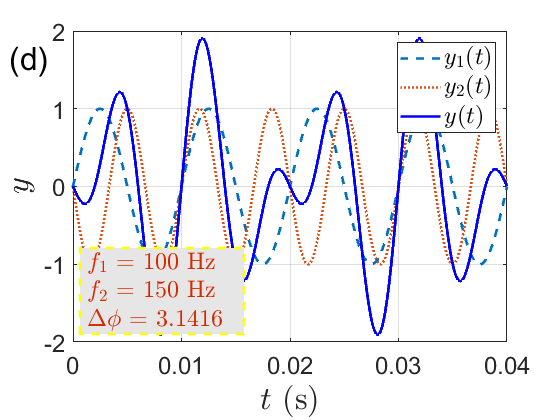}
\vspace{-0.2cm}
\includegraphics[width=0.24\textwidth]{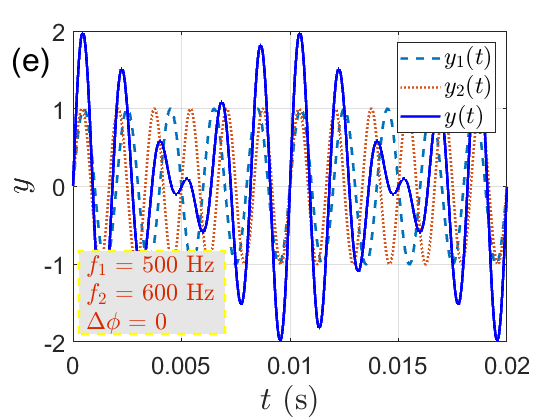}
\includegraphics[width=0.24\textwidth]{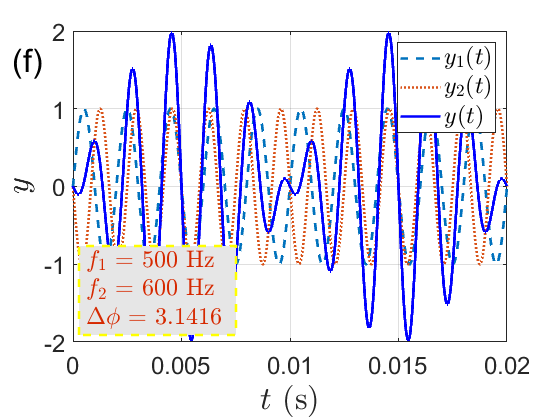}
\includegraphics[width=0.24\textwidth]{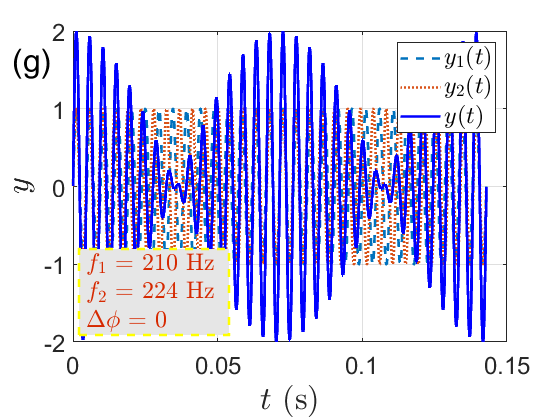}
\includegraphics[width=0.24\textwidth]{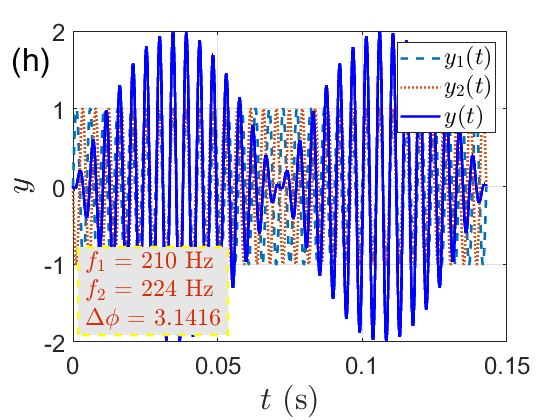}
\caption{Octave, $R=2/1$, example with 
$f_1 = 100$ Hz, $f_2 = 200$ Hz, $F = 150$ Hz:  
(a) $\Delta \phi = 0$, (b) $\Delta \phi = \pi$.
Fifth, $R=3/2$, example with 
$f_1 =  100$ Hz, $f_2 = 150 $ Hz, $F = 125$ Hz: 
(c) $\Delta \phi = 0$, (d) $\Delta \phi = \pi$.
Minor third, $R=6/5$, example with 
$f_1 = 500$ Hz, $f_2 = 600 $ Hz, $F = 550$ Hz: 
(e) $\Delta \phi = 0$, (f) $\Delta \phi = \pi$.
Minor second, $R=16/15$, example with 
$f_1 = 210$ Hz, $f_2 = 224 $ Hz, $F = 217$ Hz: 
(g) $\Delta \phi = 0$, (h) $\Delta \phi = \pi$.}
\label{fig:100Hz200Hzand100Hz150Hzand500Hz600Hzand210Hz224Hz}
\end{figure*}

\section{Methods}
\label{sec:methods}

\subsection{Stimuli: Simple-tone dyads}
\label{subsec:ditonies}
We define a simple-tone dyad, $y$, as a superposition of two simple (sine) tones, $y_1$ and $y_2$, with frequencies and periods $f_1 = 1/T_1$ and $f_2 = 1/T_2$, respectively, i.e.,
\begin{eqnarray}
	& y_1(t) = A_1 \sin(2 \pi f_1 t + \phi_1), \\
	& y_2(t) = A_2 \sin(2 \pi f_2 t + \phi_2), \\
	& y(t) = y_1(t) + y_2(t),
	\label{eq:ditonia}
\end{eqnarray}
where $t$ is time, $\phi_1$, $\phi_2$ are initial phases, $A_1$, $A_2$ amplitudes. In case of equal frequencies, amplitudes and initial phases, the composed wave has double amplitude than the components and hence quadruplicate loudness. In case of equal frequencies and amplitudes but with initial phase difference $\pi$, the composed wave has zero amplitude. We call $R$ the frequency ratio, $F$ the mean frequency, and $\Delta \phi$ the phase difference, i.e., 
\begin{equation}
	R = \frac{f_2}{f_1}, \quad F = \frac{f_1+f_2}{2}, \quad \Delta \phi = \phi_2 - \phi_1.
	\label{eq:RFDeltaphi}
\end{equation}
In Fig. \ref{fig:100Hz200Hzand100Hz150Hzand500Hz600Hzand210Hz224Hz}
we give examples of characteristic superpositions of simple-tone dyads. 
An octave, $R=2/1$, is shown in (a) and (b). During one period, the waveform of the first (second) tone makes one (two) complete movement(s). The composed waveform is periodic with period, $T = 1 T_1 = 2  T_2$. 
A pure fifth, $R=3/2$, is illustrated in (c) and (d).  During one period the waveform of the first (second) tone makes two (three) complete movements. The composed waveform is periodic with period, $T = 2 T_1 = 3 T_2$. 
A minor third, $R=6/5$, is shown in (e) and (f). During one period the waveform of the first (second) tone makes five (six) complete movements. The composed waveform is periodic with period, $T = 5 T_1 = 6  T_2$. 
Finally, a minor second, $R=16/15$, is depicted in (g) and (h). During one period the waveform of the first (second) tone makes fifteen (sixteen) complete movements. The composed waveform is periodic with period, $T = 15 T_1 = 16  T_2$. 
We observe the increasing complexity of the composed waveform from the octave ($R=2/1$) to the minor second ($R=16/15$) where a beat is formed.
A complete list of simple-tone dyads used in our experiment is shown in 
Table~\ref{Table:ListDitonies} in the Appendix. 
In our experiment, the simple-tone dyads were constructed with software \textit{Audacity}. Our simple tones, constituting a dyad, do not correspond to usual notes, but we keep the ratio of frequencies. For example, to construct an octave we do not take 220 Hz (A3) and 440 Hz (A4) but 200 Hz and 400 Hz.

\begin{table*}[htb]
\centering
\footnotesize{
\begin{tabular}{|c|c|c|c|c|c|c|c|c|} \hline
interval & $R$       & cents & $R$  & cents & $R$ & cents \\ 
& (natural) &  (natural) & (equal) & (equal) & (Pythagorean) & (Pythagorean) \\ \hline
2m & $16/15 = 1.0\overline{6}$ & 111.73129 & 1.05946 & 100 &$256/243 = 2^8 / 3^5 \approx 1.053$ 
& 90.225 \\ \hline
2M1& $10/9 = 1.\overline{1}$ & 182.40371 & & &  & \\ \hline
2M2& $9/8 = 1.125$   & 203.91    & 1.12246 & 200 & $9/8 =
3^2 / 2^3 = 1.125$  & 203.91 \\ \hline
2M3& $8/7   = 1.\overline{142857} $ & 231.17409 & & & & \\ \hline
3m & $6/5 = 1.2$     & 315.64129 & 1.18921 & 300 & 
$ 32/27 = 2^5/3^3 = 1.\overline{185}$ & 294.135 \\ \hline
3M & $5/4 = 1.25$  & 386.31371 & 1.25992 & 400 & $81/64 = 3^4 / 2^6 = 1.265625$ & 407.82 \\ \hline
4  & $ 4/3  = 1.\overline{3}$ & 498.045   & 1.33484 & 500 & $ 4/3 = 2^2 / 3^1 = 1.\overline{3}$ & 498.045 \\ \hline
5d &                         &           &         &     &
$ 1024/729 = 2^{10} / 3^6 \approx 1.405$ 
& 588.26999 \\ \hline
tritone & $7/5 = 1.4 $   & 582.51219 & 1.41421 & 600 & & \\ \hline
4a & & & & & $ 729/512 = 3^6 / 2^9 = 1.423828125$ & 611.73001 \\ \hline
5  & $3/2 = 1.5 $   & 701.955   & 1.49831 & 700 & $3/2 = 3^1 / 2^1 = 1.5 $    & 701.955 \\ \hline
6m & $8/5 = 1.6 $     & 813.68629 & 1.5874  & 800 & 
$ 128/81 = 2^7 / 3^4 \approx 1.580$ 
& 792.18 \\ \hline
6M & $5/3 = 1.\overline{6}$ & 884.35871 & 1.68179 & 900 &
$ 27/16 = 3^3 / 2^4 = 1.6875 $ & 905.865 \\ \hline
7m1& $7/4 = 1.75 $   & 968.82591 & & & & \\ \hline
7m2& $16/9 = 1.\overline{7}$ & 996.09 & 1.7818  &1000 &
$ 16/9 = 2^4 / 3^2 = 1.\overline{7} $ & 996.09 \\ \hline
7m3& $9/5 = 1.8 $   & 1017.59629& & & & \\ \hline 
7M & $15/8 = 1.875$ & 1088.26871& 1.88775 &1100 & 
$ 243/128 = 3^5 / 2^7 = 1.8984375$ & 1109.775 \\ \hline
8  & $2/1  = 2$       & 1200      & 2 /1      &1200 & 
$2 /1  = 2^1/ 3^0$ =2  & 1200 \\ \hline
\end{tabular}}
\caption{Intervals for various types. M means major, m minor, d diminished, a augmented, and pure intervals are shown without additional symbol. 
}
\label{table:Intervals}
\end{table*}

\begin{figure}[ht]
\centering
\includegraphics[width=0.5\textwidth]{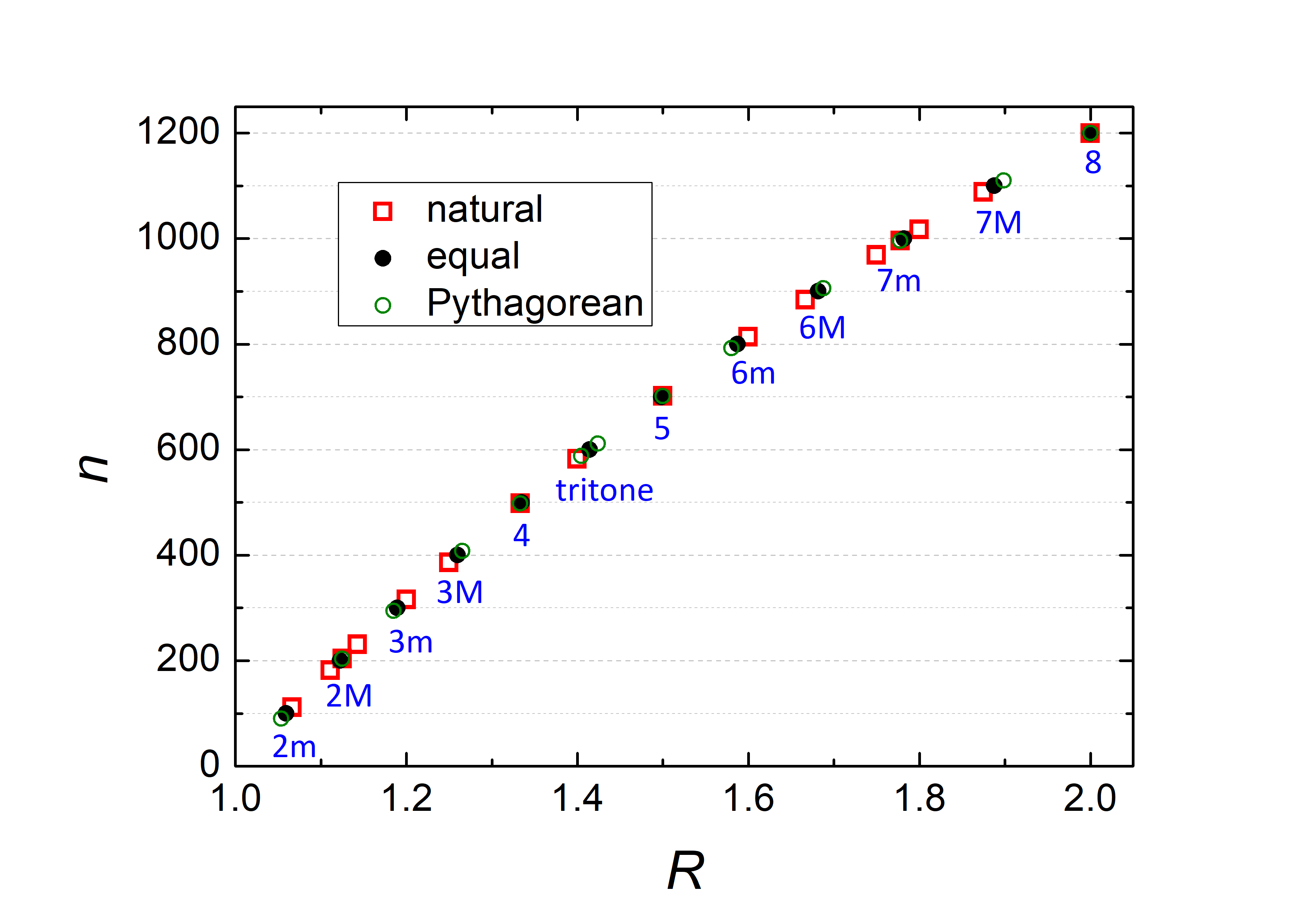} 
\vspace{-10pt} 
\caption{Natural, equal and Pythagorean intervals in cents ($n$) vs. frequency ratio ($R$). We used natural intervals.}
\label{fig:CentsVSRatio}
\end{figure}

Our simple-tone dyads correspond to \textit{natural intervals}: $R$ is a fraction of natural numbers. We list below our natural intervals together with relevant equal temperament and Pythagorean intervals.
An equal temperament interval is defined with the help of cents.
An octave has 1200 cents and is divided according to the equal temperament into 12 semitones of a 100 cents each.
The number of cents $n$ is given by 
$	n = 1200 \; \log_2 (f_2/f_1)$.
A Pythagorean interval~\cite{ BenwardSaker:2003,Benson:2003} has $R$ equal to an \textit{integer} (positive or negative) power of 2 divided by another \textit{integer} power of 3. For example, 
the octave with $ R = 2 = 2^1 / 3^0 $,
the fifth  with $ R = 3/2 = 3^1 / 2^1$, and 
the fourth with $ R = 4/3 = 2^2 / 3^1$ 
are Pythagorean intervals.
These Pythagorean intervals coincide with the corresponding natural intervals and equal temperament intervals. However, other Pythagorean intervals do not exactly coincide. For example, the minor second, 2m, for which we take as natural interval as having $R = 16/15 = 1.0\overline{6}$, corresponds to Pythagorean interval having the closest fraction of the form $2^k / 3^l$, that is $256/243 = 2^8 / 3^5 \approx 1.053$.
Table~\ref{table:Intervals} and
Fig.~\ref{fig:CentsVSRatio} illustrate the natural, equal and Pythagorean intervals in terms of cents ($n$) vs. frequency ratio ($R$).

\subsection{Procedure}
\label{subsec:Procedure}
\begin{figure}[htb]
\centering
\includegraphics[width=0.5\textwidth]{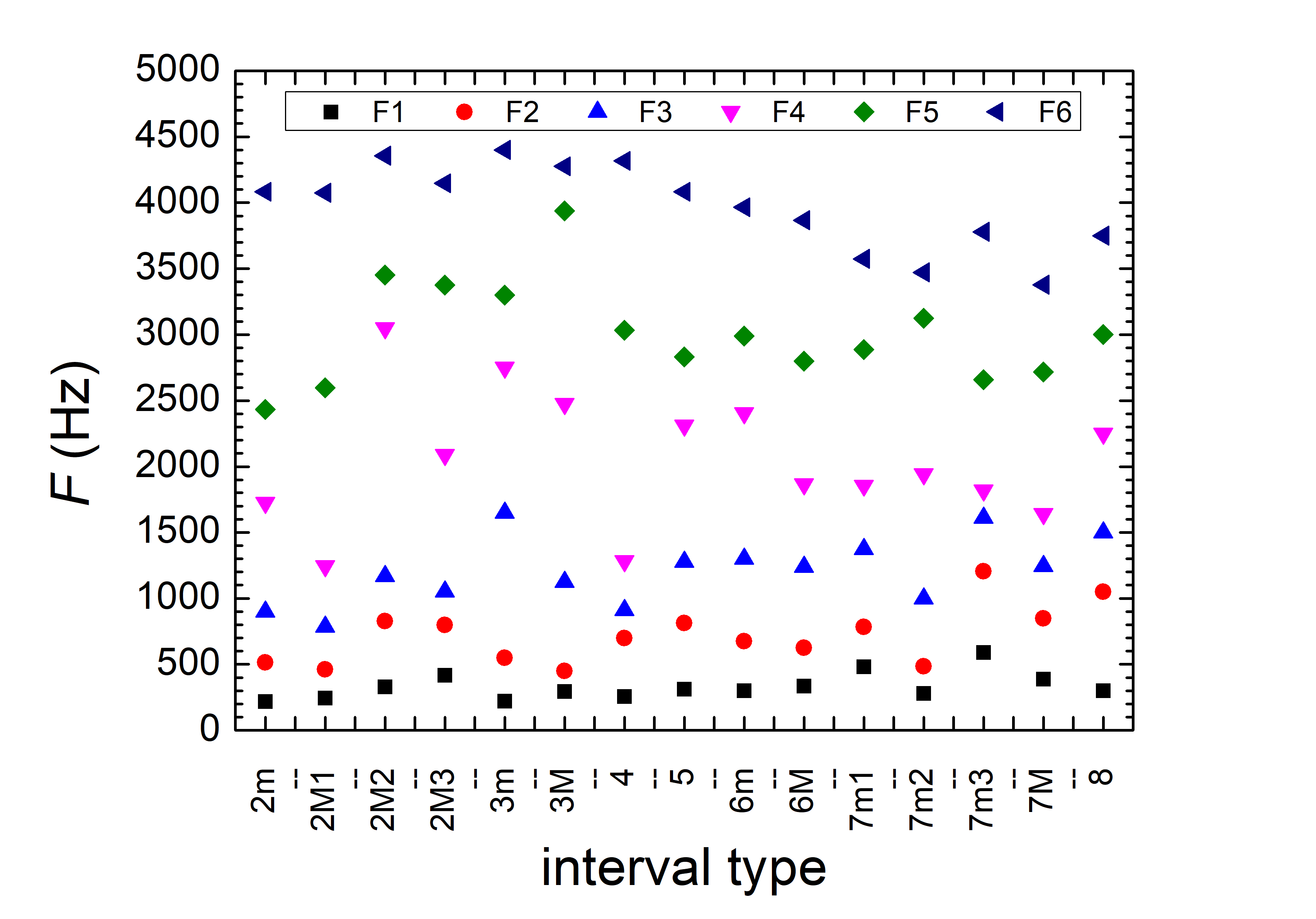}
\vspace{-10pt} 
\caption{Mean frequencies of the simple-tone dyads for each type of interval used in our experiment.}
\label{fig:F}
\end{figure}

To present the set of stimuli, we employed the open software \textit{PsychoPy3}, written in Python. We used headphones, to reduce outer noise. The experiment was conducted in two quiet rooms at the Departments of Physics and Musical Studies of the National and Kapodistrian University of Athens. The experiment had a duration of $\approx$ 5-10 minutes. Each subject listened to 90 simple-tone dyads, in random order. 
In Table~\ref{Table:ListDitonies}, in the Appendix, 
we present the full list of simple-tone dyads, created by the software \textit{Audacity}. We show the name of musical interval (the number of semitones), a codename, the ratio of frequencies, $R = \frac{f_2}{f_1}$, the mean frequency,  $F=\frac{f_1+f_2}{2}$, and the couple of frequencies. Tritone (6 semitones) with $R = 7/5 = 1.4$, was not included in the experiment. For cases of 2nd major and 7th minor we have included three different $R$ fractions. Fig.~\ref{fig:F} shows the mean frequencies $F$ of these simple-tone dyads for each type of interval. Looking now at Fig.~\ref{fig:F} we may think that we could use equal $F$ values for all types of intervals, but maybe this way we would have investigated only six specific $F$ values. 

For each subject a new random sequence was produced. Subjects had to answer whether they liked or not each simple-tone dyad by pressing two buttons in the keyboard, \textit{Yes} ($\rightarrow$) or \textit{No} ($\leftarrow$). 
A binary like/dislike format was chosen to reflect intuitive, categorical judgments of dyadic pleasantness, which align with real-world listening behavior and reduce cognitive load in large stimulus sets. While continuous scales allow finer gradations, binary responses can reduce intra-subject variability and simplify interpretation, especially in studies emphasizing group-level patterns.
During the experiment, the listener heard a simple-tone dyad of duration 2s and then had as much time as necessary to decide whether they liked it or not and press the relevant button. By pressing the button, the program continues to the next dyad, and again the listener had time to evaluate it as pleasant or unpleasant. This way the experiment continued until the list of all dyads had been presented. With the last response, the experiment ends. The responses were automatically registered in a spreadsheet, together with  relevant information, e.g. response time, the time the first \textit{Yes} or \textit{No} was recorded, etc. 
A presentation of responses, $\rho$, of a random subject, for all simple-tone dyads, i.e., for each $F$ category (1 $\to$ 6 corresponds to increasing frequency) and for each $R$ case, is given in 
Fig.~\ref{fig:SampleGraph} in the Appendix. 
Participants listened to the stimuli through headphones connected to a calibrated computer system and were instructed to adjust the playback volume to a comfortable level before beginning the experiment. Although absolute sound pressure levels (SPL) were not precisely measured, all participants were tested in the same controlled laboratory setting, using identical equipment in a quiet environment. This ensured consistent relative stimulus levels and stable listening conditions across participants.

\subsection{Participants}
\label{subsec:Participants}
The set of subjects consists of 70 persons. Subjects originate from different environments (students and professors from the Department of Physics, School of Natural Sciences and the Department of Musical Studies, Philosophical School, National and Kapodistrian University of Athens, the authors' friends and relatives as well as persons who responded to our call in social media). The set of subjects can be divided to 39 women and 31 men. Their age is between 19 and 86 years. 
Of the 70 participants, 47 reported no formal musical training or significant engagement with music, whereas 23 were either professional musicians or had played a musical instrument for at least seven years.
All participants were from Greece and primarily exposed to Western musical traditions. While this study centers on Western interval classes, future work should examine cross-cultural variation in consonance judgments. Greek popular and traditional music includes harmonic structures that sometimes differ from Western classical norms, which may have subtly influenced listeners’ responses—though all stimuli used culturally neutral sine tones.

\textit{The Ethics Committee of the National and Kapodistrian University of Athens approved the experiments (decision 134/2024, 04-03-2024, protocol number 19930). All experiments were performed in accordance with relevant guidelines and regulations. A signed consent was obtained from all participants.}


\subsection{Data Analysis: Response, preference, statistics}
\label{subsec:ResponsePreferenceStatistics}
Subjects responded with Yes or No that was translated by us to 1 or 0, respectively. These are the \textit{responses}, $\rho$. We defined the \textit{preferences}, $P$, as mean values of the responses. Responses and preferences were analyzed. We present these mean values and mean errors (standard deviations / square root of number of measurements). For the total set of 70 persons, we created relevant $P$ diagrams. When we compared between two groups of subjects (e.g. musicians vs. non-musicians, men vs. women, or age groups), in addition to the $P$ diagrams, we compared the two groups via statistical methods. Our samples are independent, hence, we had to employ an adequate criterion. To check whether these mean values were extracted from the normal distribution or not, we performed the \textit{Shapiro-Wilk test} \cite{ShapiroWilk:1965}. It occurred that  preferences are not extracted from the normal distribution, a basic prerequisite to use a parametric method like the \textit{t-test} \cite{Student:1908}.
Therefore, we employed the Mann-Whitney (MW) test \cite{MannWhitney:1947}, a non parametric rank sum criterion for comparison between two independent samples, where normality is not a prerequisite. 
The Mann–Whitney test is not identical to the Wilcoxon test~\cite{Wilcoxon:1945}, although both are non-parametric and involve summation of ranks. The Mann–Whitney test refers to independent samples; the Wilcoxon test refers to dependent samples. 
The MW criterion is reliable for samples with population, $\nu>6$. Hence, in some cases, for preferences, we had to construct larger groups to increase the population well above $\nu=6$. These Larger Groups were constructed based on the simplicity of ratio $R$, cf. Table~\ref{Table:LargerGroupsDitonies}. Fourths, fifths and octaves with irreducible $R$ of small nominator and denominator belong to the 3rd Larger Group. Seconds with irreducible $R$ of large nominator and denominator belong to the 1st Larger Group. Sevenths constitute the 4th Larger Group. Intermediate cases, thirds and sixths, belong to the 2nd Larger Group. The standard $p$-value~\cite{Fisher:1992} is usually considered 0.05. For the MW text we used different implementations, a home-made Excel procedure written by us and a free code in MATLAB written by G. Cardillo~\cite{Cardillo:2023}.

We realized that approximatively, preferences fall, increasing pitch. Although these falls of preference as a function of mean frequency, $P(F)$, are not linear, to have a crude idea of how steep the falls are, we performed linear fits of $P(F)=AF+B$. We estimated slope $A$ and its error $\delta A$, intercept $B$ and its error $\delta B$ as well as adjusted R squared $\mathcal{R}^2$ to evaluate the quality of fits.
Finally, we calculated (a) $\overline{P}(R)$, i.e., the mean over all $F$ categories for each $R$ case, and (b) $\overline{P}(F)$, i.e., the mean over all $R$ cases for each $F$ category.

\begin{table}[htb]
\centering
\begin{tabular}{|c|c|c|} \hline
Larger Group & $R$ cases & $R$  \\ \hline
1st & 2m, 2M1, 2M2, 2M3 & 16/15, 10/9, 9/8, 8/7 \\ \hline
2nd & 3m, 3M, 6m, 6M    & 6/5, 5/4, 8/5, 5/3 \\ \hline
3rd & 4, 5, 8 & 4/3, 3/2, 2/1 \\ \hline
4th & 7m1, 7m2, 7m3, 7M & 7/4, 16/9, 9/5, 15/8 \\ \hline
\end{tabular}
\vspace{0.3cm}
\caption{Simple-tone dyads larger groups for the preferences MW test.}
\label{Table:LargerGroupsDitonies}
\end{table}

\section{Results and Discussion}
\label{sec:ResultsDiscussion}

\subsection{Overall Preference Patterns}
\label{subsec:TotalSet}
\begin{figure*}[t!]
	\centering
	\includegraphics[width=0.32\textwidth]{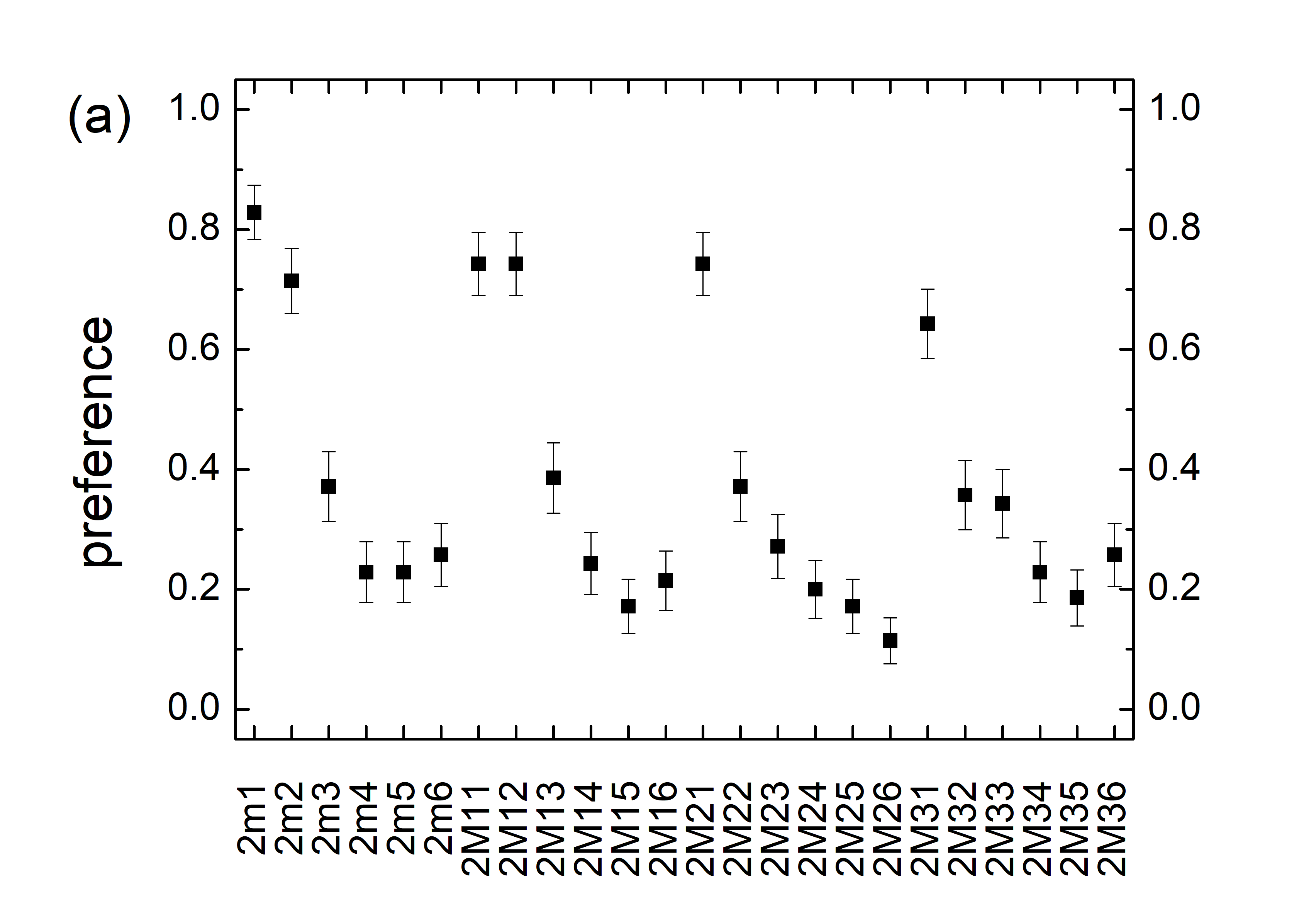}
	\includegraphics[width=0.32\textwidth]{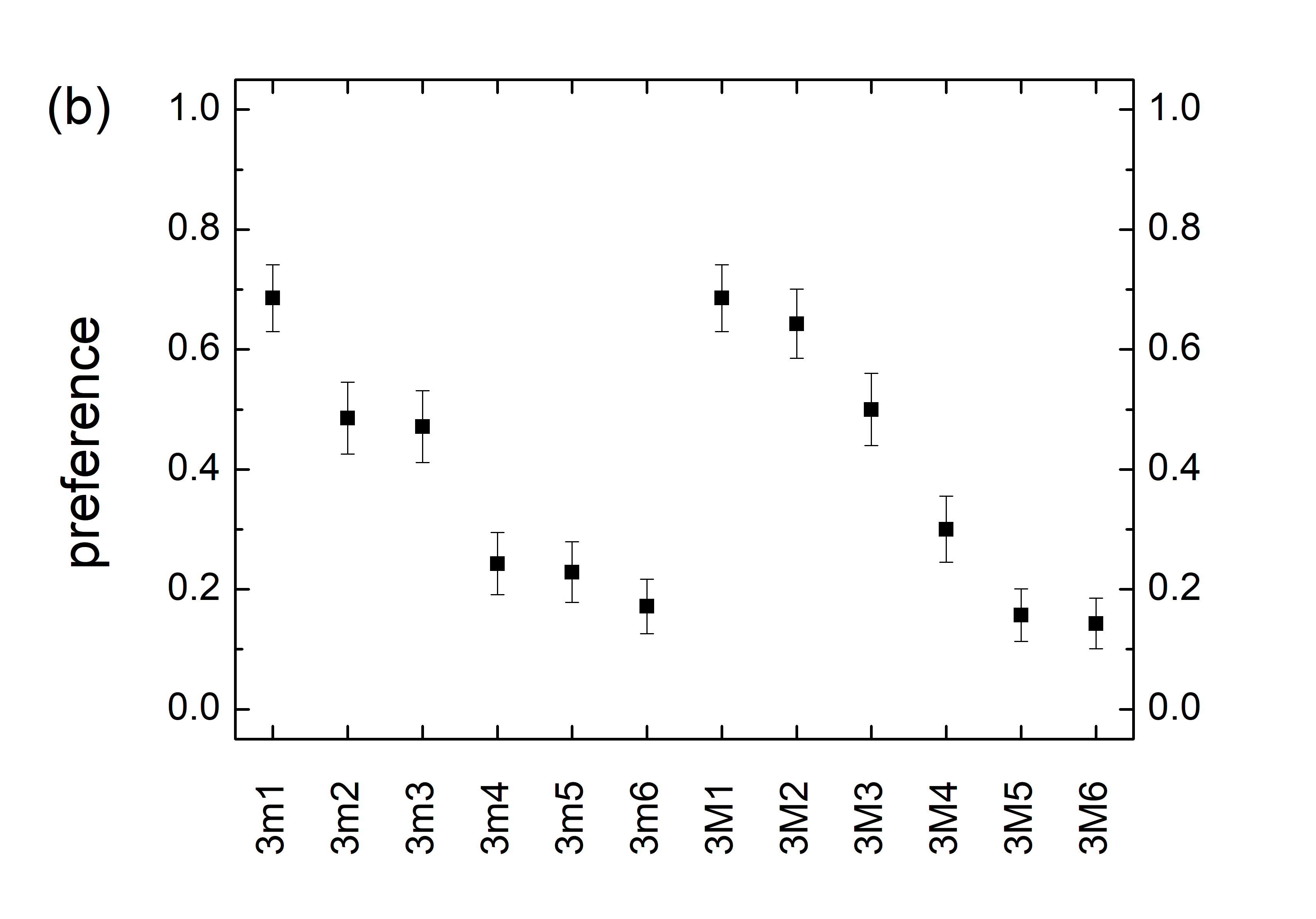}
	\includegraphics[width=0.32\textwidth]{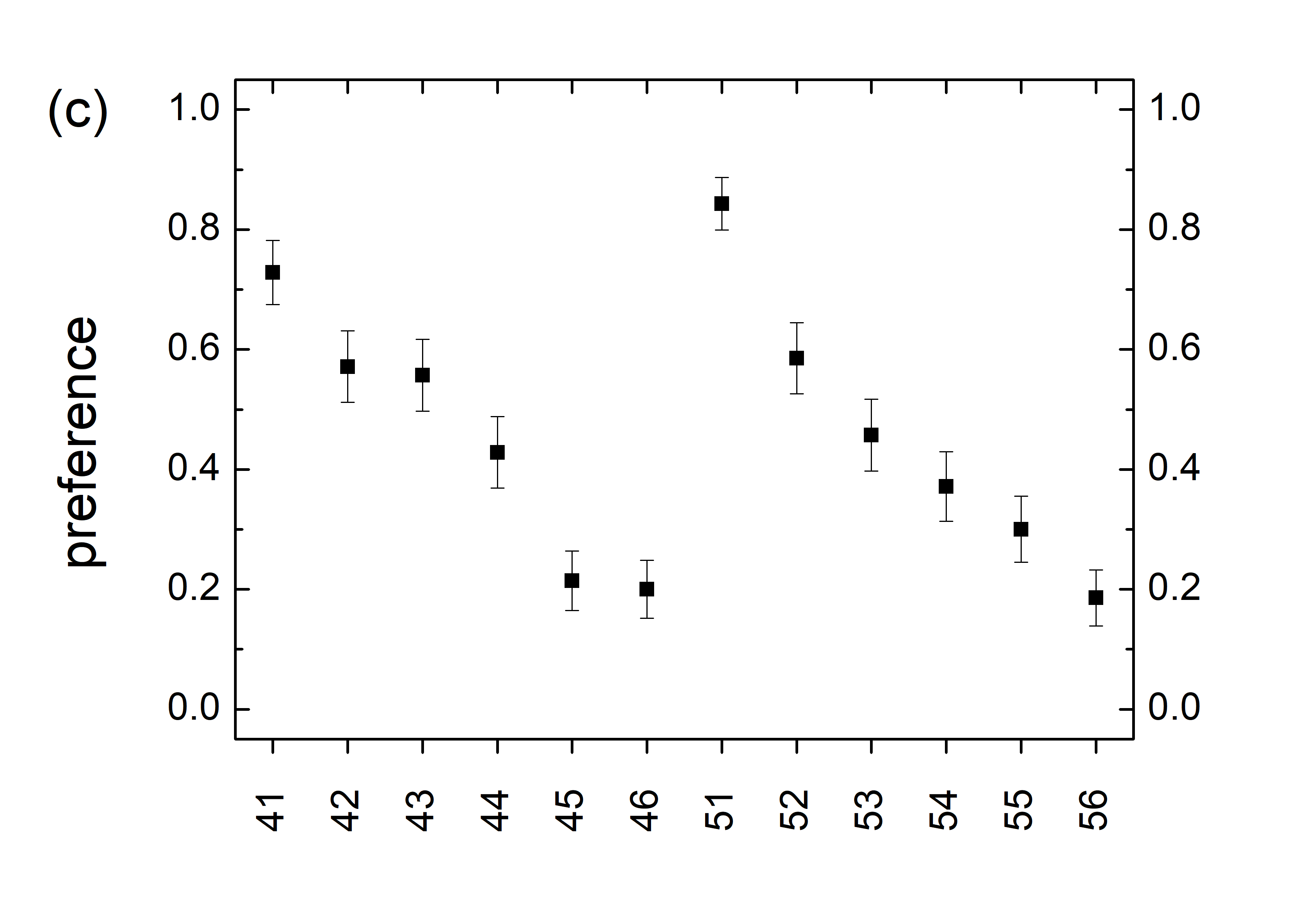}\\
	\vspace{-0.2cm}
	\includegraphics[width=0.32\textwidth]{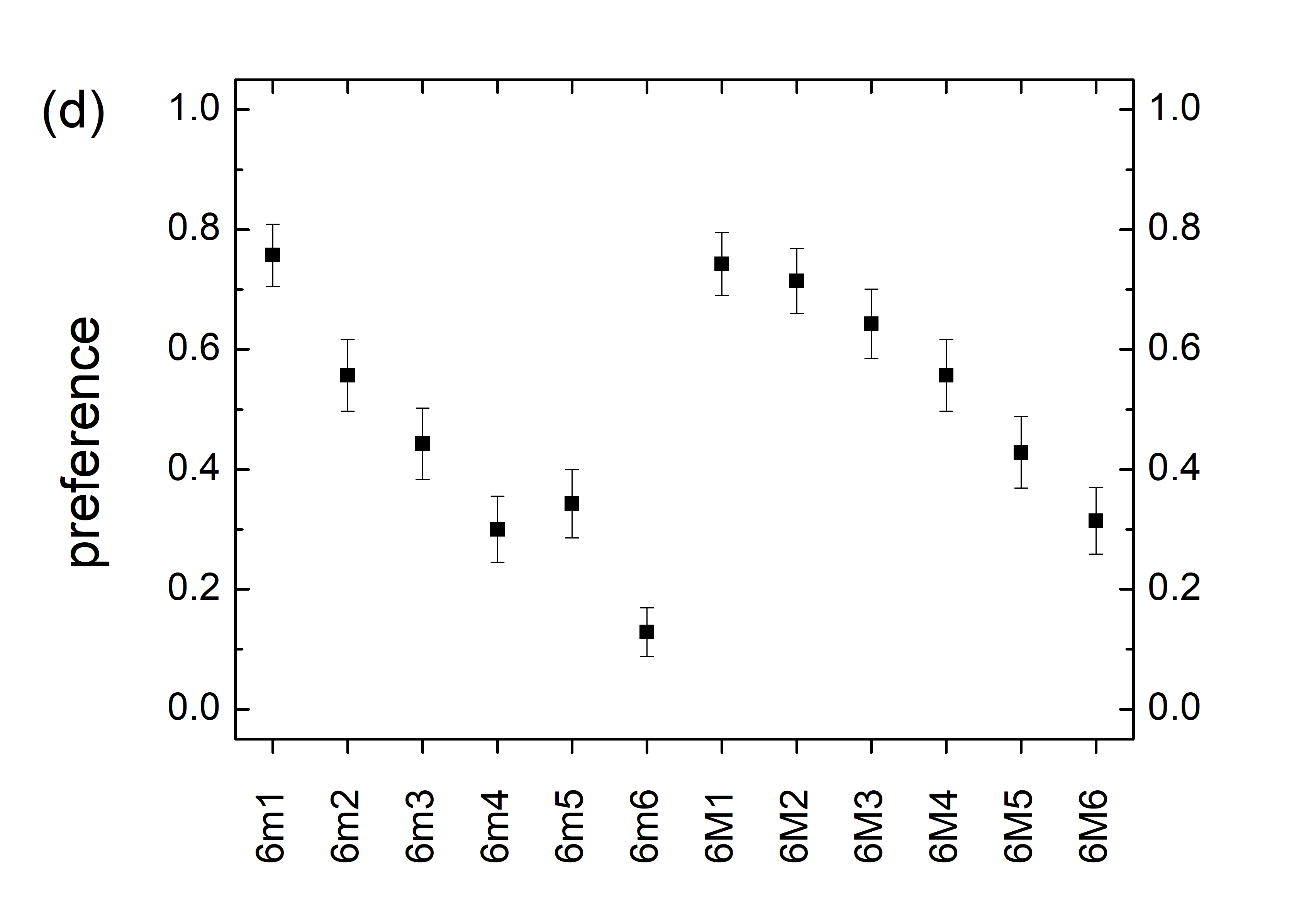}
	\includegraphics[width=0.32\textwidth]{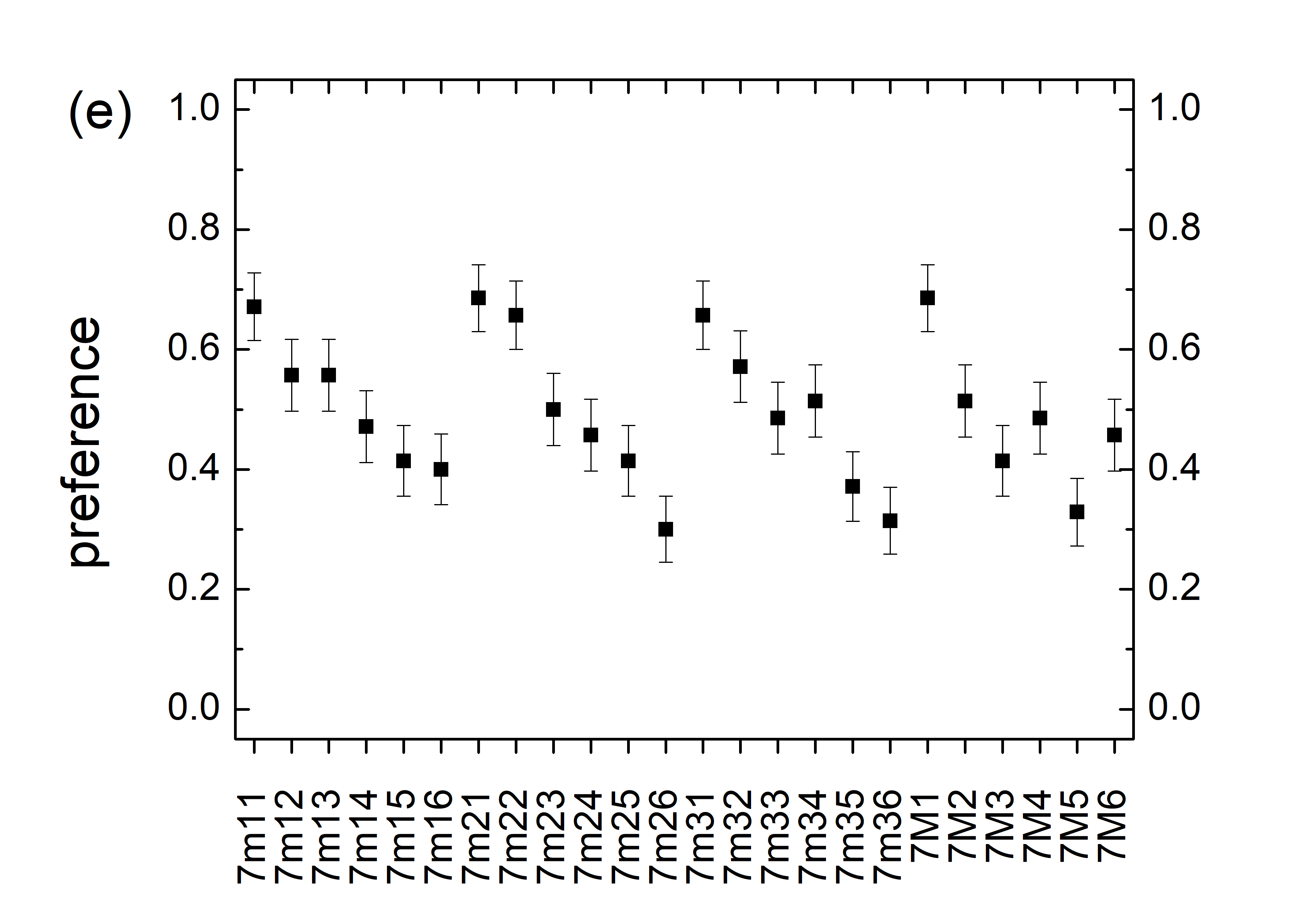}
	\includegraphics[width=0.32\textwidth]{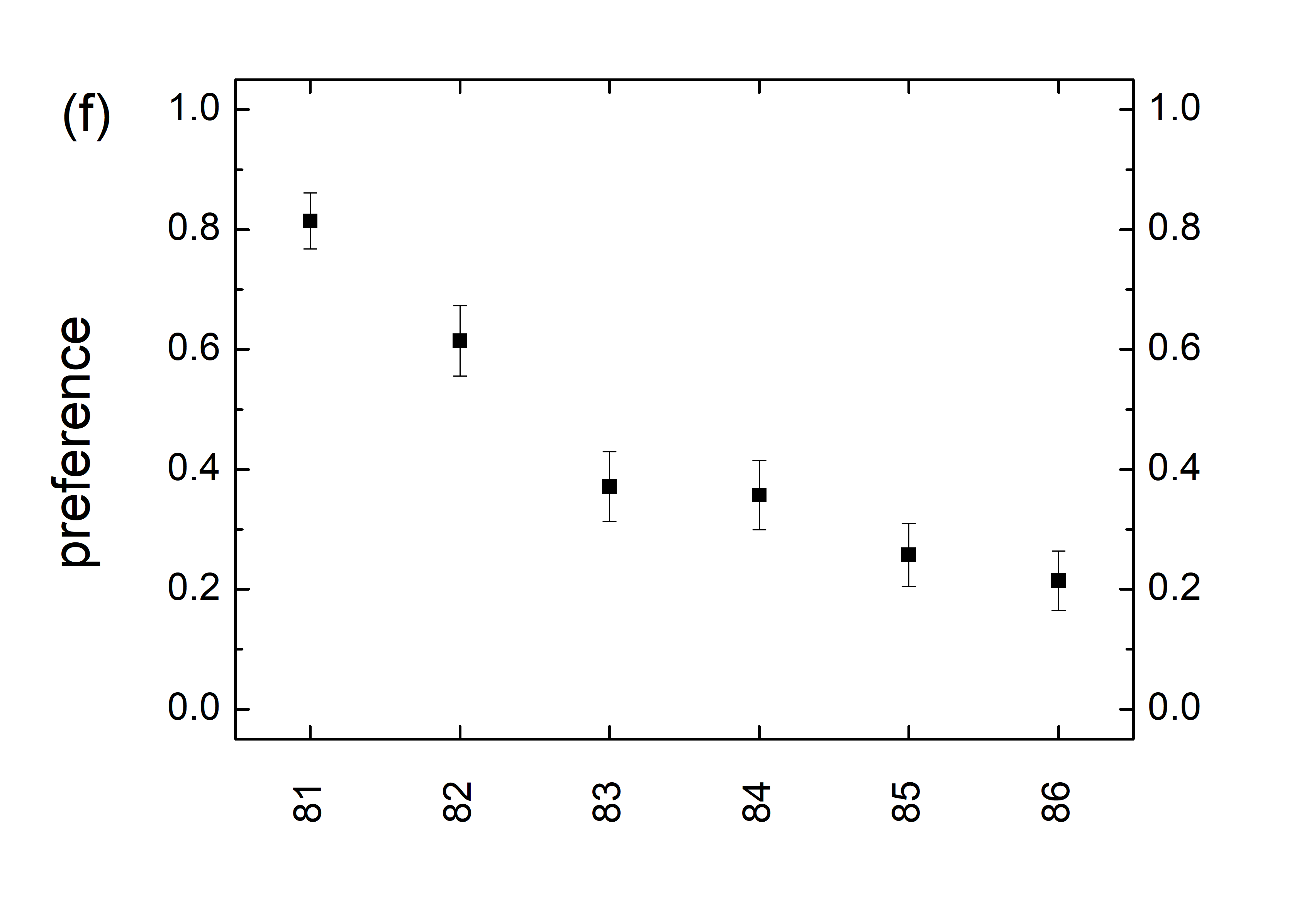}
	\caption{Mean values and errors of responses or preference, $P$, for the total set of subjects (70 persons), for each simple-tone dyad: (a) seconds, (b) thirds, (c) fourths and fifths, (d) sixths, (e) sevenths, (f) octaves.}
	\label{fig:meanvaluestotalset}
\end{figure*}

In Fig.~\ref{fig:meanvaluestotalset} we depict preference, $P$, diagrams (mean values and mean errors) for the total set of subjects, for each $R$ case. 
The general trend is a decrease in $P$ as $F$ increases.
The dyads with larger $P$ are mainly those of $F$ category 1 (the dyad of lower $F$ for each $R$ case) and then $F$ category 2 follows. $F$ category 1 for all $R$ cases has the highest $P$. For fifths ($R=3/2$), octaves ($R=2/1$) and for 2nd minors ($R=16/15$), for $F$ category 1, $P$ is above 0.8. Then follow fourths ($R=4/3$), sixths [majors ($R=5/3$) and minors ($R=8/5$)], where the $F$ category 1 has $P$ between 0.7 and 0.8. In other words, the majority of listeners have pointed $F$ category 1 as pleasant. 
Unexpectedly, certain dyads such as 2m1, 2M11, and 2M21 exhibit relatively high $P$ values.
For example, 2m1 corresponds to the lowest $F = 217$ Hz ($f_1 = 210$ Hz, $f_2 = 224$ Hz, $\Delta f = 14$ Hz) and has clear beats. However, it is not characterized unpleasant by most of the listeners.
We observe in Fig.~\ref{fig:meanvaluestotalset} that increasing $F$, $P$ show an almost ``linear'' fall, for thirds, fourths, fifths, sixths  and octaves, an effect less pronounced for seconds and sevenths. Increasing $F$, $P$ fall towards 0.2 or 0.1. For seconds and sevenths, $P$ are more dispersed, however $P$ still fall, as a general trend, increasing $F$. 
Overall, seconds appear to be rated as the most unpleasant among the dyads.

\begin{figure}[h]
\centering
\includegraphics[width=0.45\textwidth]{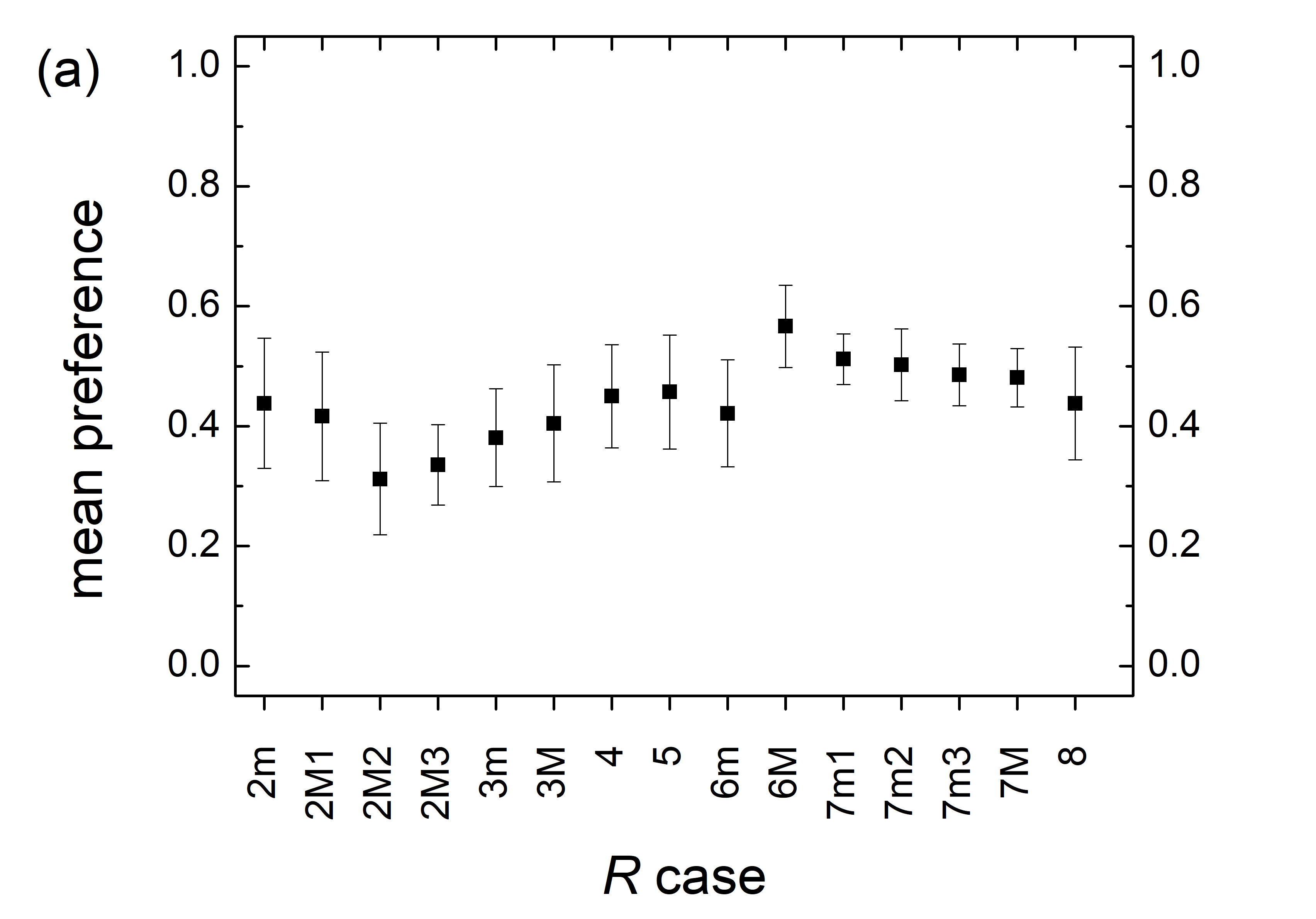} 
\includegraphics[width=0.45\textwidth]{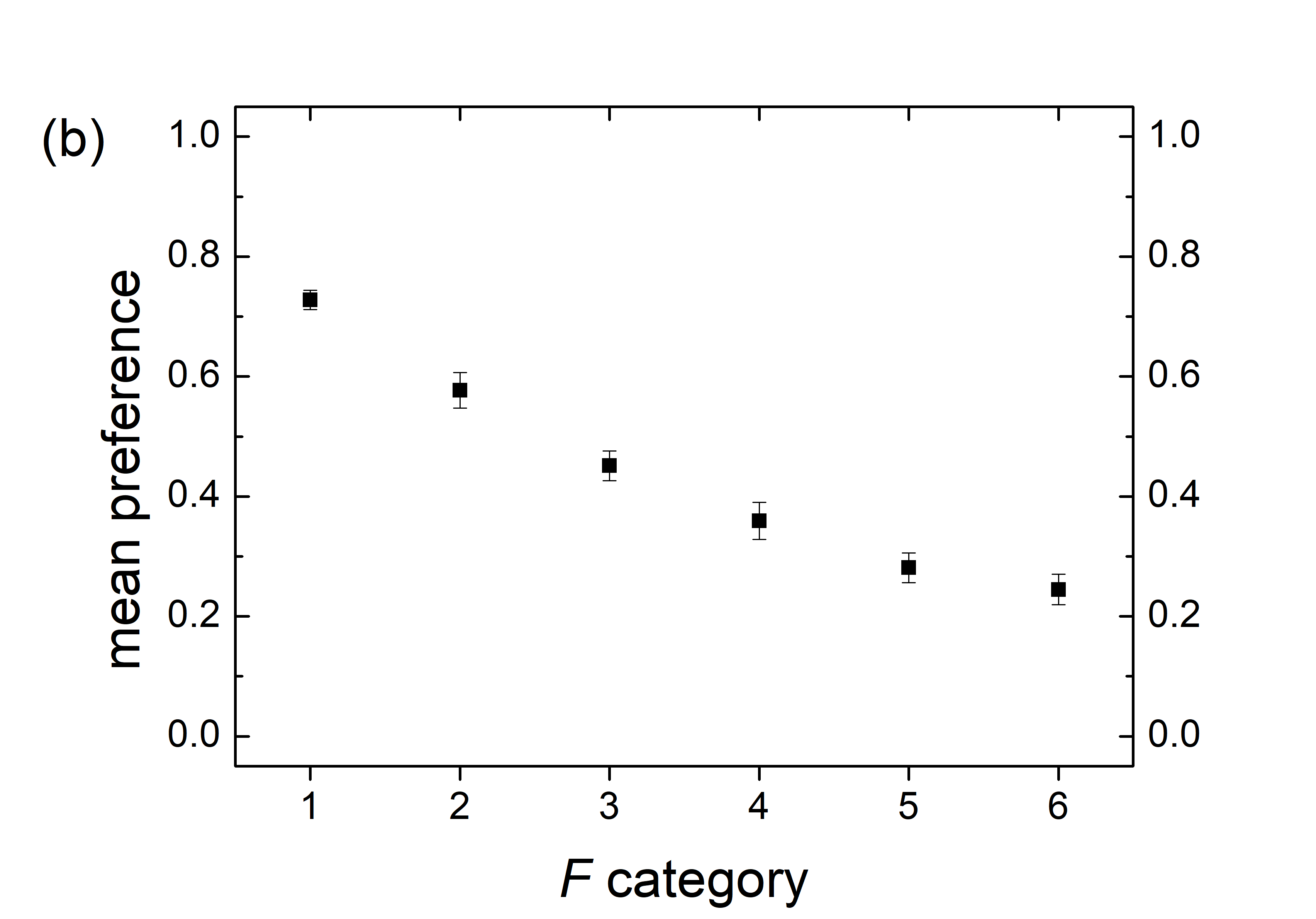} 
\caption{Total set of subjects (70 persons): Mean values and errors of $P$,  (a) Mean over all $F$ categories for each $R$ case, $\overline{P}(R)$. (b) Mean over all $R$ cases for each $F$ category, $\overline{P}(F)$.}
\label{fig:meanvaluestotalsetRF}
\end{figure}

Preference, $P$, is itself a mean value, cf. Subsec.~\ref{subsec:ResponsePreferenceStatistics}. 
In Fig.~\ref{fig:meanvaluestotalsetRF}(a) we present the mean $P$ over all $F$ categories for each $R$ case, $\overline{P}(R)$. The $R$ case with the greater $\overline{P}(R)$ is 6M ($\overline{P} \approx 0.57$), while 7m2 and 7m1 follow. On the other hand, the $R$ cases with the lower $\overline{P}(R)$ are 2M2 and 2M3 ($\overline{P} \approx 0.31$ and $0.34$, respectively). Most $\overline{P}(R)$ have large errors of mean value, of the order of $0.1$ (between $\approx 0.04$ and $0.11$), therefore, the picture is not very clear. In Fig.~\ref{fig:meanvaluestotalsetRF}(b) we present the mean $P$ over all $R$ cases for each $F$ category, $\overline{P}(F)$. Clearly, increasing $F$, $\overline{P}(F)$ falls and the errors of mean values are of the order of $0.01$ 
(between $\approx 0.01$ and $0.03$), almost invisible in that scale, while mean values are of order of $0.5$. The lower $F$ category 1 simple-tone dyads, have much higher $\overline{P}$ ($\approx 0.73$) compared to the rest, e.g., $F$ categories 4, 5, 6 have $\overline{P} < 0.4$.

Terhardt~\cite{Terhardt:1984} conducted experiments presenting dyads consisting of two tones—one fixed at La4 and the other varying gradually up to an octave (La5)~\cite{Terhardt:1984}. Listeners rated these dyads for ``consonance'' and ``roughness''. Concerning consonance, our results agree with the minimum at seconds shown by Terhardt~\cite{Terhardt:1984}, but do not confirm the monotonous increase shown in his idealized curve. In Figure 1 of this famous work~\cite{Terhardt:1984}, an evaluation of ``consonance'' and ``roughness'' of single, isolated dyads as a function of interval width was presented. The presentation, according to Terhardt was ``slightly idealized'' and included data obtained from Plomp and Levelt~\cite{PlompLevelt:1965}, Kameoka and Kyriyagawa~\cite{KameokaKuriyagawa:1969I}-\cite{KameokaKuriyagawa:1969II} and Terhard~\cite{Terhardt:1977}. The interval width was represented by the higher note, whereas the lower was kept at La4 (440 Hz). Terhardt also noted: ``Although only a limited number of dyads was involved (i.e., interval changed in discrete steps), the results are represented by continuous curves. The ``most dissonant'' interval of pure-tone dyads corresponds to the minimum of the solid curve.'' In other words, this was a curve of ``evaluation'' versus the highest note of the dyad, from La4 to La5. This evaluation begins with a value of 1 at La4, which we interpret as either a dyad consisting of two simultaneous La4 notes or a single La4. In any case, Terhardt's curve falls rapidly at minor second and shows a clear minimum at major seconds, as also in our results. But then, Terhardt shows a monotonous increase of ``evaluation'' up to the octave, while, in our results the increase is certainly not monotonous for the general public (this Section~\ref{subsec:TotalSet}, Fig.~\ref{fig:meanvaluestotalsetRF}(a)) and musicians show two local maxima at major sixth and octave, cf. Section~\ref{subsec:MusiciansNonmusicians},
Fig.~\ref{fig:meanvaluesMNMRF}(a).
In the same work~\cite{Terhardt:1984}, when composite tones including harmonics were considered, the interactions became more complex, resulting in perceptual patterns that diverge from those observed with simple tones. These differences likely arise from the harmonic interactions and resulting roughness, highlighting the nuanced nature of consonance perception in complex sounds.
Regarding Terhardt’s Figure 2 where he evaluated ``the width of the most dissonant interval’’, we have not run a similar experiment, but we find that increasing frequency, the feeling of consonance decreases, as shown in Fig.~\ref{fig:meanvaluestotalsetRF}(b) and in all similar figures below concerning comparison of nominal groups.

Inference: In all, for the general public, increased mean frequency ($F$) makes simple-tone dyads less pleasant. For octaves, fifths, fourths, and sixths the feeling of consonance shows an almost ``linear'' fall, increasing $F$. Seconds and sevenths are more dispersed but still increased $F$ increases the feeling of dissonance. Seconds seem the most unpleasant simple-tone dyads.


\subsection{Effects of Musical Training}
\label{subsec:MusiciansNonmusicians}
\begin{figure*}[htb]
\centering 
\includegraphics[width=0.32\textwidth]{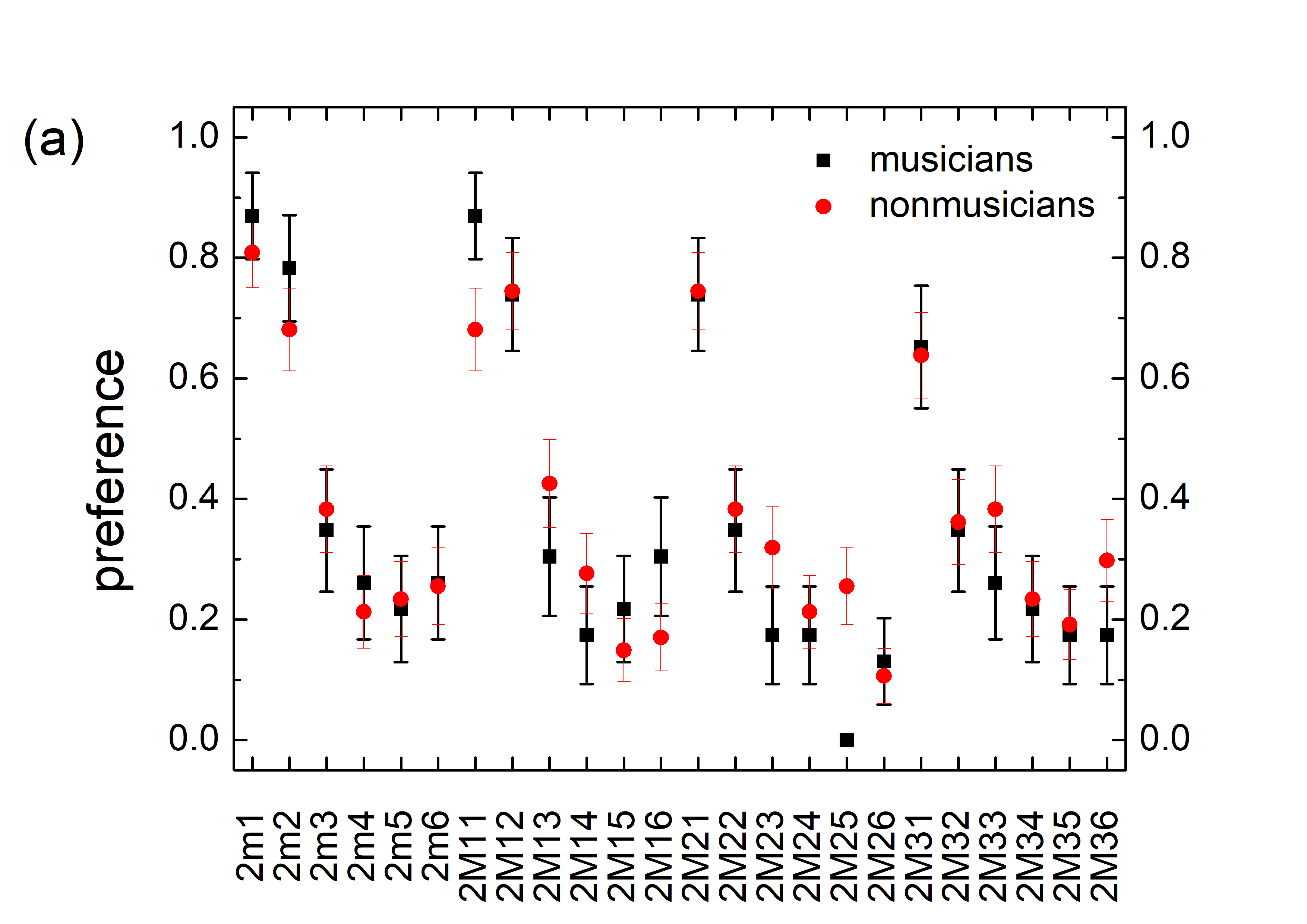}
\includegraphics[width=0.32\textwidth]{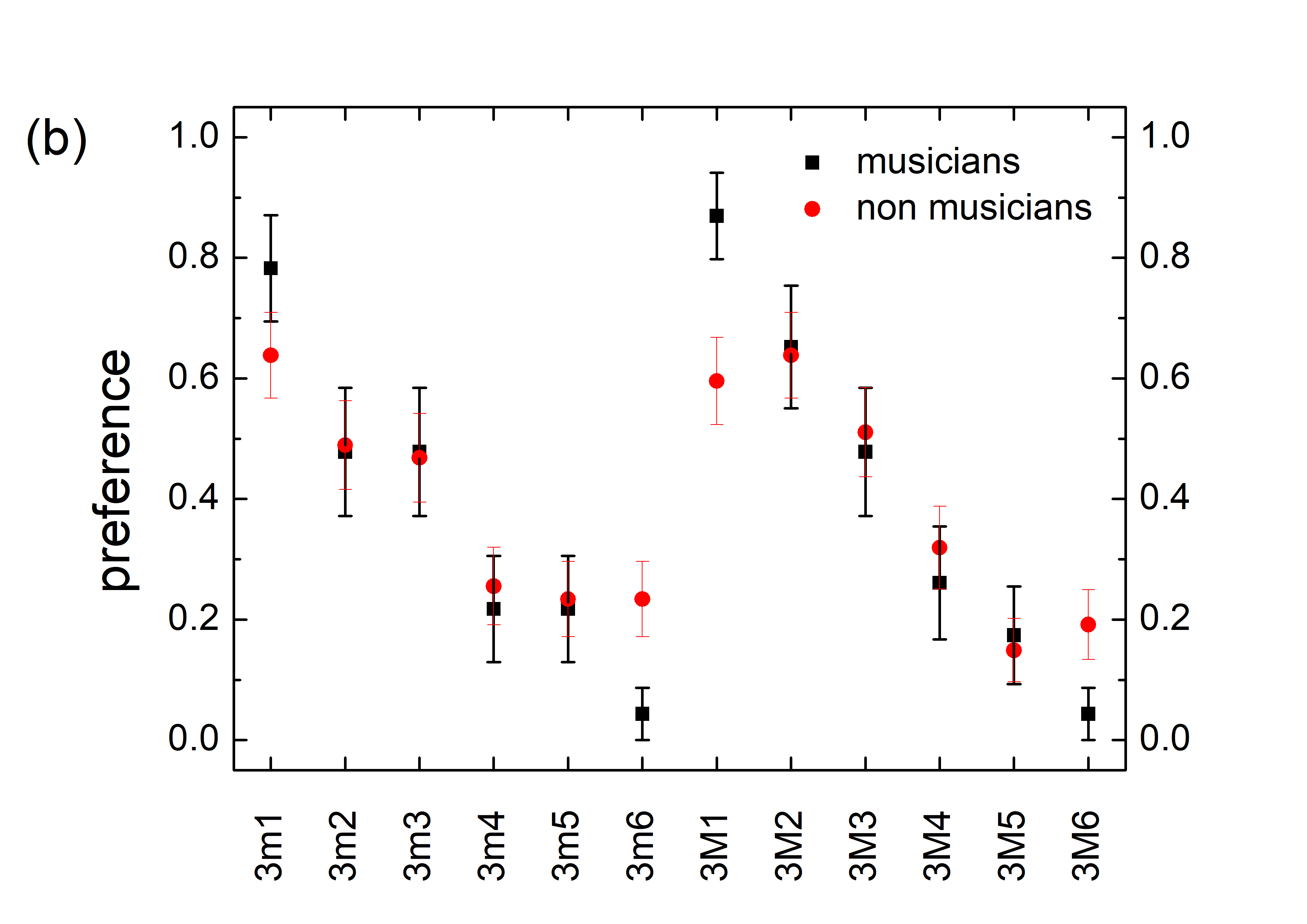} 
\includegraphics[width=0.32\textwidth]{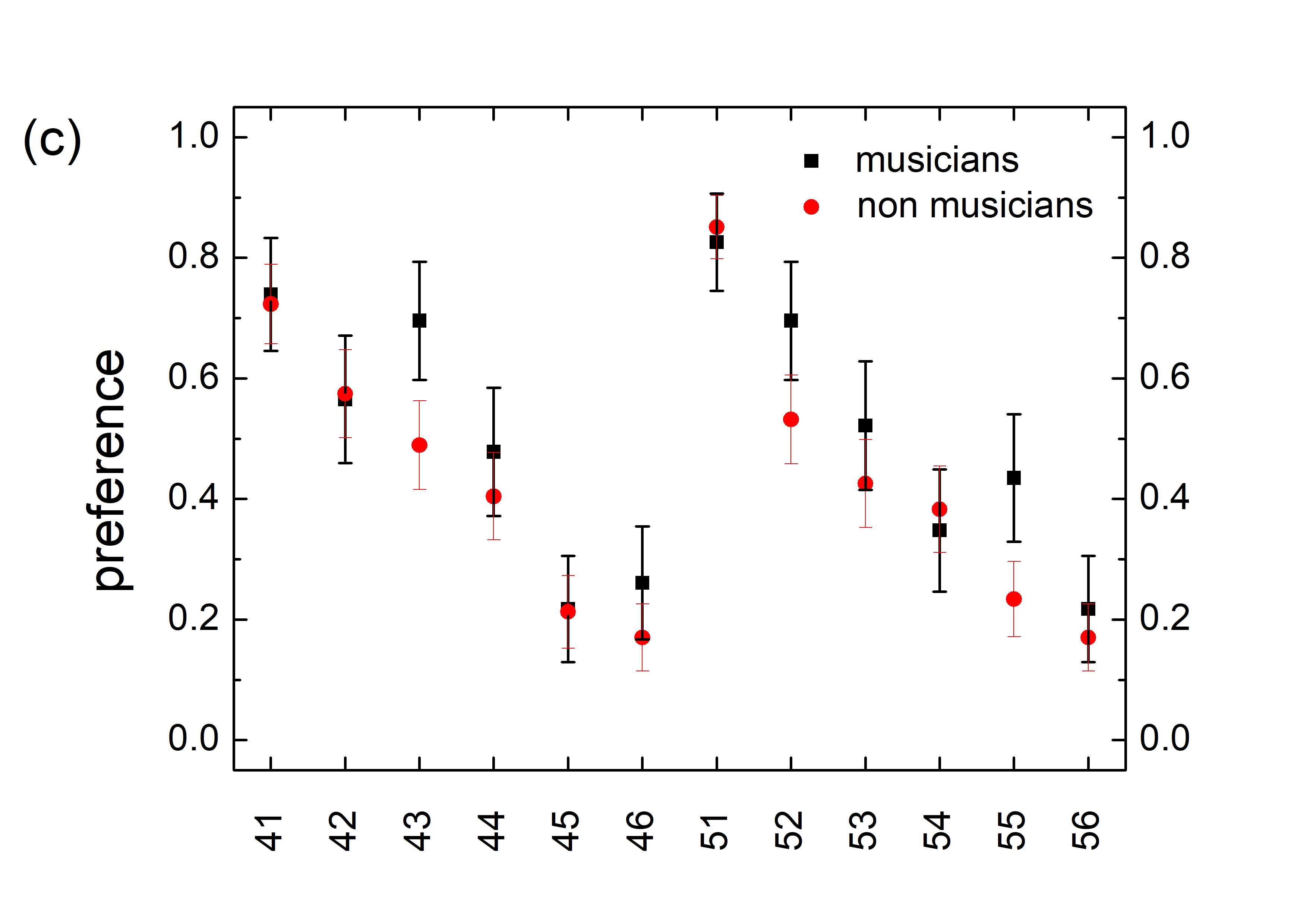}\\
\vspace{-0.2cm} 
\includegraphics[width=0.32\textwidth]{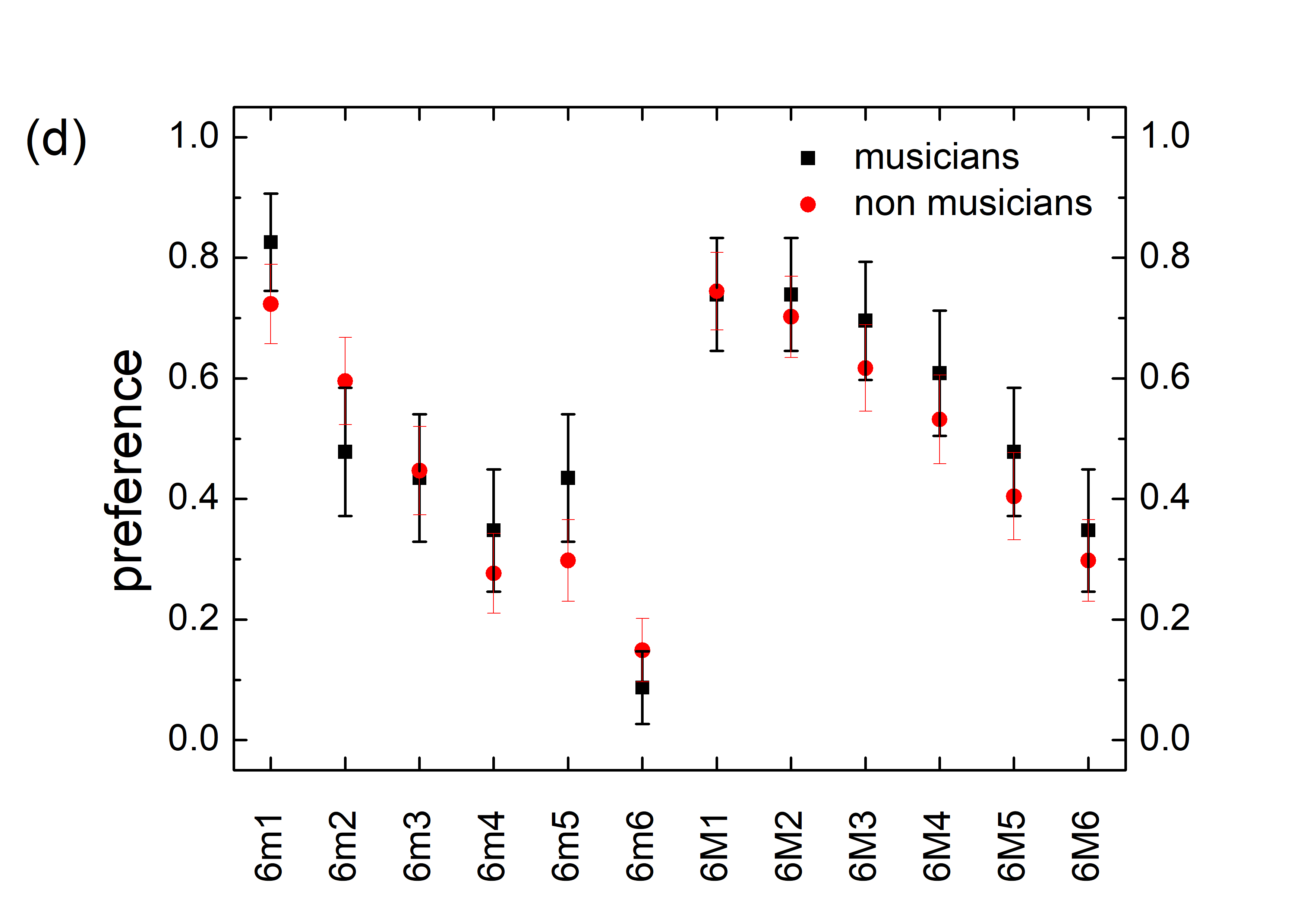} 
\includegraphics[width=0.32\textwidth]{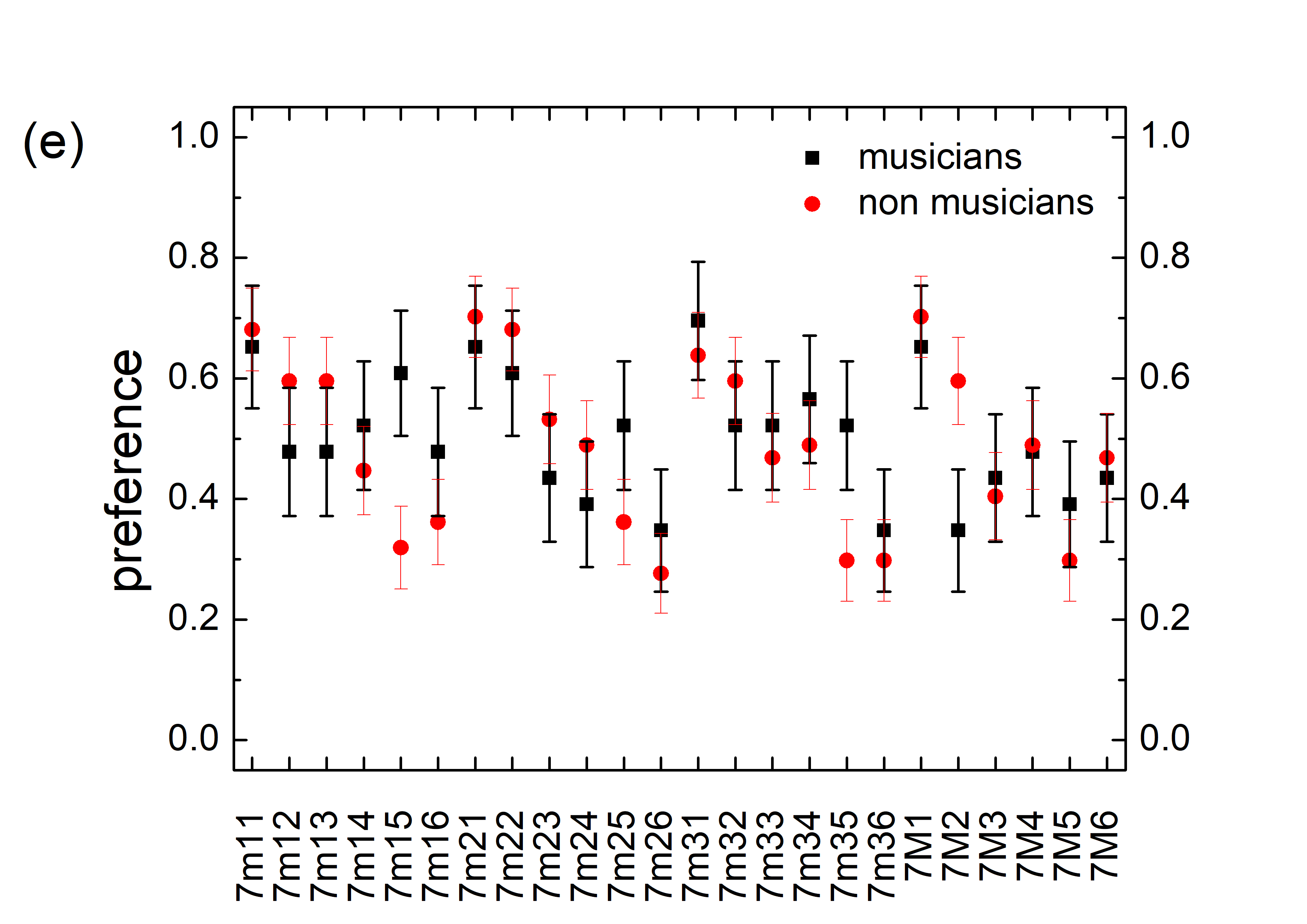}
\includegraphics[width=0.32\textwidth]{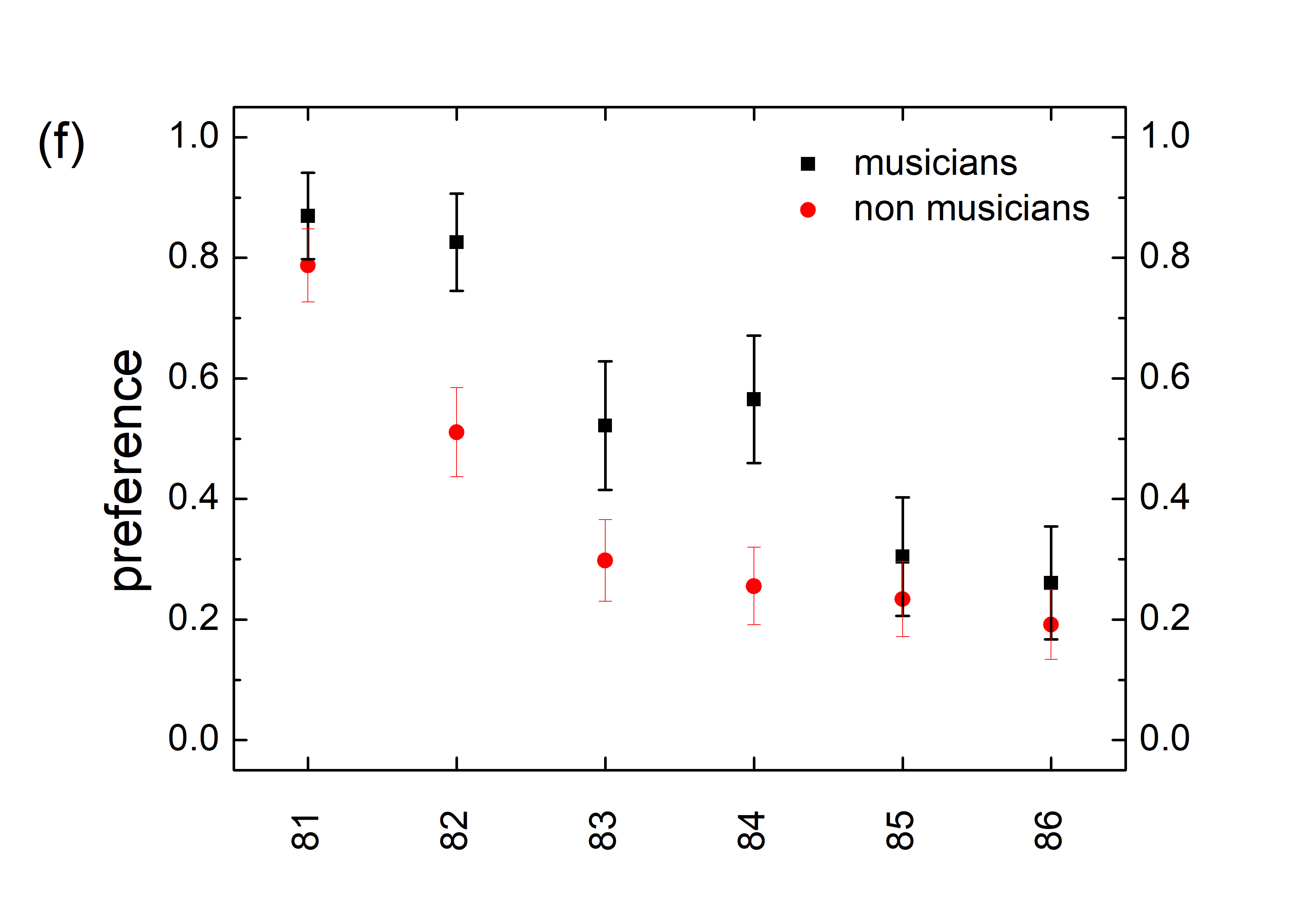} 
\caption{Mean values of responses or preference, $P$, for groups \textit{musicians} (black squares, 23 persons) and \textit{non-musicians} (red circles, 47 persons), for each simple-tone dyad: (a) seconds, (b) thirds, (c) fourths and fifths, (d) sixths, (e) sevenths, (f) octaves. Errors are larger for musicians compared to those for non-musicians.}
	\label{fig:MusiciansNonMusicians}
\end{figure*}

In Fig.~\ref{fig:MusiciansNonMusicians} we compare preferences of musicians (black squares) with those of non-musicians (red circles). For musicians, the errors are larger because the number of musicians who participated in the experiment was 23 (compared to 47 non-musicians).
Octaves show the most pronounced divergence in preferences between musicians and non-musicians: musicians seem to like more octaves ($R=2/1$) than non-musicians. 
For musicians, for the lower $F$ categories (1 to 4), $P > 0.5$. For $F$ categories 2, 3, 4, we have the largest separation; for the $F$ categories 1, 5, 6, the separation remains high. 
For octaves, linear fits $P(F)=AF+B$ give for musicians
$B \approx 0.93	\pm 0.07$, 
$A \approx (-1.9 \pm 0.3) \times 10^{-4}$ Hz$^{-1}$ 
(with adjusted $\mathcal{R}$ squared,  
$\mathcal{R}^2 \approx 0.87$) 
and for non-musicians
$B \approx 0.69	\pm 0.10$, 
$A \approx (-1.6 \pm 0.4) \times 10^{-4}$ Hz$^{-1}$
($\mathcal{R}^2 \approx 0.71$).
Hence, for octaves, the falls $P(F)$ have similar slope, but preferences of musicians are higher. For fifths ($R=3/2$), this observation is not so strong, i.e., for musicians
$B \approx 0.80	\pm 0.06$,
$A \approx (-1.5 \pm 0.3) \times 10^{-4}$ Hz$^{-1}$ ($\mathcal{R}^2 \approx 0.86$) and 
for non-musicians $B \approx 0.74 \pm 0.09$, 
$A \approx (-1.6 \pm 0.4) \times 10^{-4}$ Hz$^{-1}$ ($\mathcal{R}^2 \approx 0.77$). 
All calculated slopes $A$, intercepts $B$, and adjusted R squared $\mathcal{R}^2$ for the groups musicians - non-musicians can be found 
in Fig.~\ref{fig:ABR-MNM}, in the Appendix.  
The fits are not generally good, i.e. the relationships $P(F)$ are not generally straight lines. However, all slopes are clearly negative (but we notice that sevenths have very small negative slopes): We find again that increasing pitch, the feeling of consonance decreases. We also notice that the most negative slopes are those for octaves and major thirds; maybe in those cases dissonance comes mainly from increasing the mean frequency.

\begin{figure}[h!]
\centering
\includegraphics[width=0.45\textwidth]{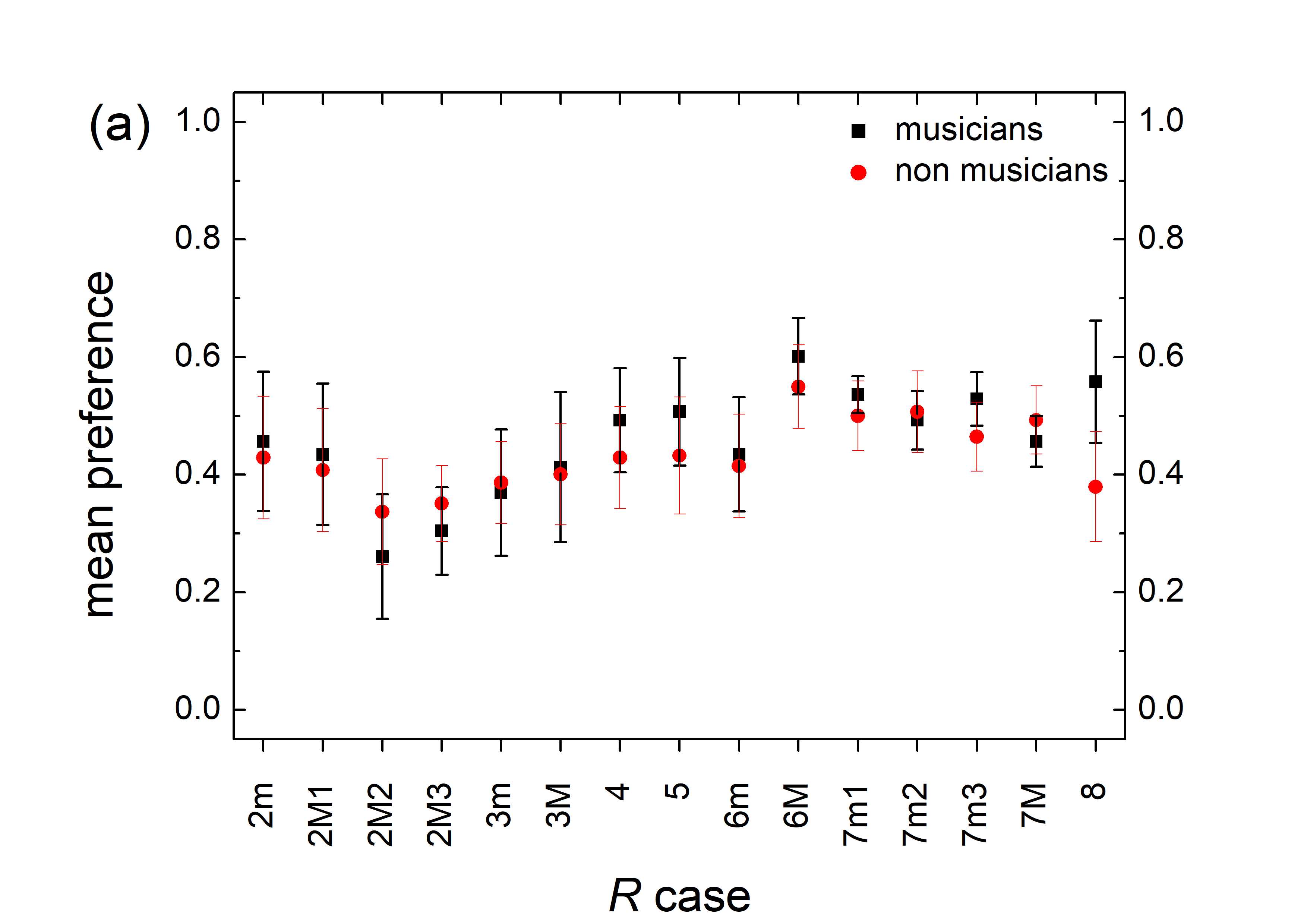}
\vspace{-10pt} 
\includegraphics[width=0.45\textwidth]{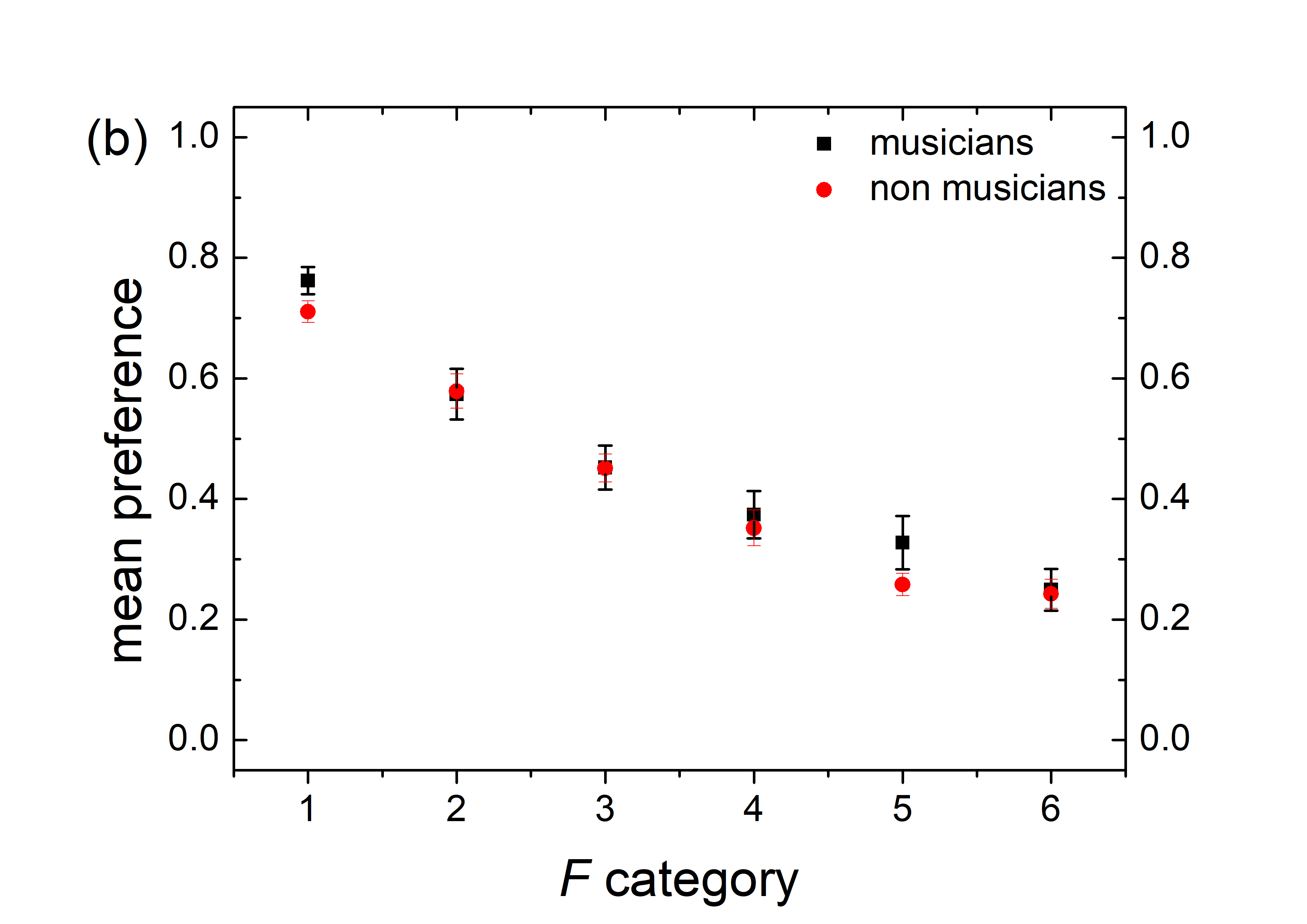}
\caption{Mean values and errors of $P$, for \textit{musicians} (black squares, 23 persons) and \textit{non-musicians} (red circles, 47 persons). (a) Mean over all $F$ categories for each $R$ case, $\overline{P}(R)$. (b) Mean over all $R$ cases for each $F$ category, $\overline{P}(F)$.}
\label{fig:meanvaluesMNMRF}
\end{figure}

In Fig.~\ref{fig:meanvaluesMNMRF}(a) we show the mean $P$ over all $F$ categories for each $R$ case, $\overline{P}(R)$. The $R$ cases with the greatest $\overline{P}(R)$ are 6M ($\overline{P} \approx 0.60$) and 8 ($\overline{P} \approx 0.56$) for musicians and 6M ($\overline{P} \approx 0.55$) for non-musicians. It seems that musicians have greater preference to 4, 5, 6M, 8 and (maybe 7m) and maybe lesser preference to 2M and 7M. 
Notice that for octaves the error bars touch each other very little. 
The $R$ cases with the lowest $\overline{P}(R)$ are 2M2 and 2M3 both for musicians ($\overline{P} \approx 0.26$ and $0.30$) and non-musicians ($\overline{P} \approx 0.34$ and $0.35$).
Most $\overline{P}(R)$ have errors of the order of $0.1$ (musicians: min $\approx 0.03$, max $\approx 0.13$; non-musicians: min $\approx 0.06$, max $\approx 0.10$). 
In Fig.~\ref{fig:meanvaluesMNMRF}(b), we present the mean $P$ over all $R$ cases for each $F$ category, $\overline{P}(F)$. Clearly, increasing $F$, $\overline{P}(F)$ falls both for musicians and non-musicians. Errors are of the order of $0.01$ (musicians: min $\approx 0.02$, max $\approx 0.04$; non-musicians: min $\approx 0.02$, max $\approx 0.03$).

We performed the Mann-Whitney test for the four Larger Groups (cf. Table~\ref{Table:LargerGroupsDitonies}). 
An illustration of the MW tests is shown 
in Fig.~\ref{fig:MW-MNM} (for all $R$ cases) 
and 
in Fig.~\ref{fig:MW-MNM-F} (for all $F$ categories). 
These MW tests confirm the inferences of 
Figs.~\ref{fig:meanvaluesMNMRF} (a) and (b). 
Each of the 1st, 2nd and 4th Larger Groups contains 24 preference values for musicians and 24 preference values for non-musicians and the 3rd Larger Group contains 18 preference values for musicians and 18 preference values for non-musicians [6 $F$ categories for each $R$ case]. These preference values are the mean values shown in Fig.~\ref{fig:MusiciansNonMusicians}.
From the results of the MW test, we deduce that for all Larger Groups the results for musicians and non-musicians are not significantly different at the 0.05 $p$-level. 
For the 3rd Larger Group (fourths, fifths, octaves), the differences between musicians and non-musicians were not statistically significant at the 0.05 level. However, when we examined each interval separately using the Mann–Whitney test,the $p$-value for octaves was smaller than for fourths or fifths. While still not significant in this group-level analysis, this suggested a potential trend. To explore this further, we conducted an additional analysis using individual responses rather than averaged preferences. In this finer-grained test, octaves yielded a statistically significant difference between musicians and non-musicians ($p \approx$ 0.0005), while the other intervals did not. These results support the conclusion that the perception of octaves differs significantly between the two groups. No significant differences were observed for the other simple-tone dyads, with $p$-values of approximately 0.15 for fifths and 0.22 for fourths. While these are not significant, they may suggest weak trends that could be examined in future research. We note that no corrections for multiple comparisons were applied, and these results should therefore be interpreted with caution.

Inference: For octaves, the falls $P(F)$ show a similar slope between groups, but the overall preference level is higher for musicians. For fifths, a similar trend is observed, though less pronounced. Across all R cases, increasing pitch, consonance ratings tend to decrease for both groups. MW tests on preference scores for the combined group of fourths, fifths, and octaves did not show a statistically significant difference at the 0.05 level, though the result approaches this threshold. When analyzing individual $R$ cases, octaves yielded the lowest $p$-value among all intervals, indicating a possible trend, although not reaching statistical significance. In contrast, the MW test on responses (i.e., raw response data, without averaging) showed a statistically significant difference between musicians and non-musicians for octaves ($p \approx$ 0.0005). This result suggests that the perception of octaves differs reliably between the two groups. The averaged preference scores $P(R)$ indicate that musicians rate octaves (8), and possibly fourths (4), fifths (5), and major sixths (6M), higher than non-musicians. Both groups rated the intervals 2M2 and 2M3 lowest. Finally, $P(F)$ is a decreasing function both for musicians and non-musicians.


\subsection{Effects of Gender}
\label{subsec:MenWomen}
\begin{figure*}[ht!]
\centering 
\includegraphics[width=0.32\textwidth]{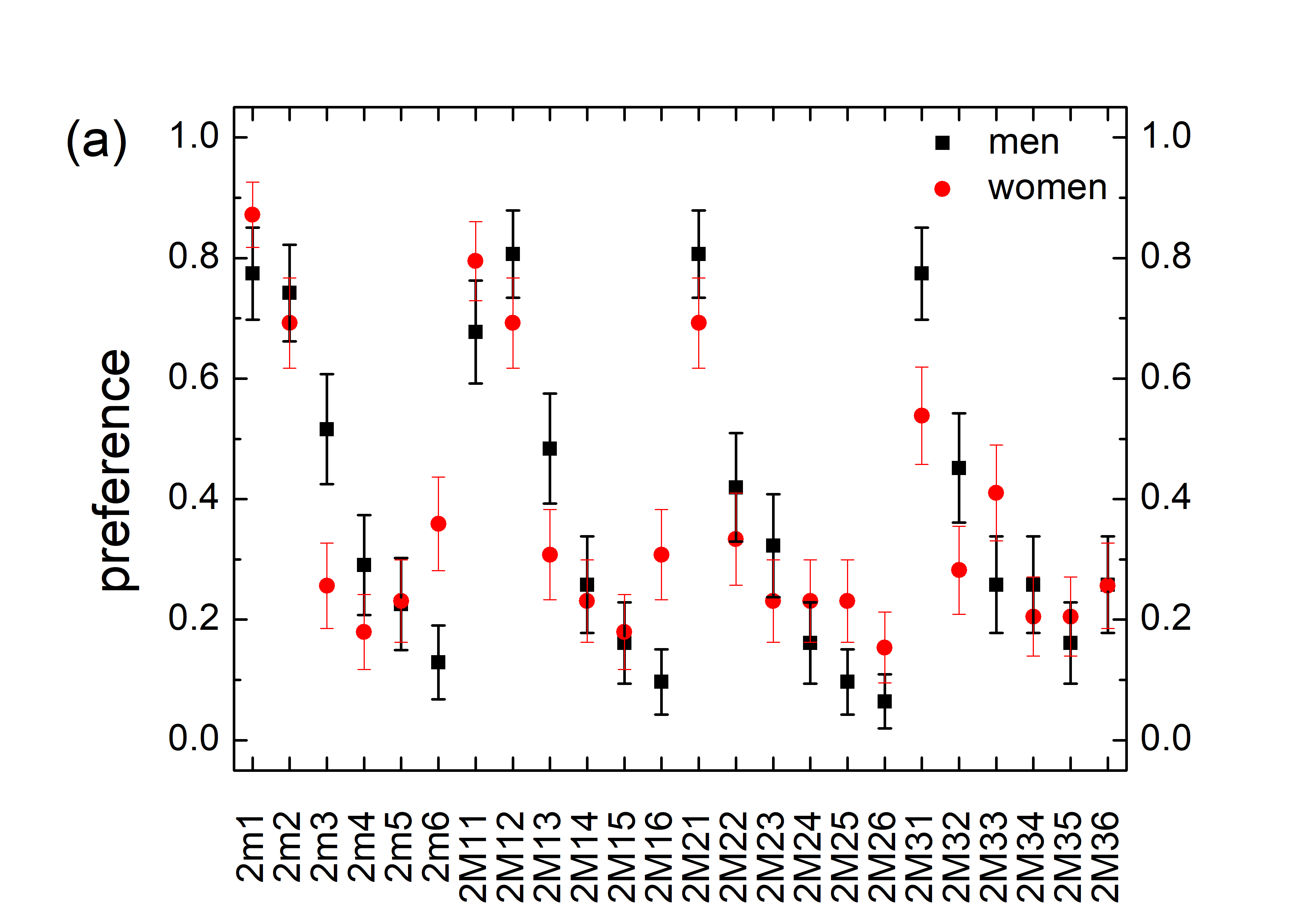}
\includegraphics[width=0.32\textwidth]{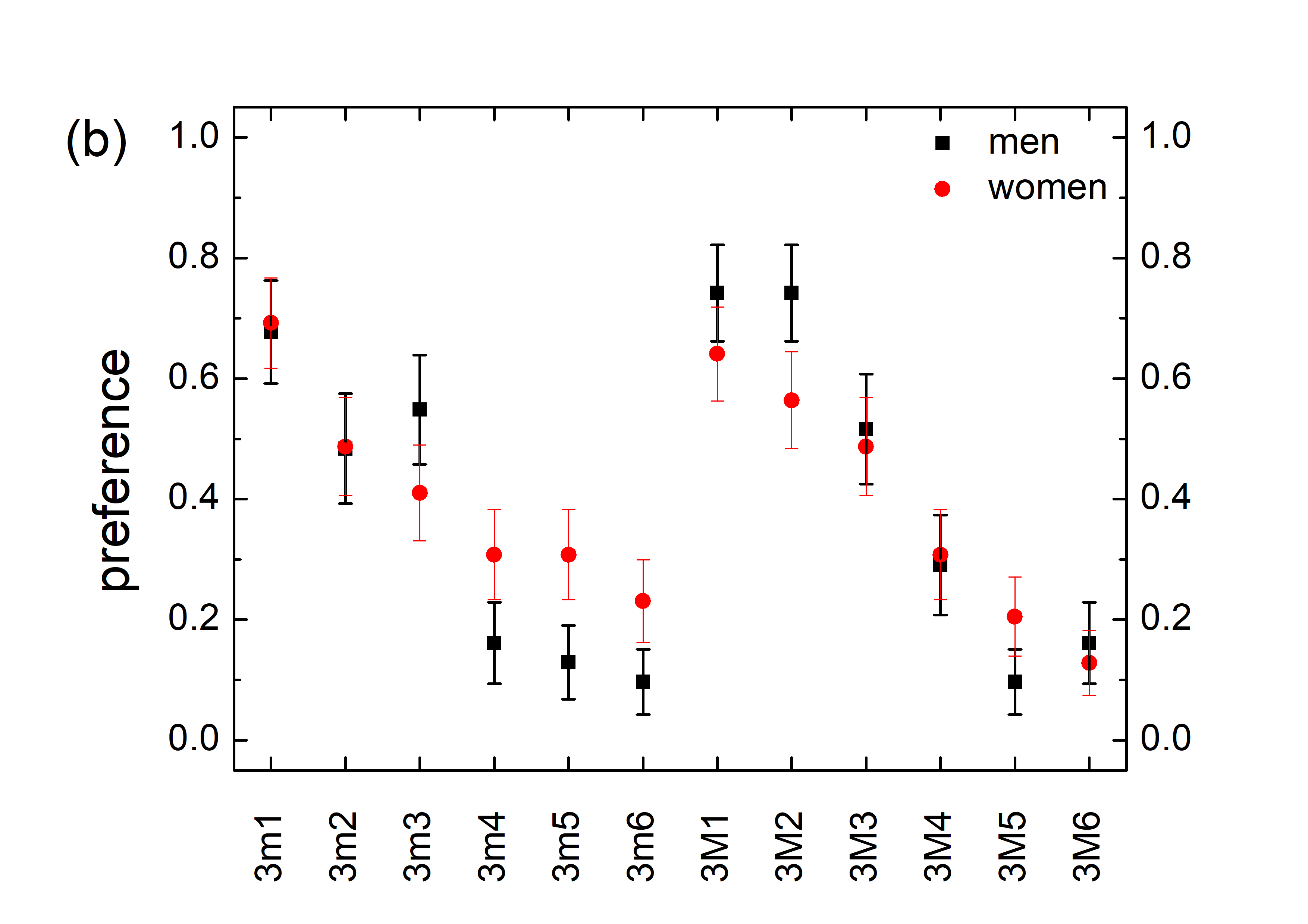}
\includegraphics[width=0.32\textwidth]{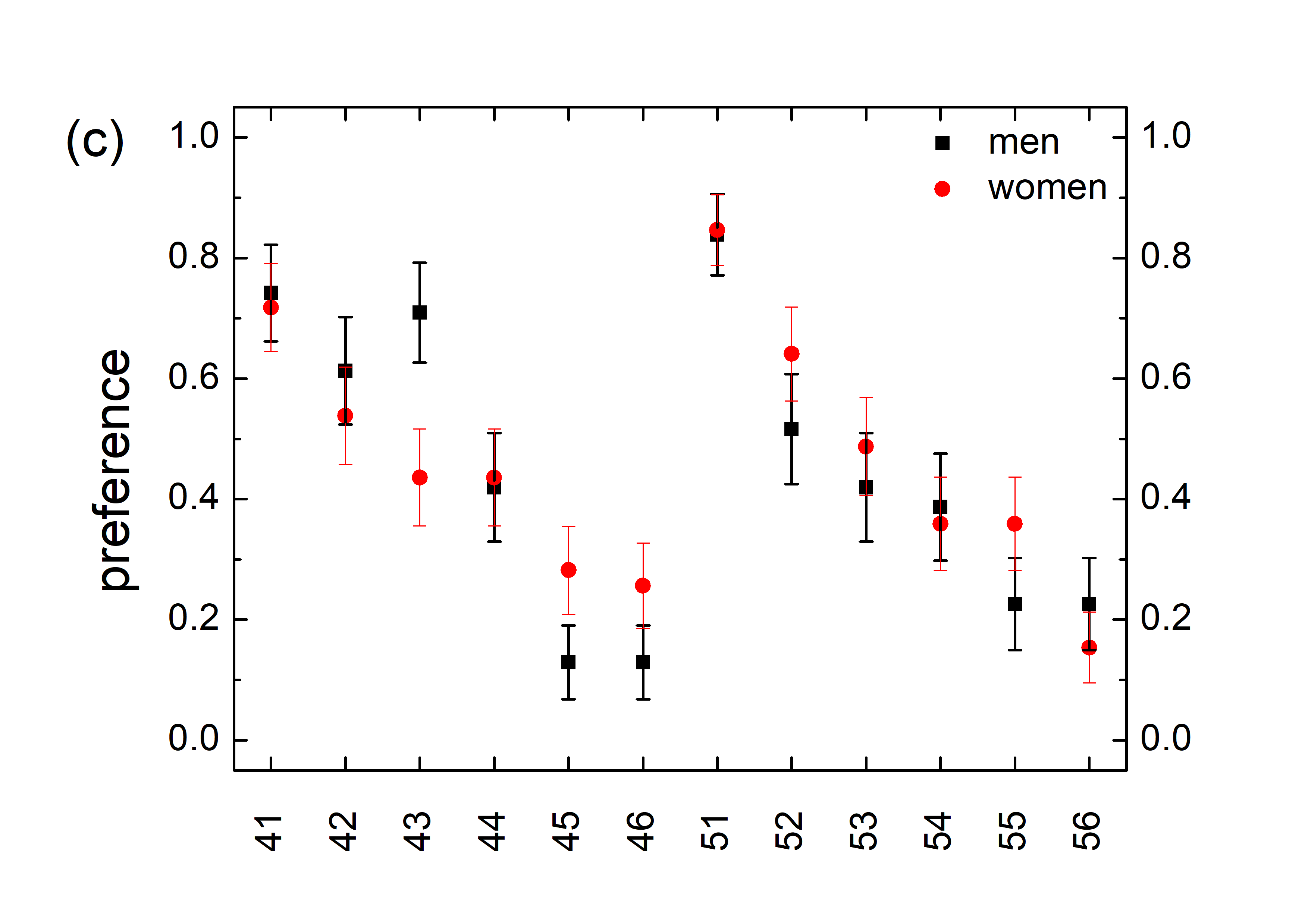}\\
\vspace{-0.2cm}	
\includegraphics[width=0.32\textwidth]{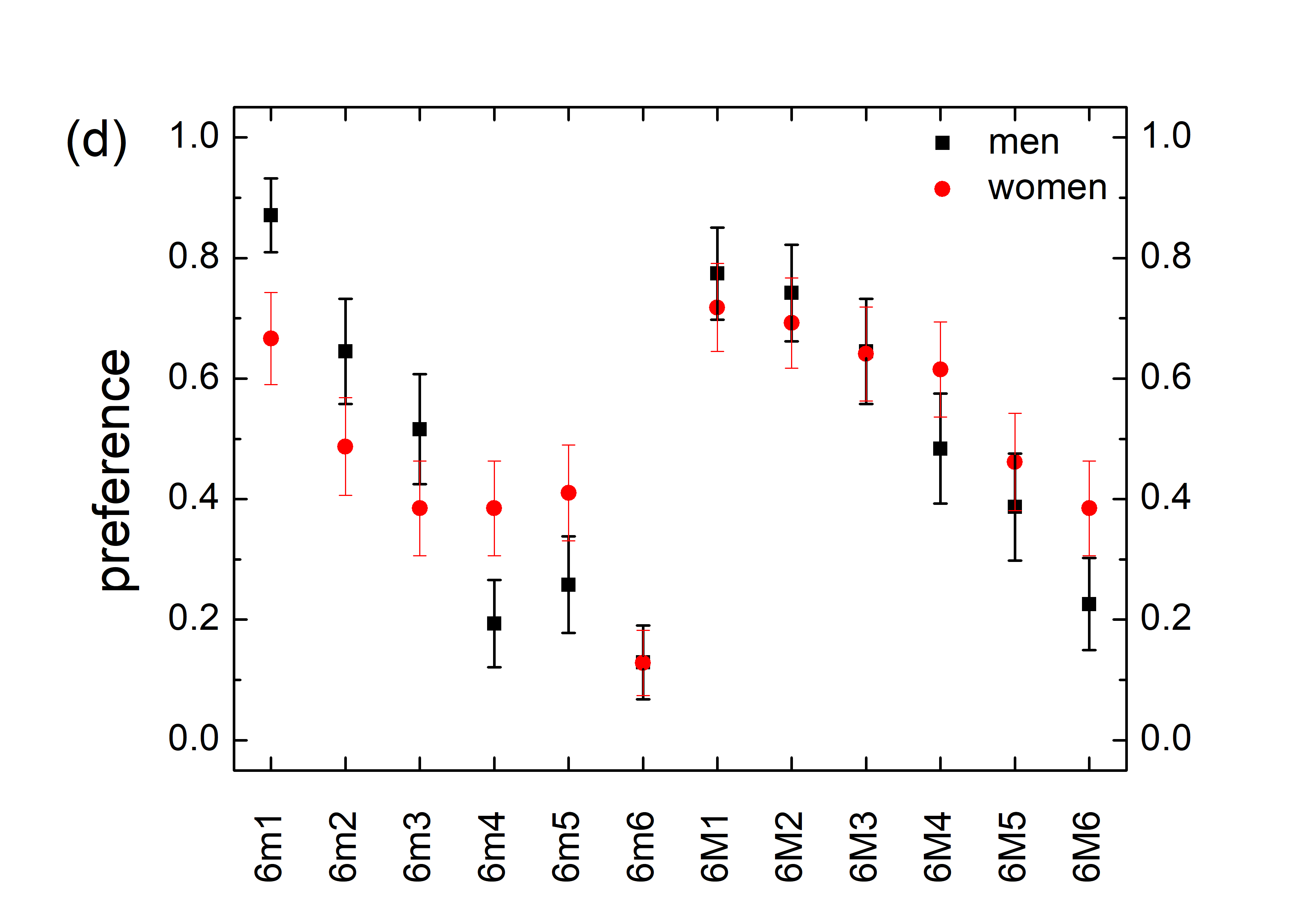}
\includegraphics[width=0.32\textwidth]{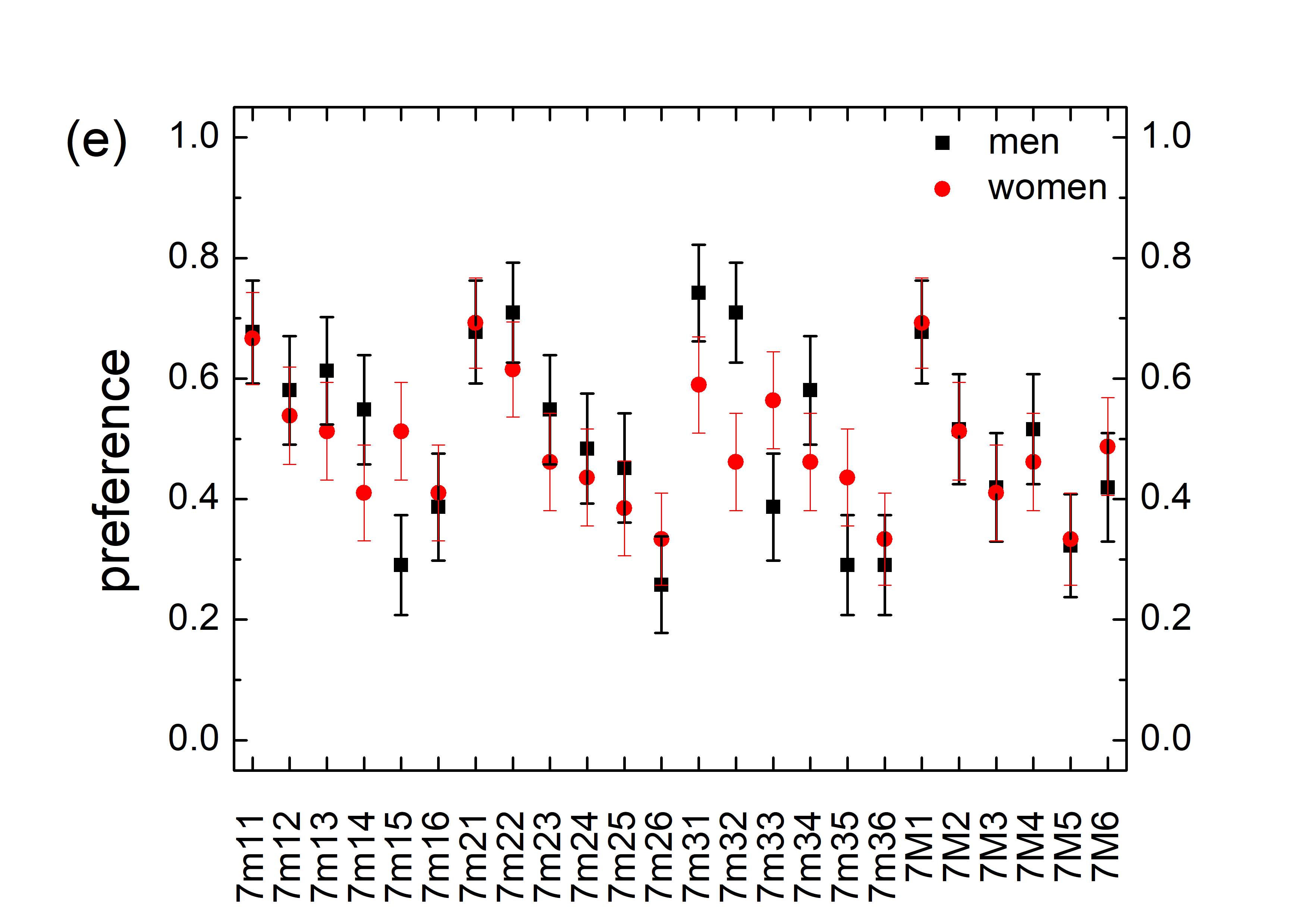}
\includegraphics[width=0.32\textwidth]{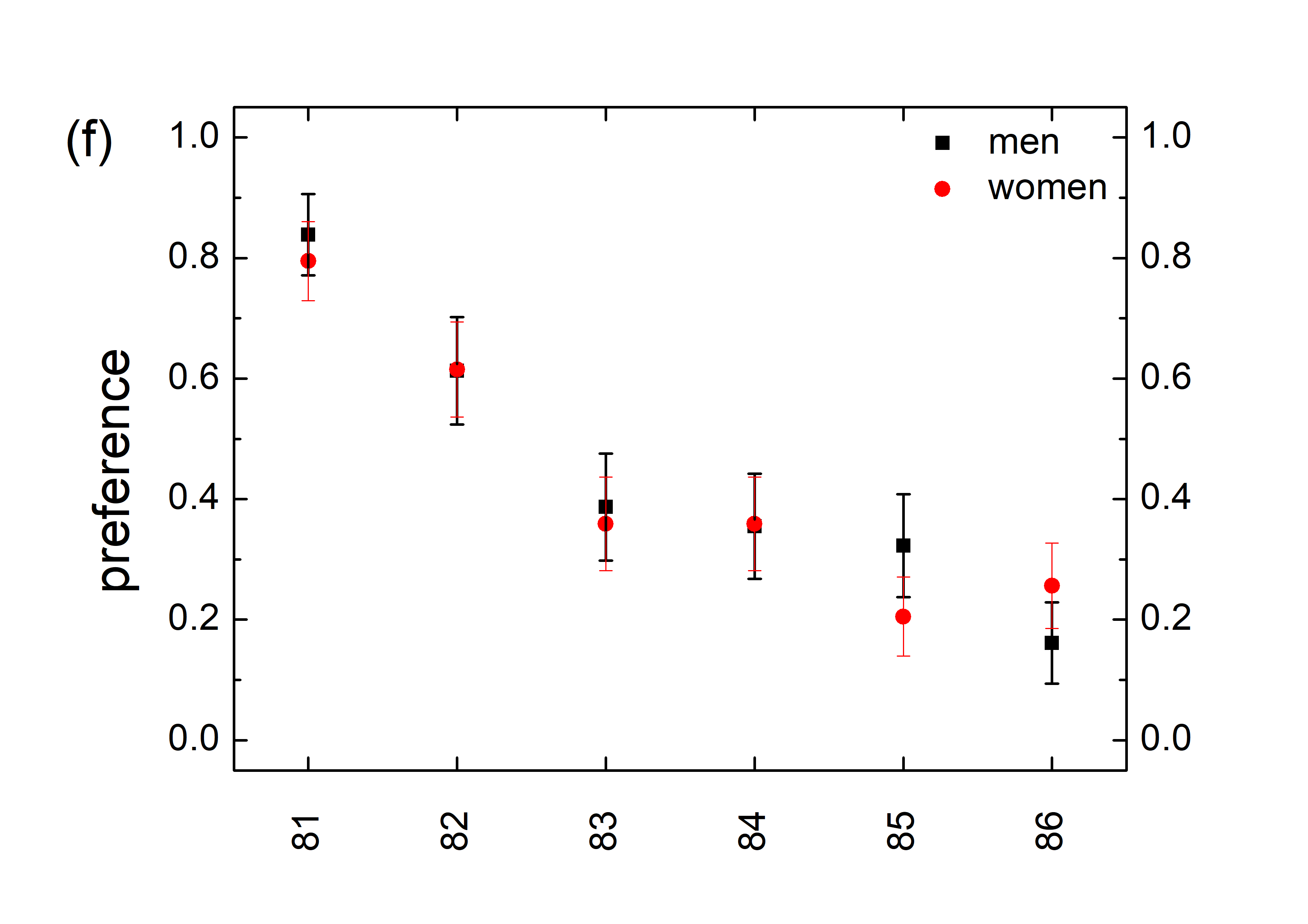}
\caption{Mean values and errors of responses or preference, $P$, for groups \textit{men} (black squares, 31 persons) and \textit{women} (red circles, 39 persons): (a) seconds, (b) thirds, (c) fourths and fifths, (d) sixths, (e) sevenths, (f) octaves.}
	\label{fig:MenWomen}
\end{figure*}

In Fig.~\ref{fig:MenWomen} we compare the preferences of men (black squares) with those of women (red circles). These do not seem to show strong differences from a first glance. We observe again the decrease of preferences, increasing pitch, both for men and women. The preferences of both groups start from 0.9 - 0.8 for lower $F$ categories and fall between 0.2 - 0.1 for the higher $F$ categories. The preferences of sevenths, for men or women, are higher than 0.2. In some $R$ cases, for lower $F$ categories men score high but for higher $F$ categories their scores fall significantly, i.e., there is a large extent of scores. In some $R$ cases women score higher for higher $F$ categories but their scores have smaller variation range. 
This pattern may suggest that men are more sensitive to or less tolerant of higher frequencies, though further investigation is needed.

\begin{figure}[h!]
\centering
\includegraphics[width=0.45\textwidth]{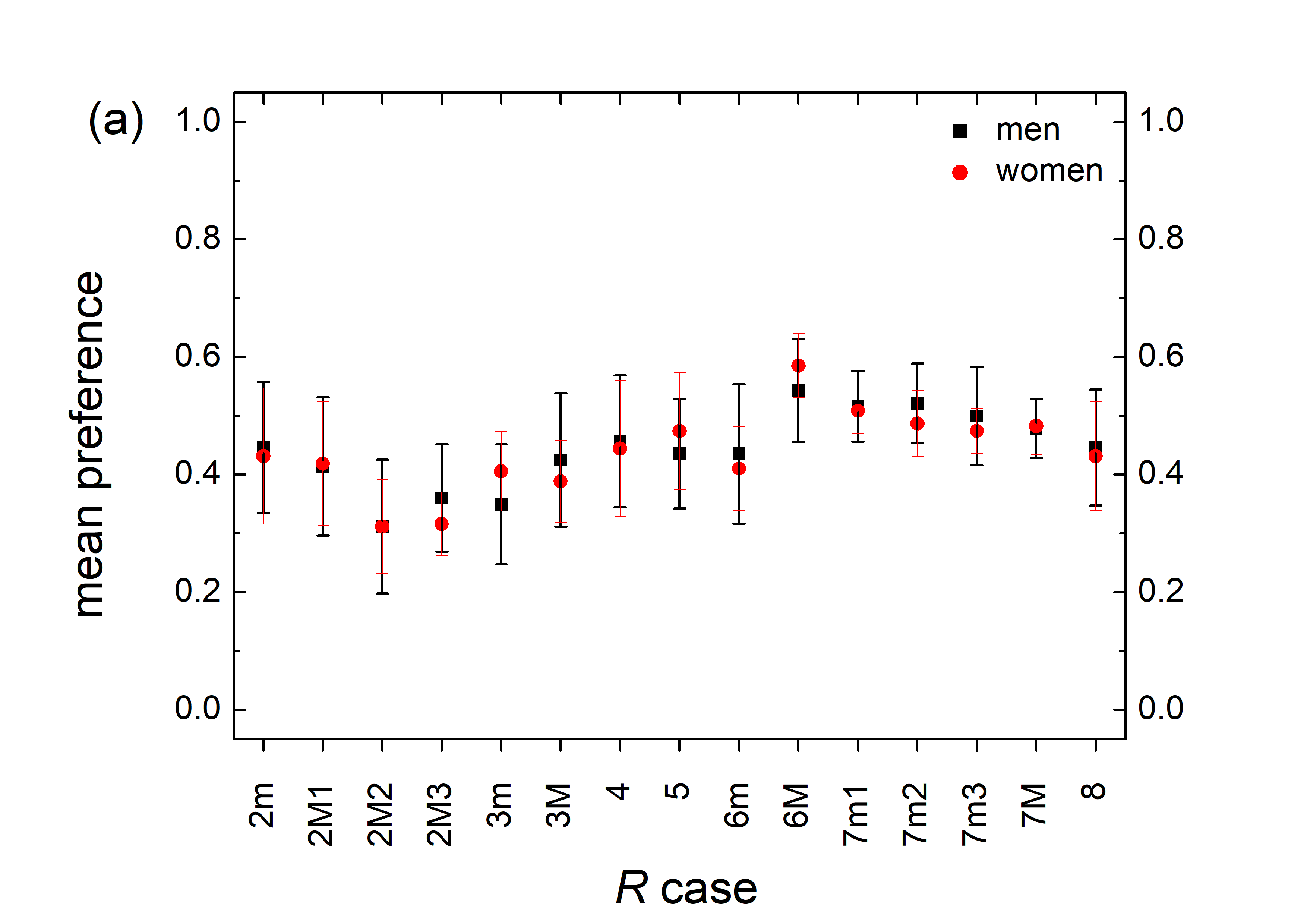} 
\includegraphics[width=0.45\textwidth]{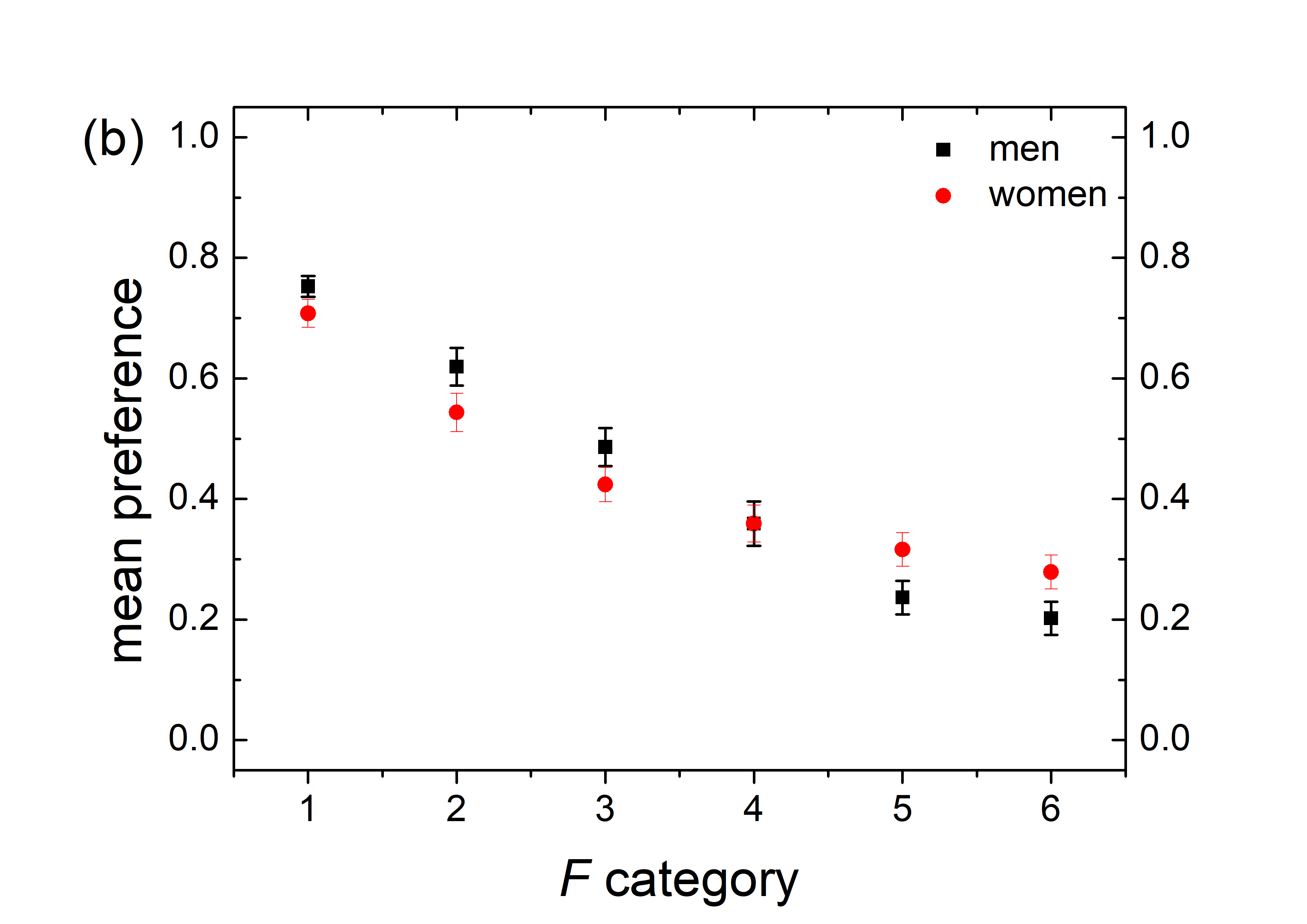}
\caption{Mean values and errors of $P$, for \textit{men} (black squares, 31 persons) and \textit{women} (red circles, 39 persons). (a) Mean over all $F$ categories for each $R$ case, $\overline{P}(R)$. (b) Mean over all $R$ cases for each $F$ category, $\overline{P}(F)$.}
\label{fig:meanvaluesMWRF}
\end{figure}

In Fig.~\ref{fig:meanvaluesMWRF}(a) we present the mean over all $F$ categories for each $R$ case, $\overline{P}(R)$. Errors are rather large. Errors of women are a little smaller (39 women vs. 31 men). The $\overline{P}(R)$ of the two groups do not show great differences. In Fig.~\ref{fig:meanvaluesMWRF}(b) we show the mean over all $R$ cases for each $F$ category, $\overline{P}(F)$. Again, we observe that high $F$ simple-tone dyads have lower $\overline{P}(F)$ than low $F$ simple-tone dyads: We find again that the feeling of consonance decreases, increasing pitch. However, there is a significant difference between men and women: Clearly, for the lower $F$ categories, $\overline{P}(F)$ are higher for men; for the higher $F$ categories, $\overline{P}(F)$ are higher for women. In other words, for lower pitches men show higher preferences, but for higher pitches women show higher preferences.

However, from the MW rank sum test for preferences between the groups  \textit{men} and \textit{women}, we realize that statistically significant differences for preferences do not exist. We repeated the MW test for responses, i.e., without averaging first. We also checked dividing the responses into the lower 3 / upper 3 $F$ cases. It seems that only in a few cases there is significant difference (at least at the 0.05 level) between the responses of men and women. Men seem to dislike more higher frequencies than women. An illustration of MW tests comparing the groups \textit{men} and \textit{women} is shown in the Appendix, 
in Fig.~\ref{fig:MW-MW} (for all $R$ cases) and 
in Fig.~\ref{fig:MW-MW-F} (for all $F$ categories).
These MW tests confirm the inferences of Figs.~\ref{fig:meanvaluesMWRF} (a) and (b).

Gender differences play a role in auditory perception. Longitudinal studies, such as Pearson et al.~\cite{Pearson:1995}, have consistently found that men experience faster and greater declines in hearing sensitivity than women, especially in the high-frequency range relevant for musical perception. This disparity may contribute to subtle differences in dyad perception and pitch discrimination between genders. McFadden~\cite{McFadden:1998} further supports this by showing that women tend to have stronger otoacoustic emissions, indicating better cochlear (outer hair cell) function, which is often linked to finer frequency resolution. Additionally, a large-scale global study by Koerner et al. \cite{KoernerZhang:2018} confirmed that women exhibit, on average, approximately 2 dB greater cochlear sensitivity than men, even when controlling for age and environment. These gender-based auditory differences may influence how pitch intervals and consonance are perceived and should be considered in psychoacoustic models.

Inference: As a general trend, for lower $F$ men show higher preferences, but this is reversed for higher $F$ where women show higher preferences. It seems that, men dislike high frequencies more than women. It also seems that the scores of men have higher variation than the scores of women.


\subsection{Effects of Age}
\label{subsec:35and50}
\begin{figure*}[htb]
\centering 
\includegraphics[width=0.32\textwidth]{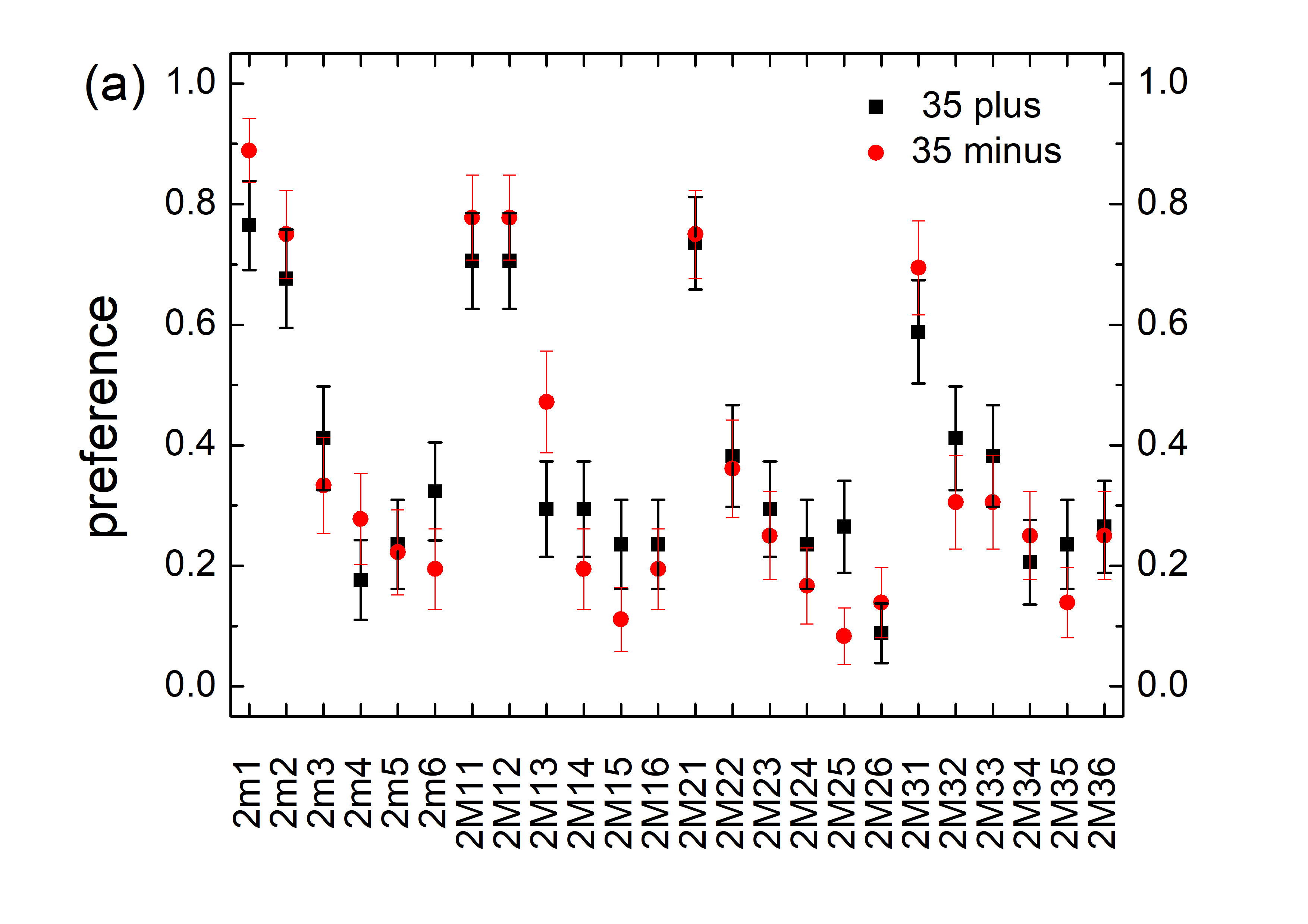}
\includegraphics[width=0.32\textwidth]{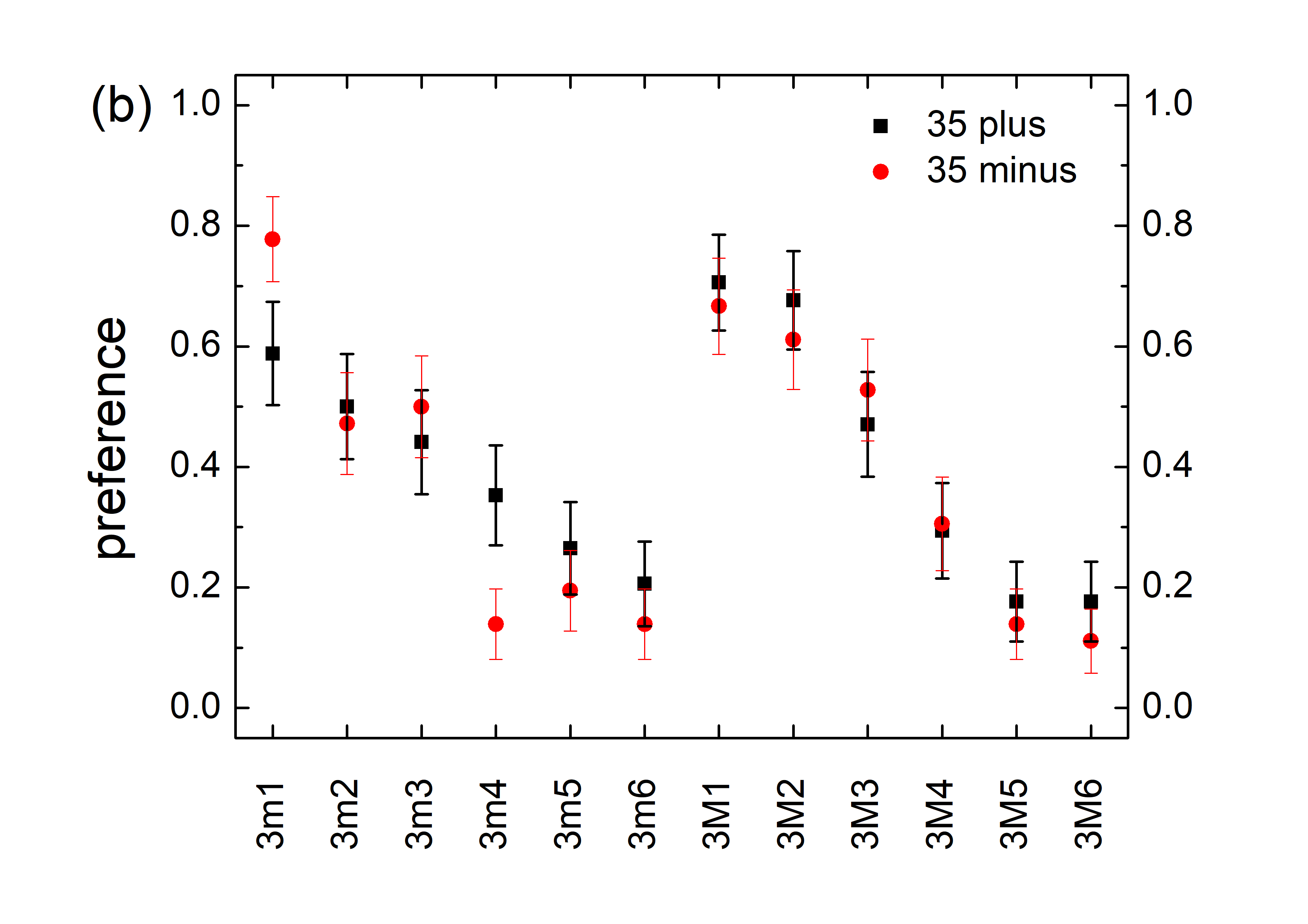}
\includegraphics[width=0.32\textwidth]{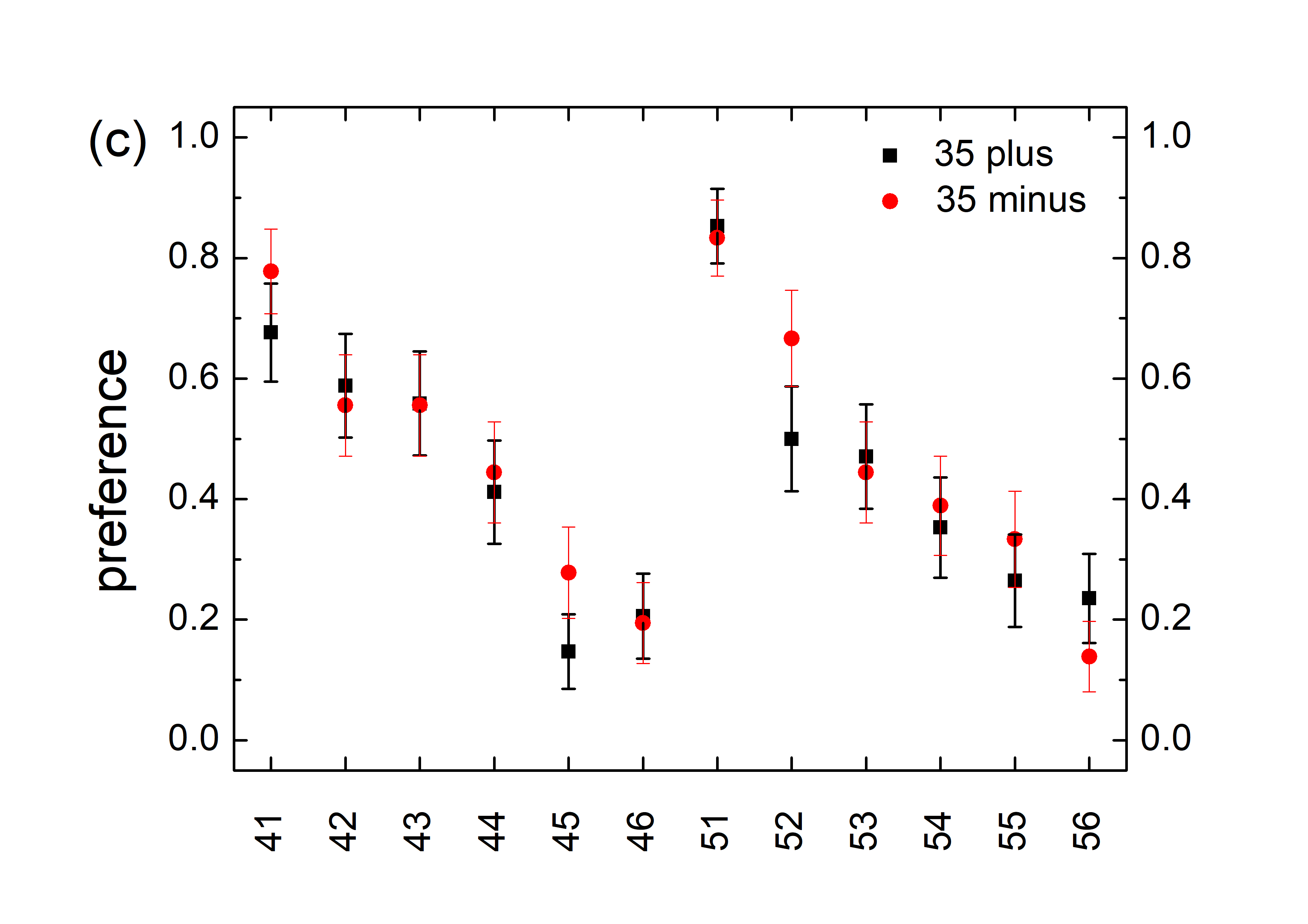} \\
\vspace{-0.2cm}
\includegraphics[width=0.32\textwidth]{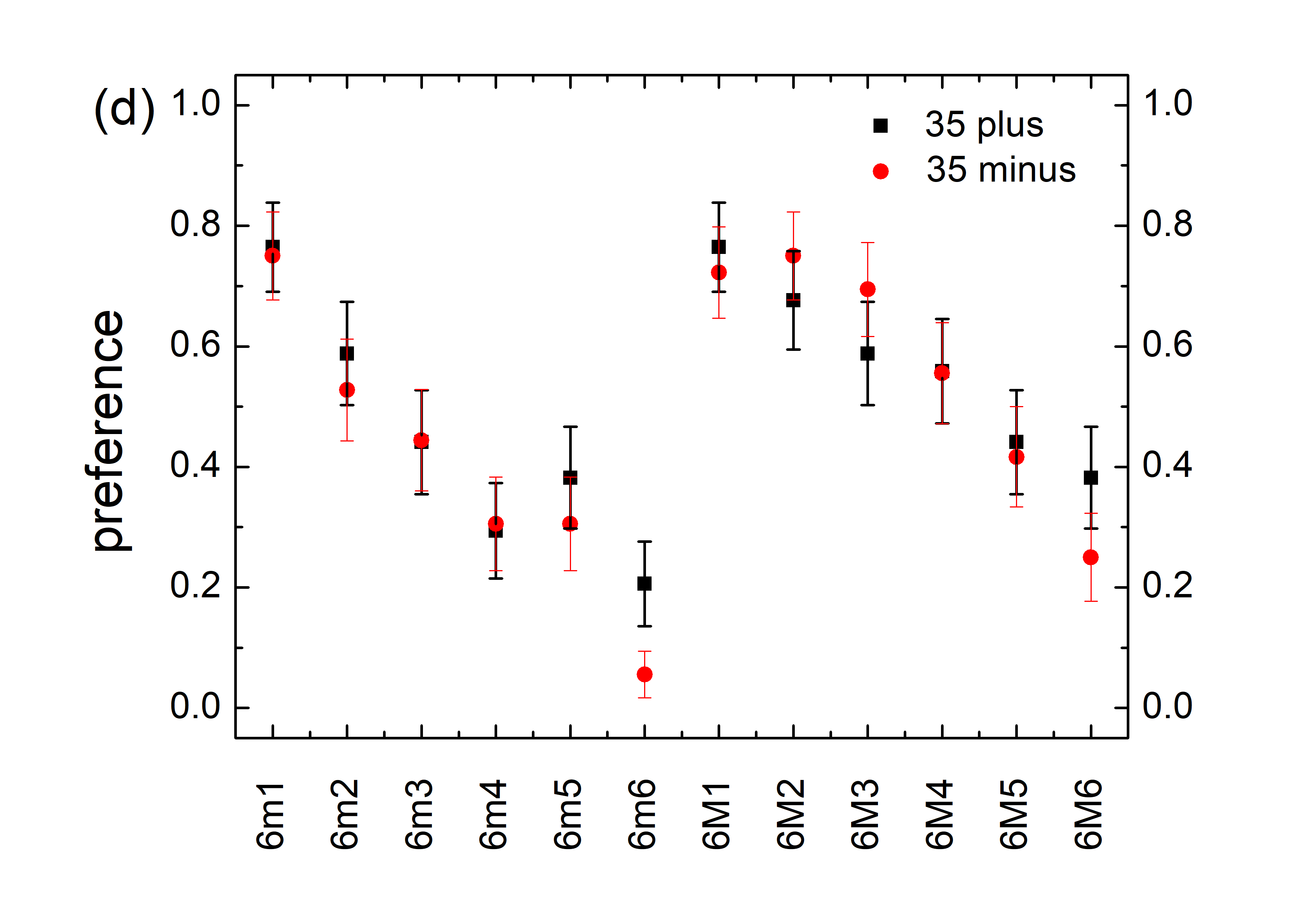}
\includegraphics[width=0.32\textwidth]{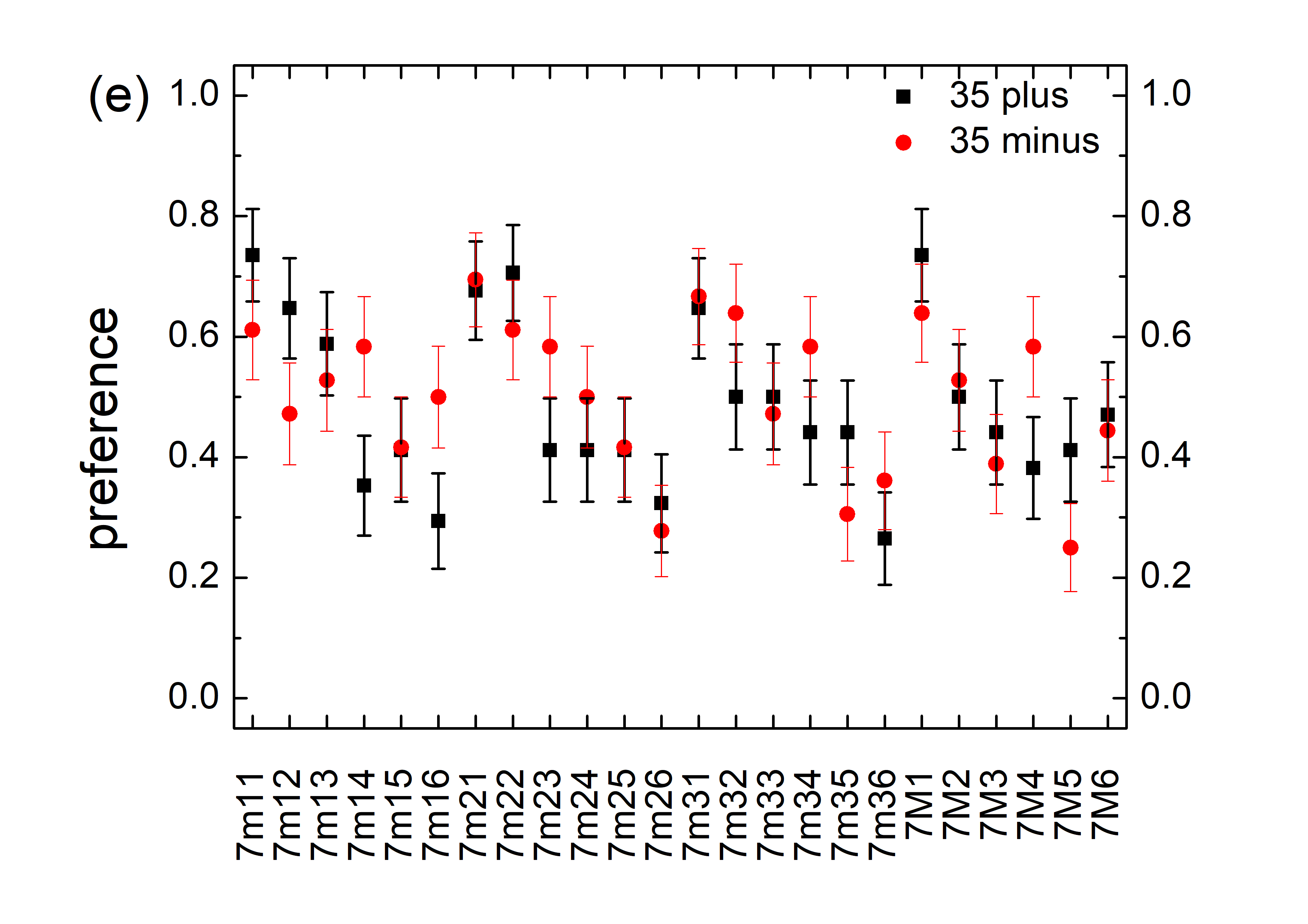}
\includegraphics[width=0.32\textwidth]{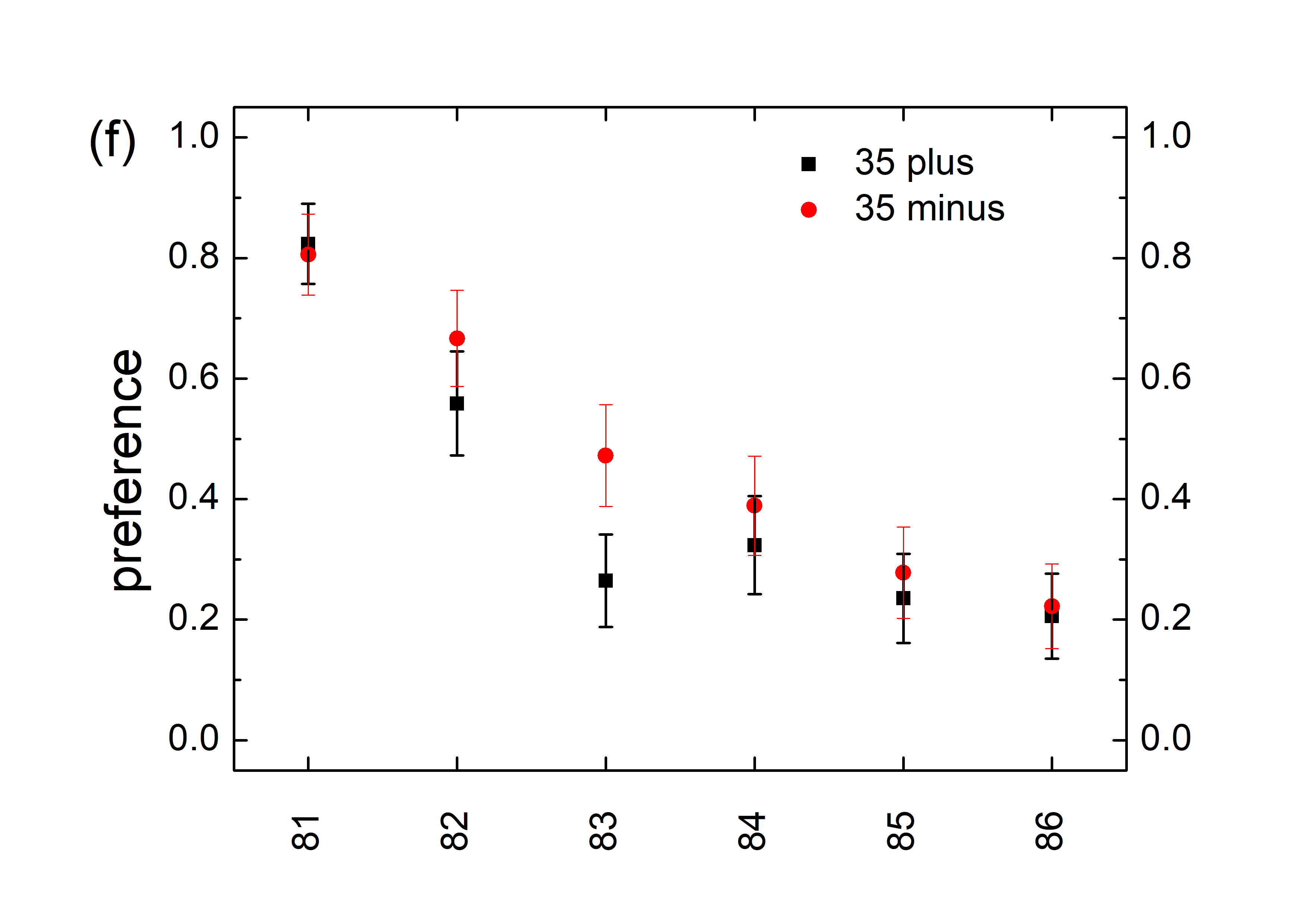}
\caption{Mean values and errors of responses or preference, $P$, for age groups \textit{above 35} (black squares, 34 persons) and \textit{below 35} (red circles, 36 persons): (a) seconds, (b) thirds, (c) fourths and fifths, (d) sixths, (e) sevenths, (f) octaves.}
\label{fig:PlusMinus35}
\end{figure*}

\begin{figure*}[htb]
\centering
\includegraphics[width=0.45\textwidth]{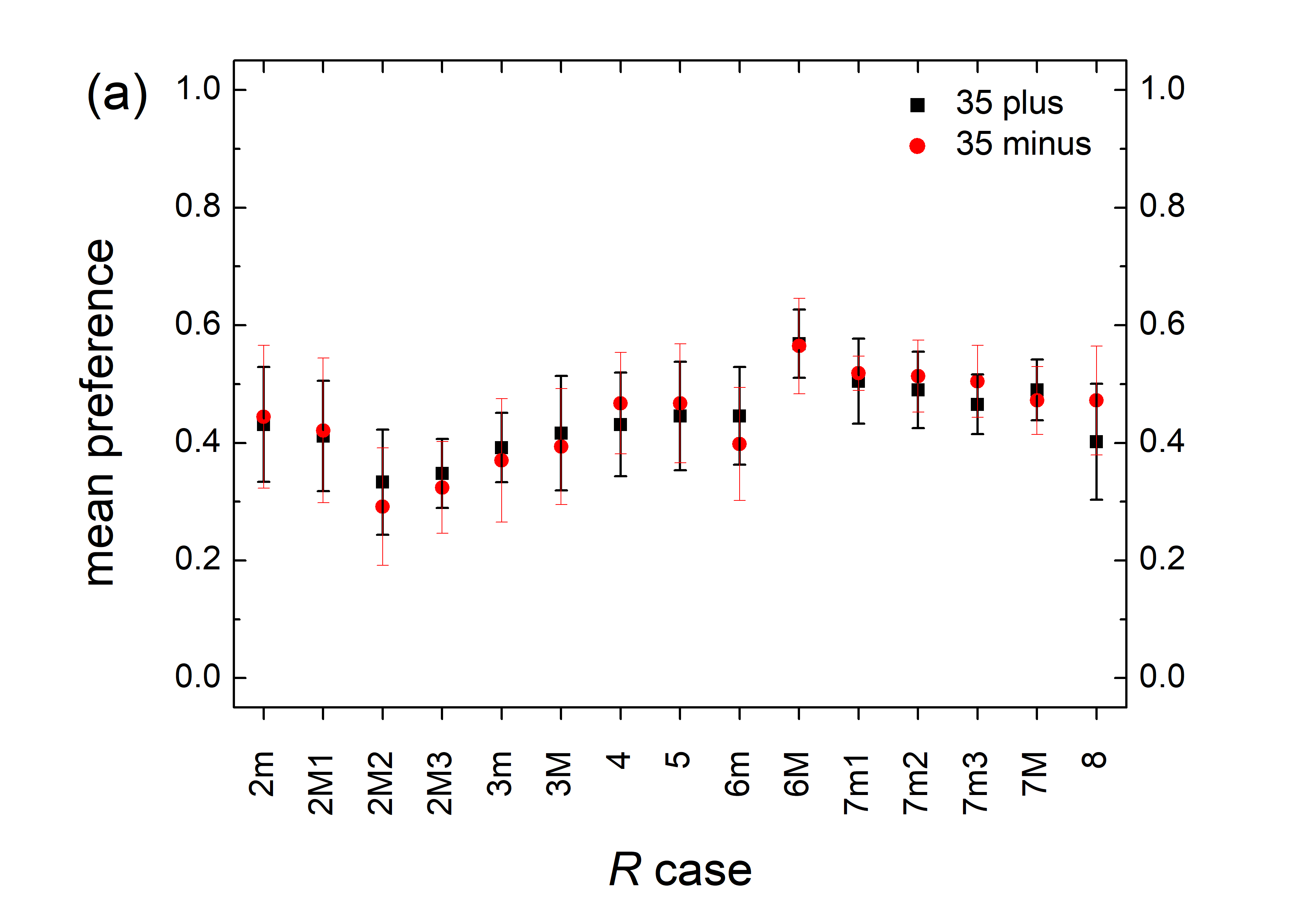} 
\includegraphics[width=0.45\textwidth]{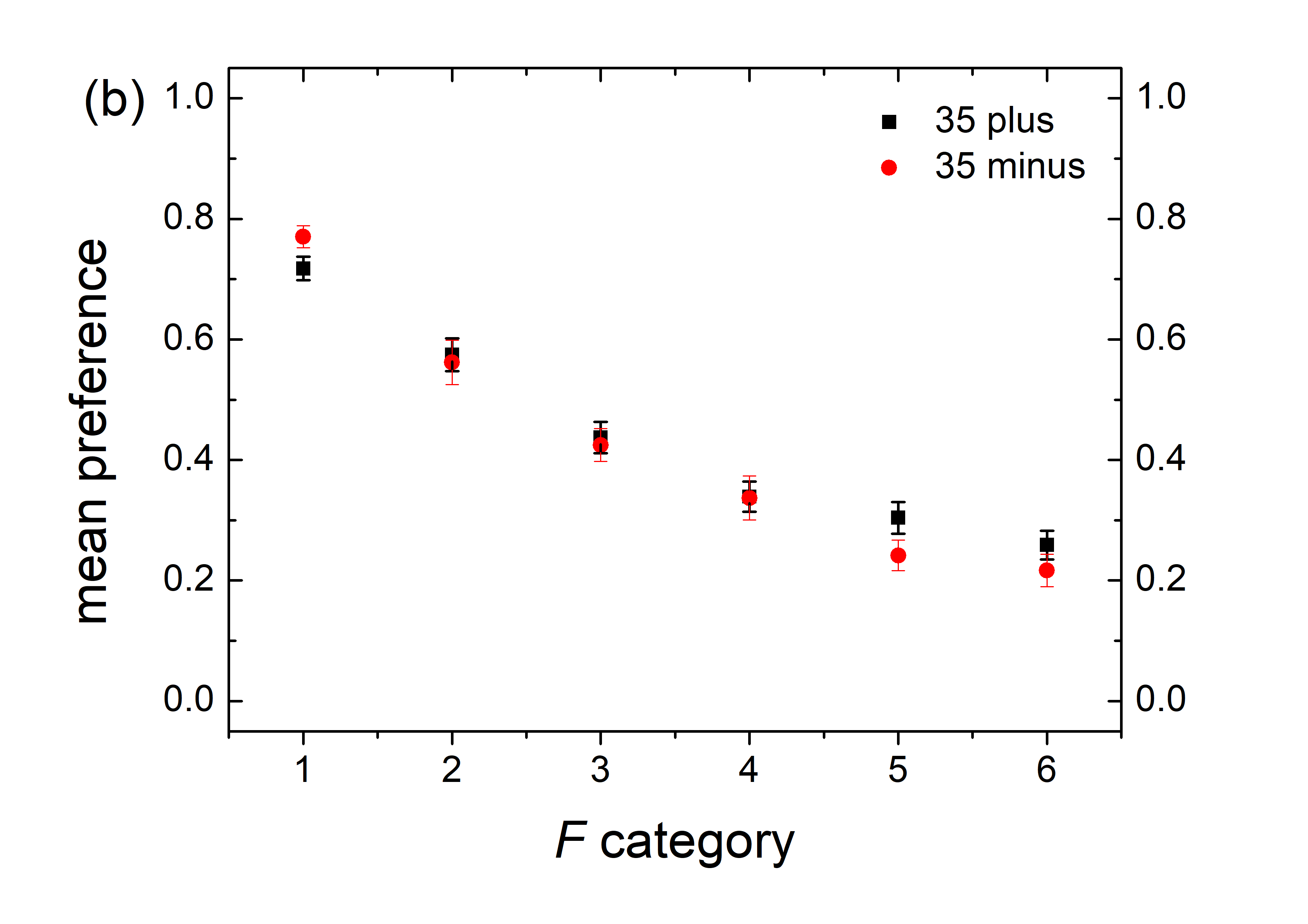}
\caption{Mean values and errors of $P$, for groups \textit{over 35} (black squares, 34 persons) and \textit{under 35} (red circles, 36 persons). (a) Mean over all $F$ categories for each $R$ case, $\overline{P}(R)$. (b) Mean over all $R$ cases for each $F$ category, $\overline{P}(F)$.}
	\label{fig:meanvalues35RF}
\end{figure*}

We now compare age groups, taking as borderlines 35 or 50 years of age. In Fig.~\ref{fig:PlusMinus35} we compare preferences of subjects above 35 years of age (34 persons, black squares) with those below 35 years of age (36 persons, red circles). Next, in Fig.~\ref{fig:meanvalues35RF} we depict, for groups \textit{over 35} and \textit{under 35}: in (a) mean over all $F$ categories for each $R$ case, $\overline{P}(R)$ and in (b) mean over all $R$ cases for each $F$ category, $\overline{P}(F)$. 
The main observation from Fig.~\ref{fig:meanvalues35RF}(a) is that there are no significant differences between the two age groups for the perception of frequency ratio $R$ cases, as the mean values of one group are within the mean errors of the other group and vice versa, which is also confirmed by the MW tests shown in the Appendix,
in Fig.~\ref{fig:MW-35} (for all $R$ cases),
The MW tests for all $F$ categories are shown 
in Fig.~\ref{fig:MW-35-F} in the Appendix. 
For the perception of mean frequency $F$ categories, the preferences of persons under 35 in Fig.~\ref{fig:meanvalues35RF}(b) seem to cover a wider range. The MW tests confirm the inferences of  Figs.~\ref{fig:meanvalues35RF} (a) and (b).

\begin{figure*}[htb]
\centering 
\includegraphics[width=0.32\textwidth]{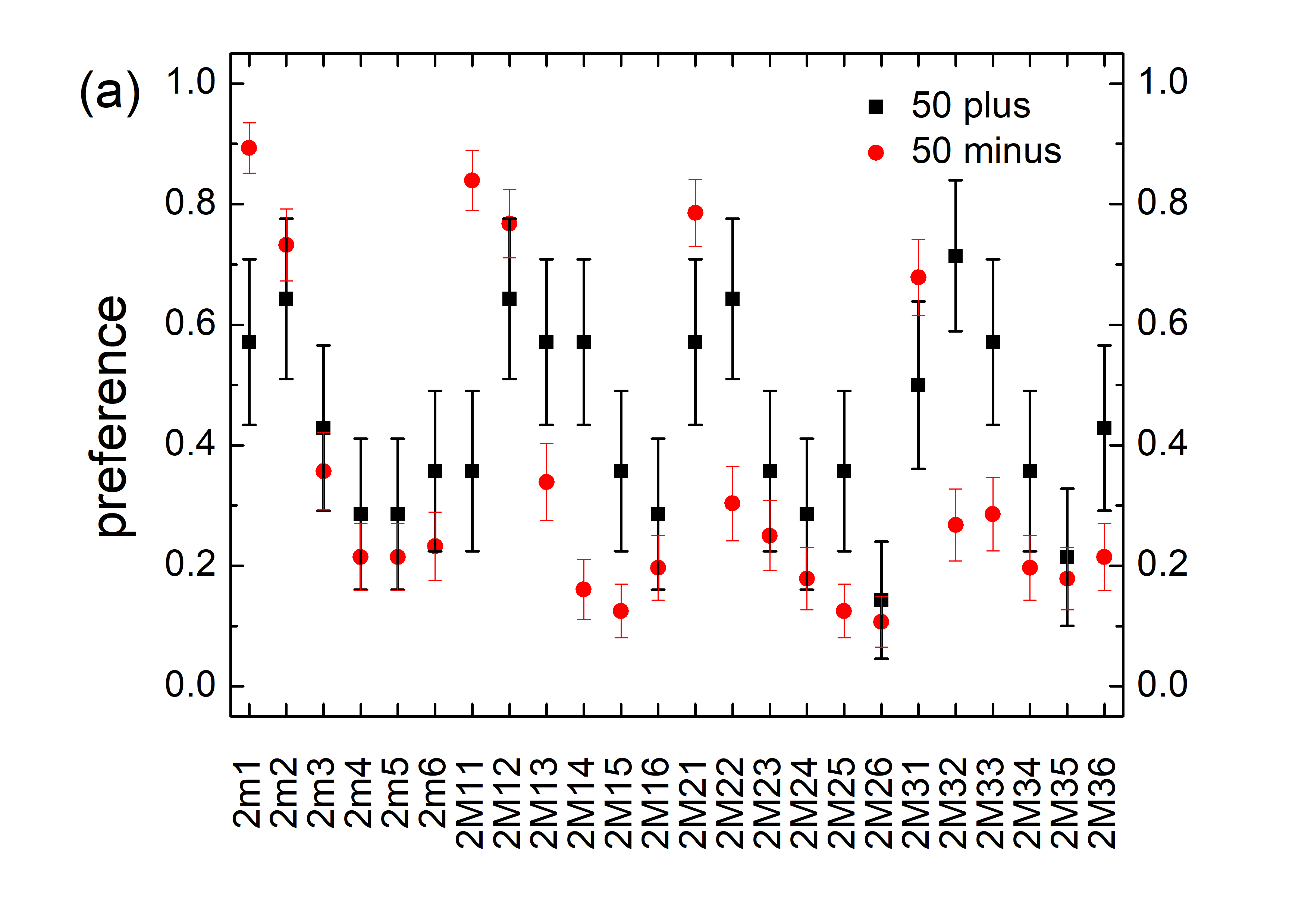}
\includegraphics[width=0.32\textwidth]{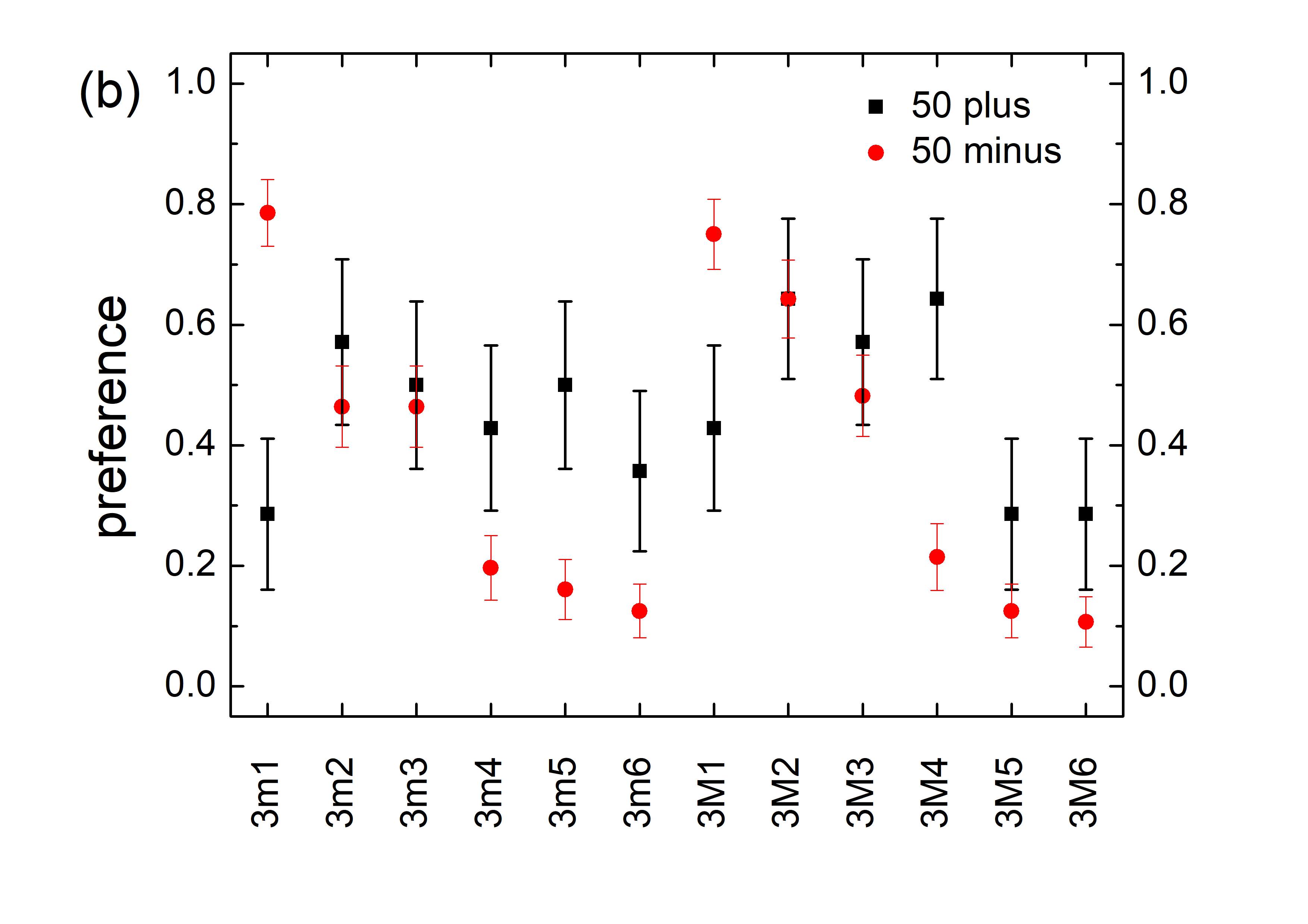}
\includegraphics[width=0.32\textwidth]{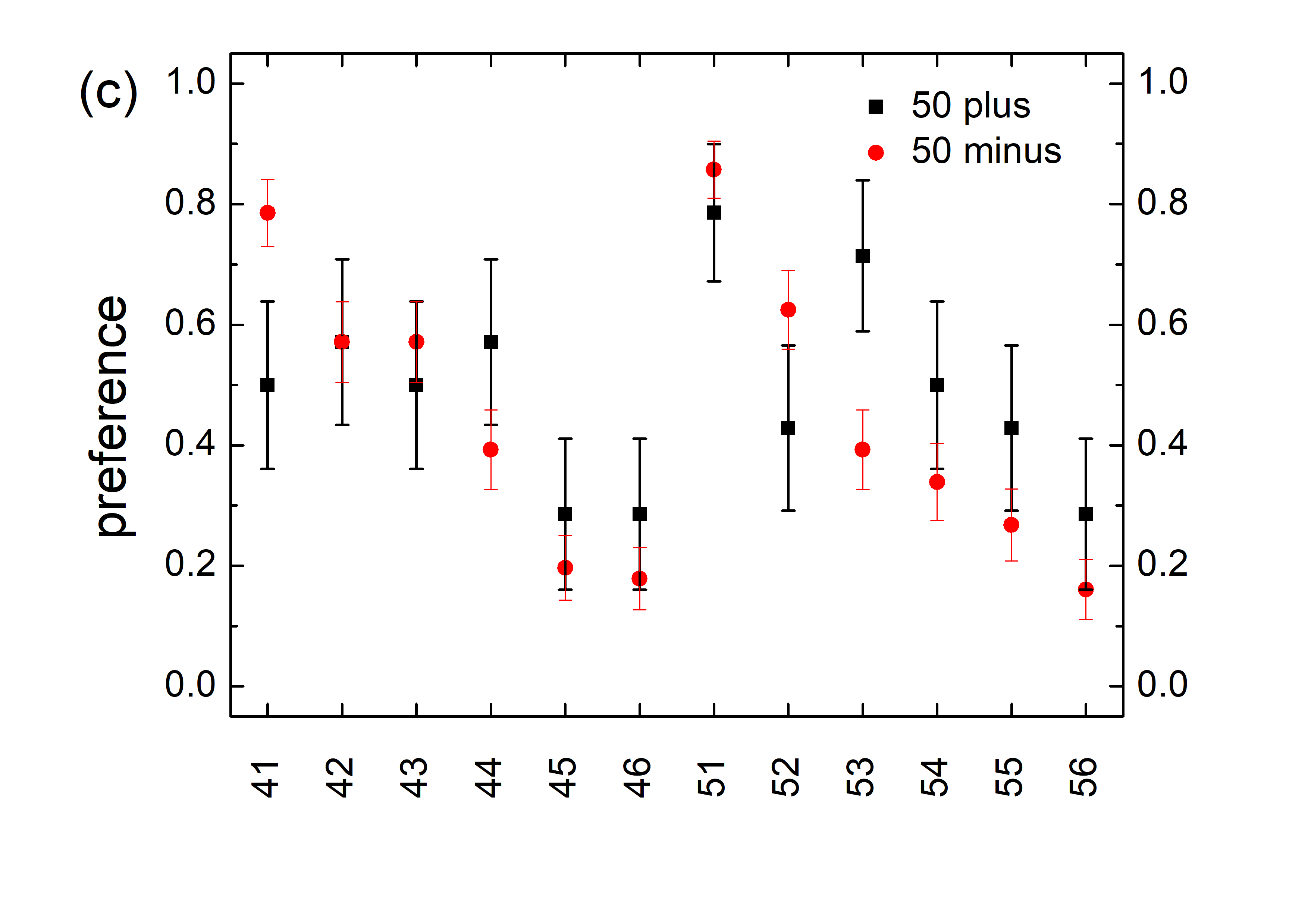} \\
\vspace{-0.2cm}
\includegraphics[width=0.32\textwidth]{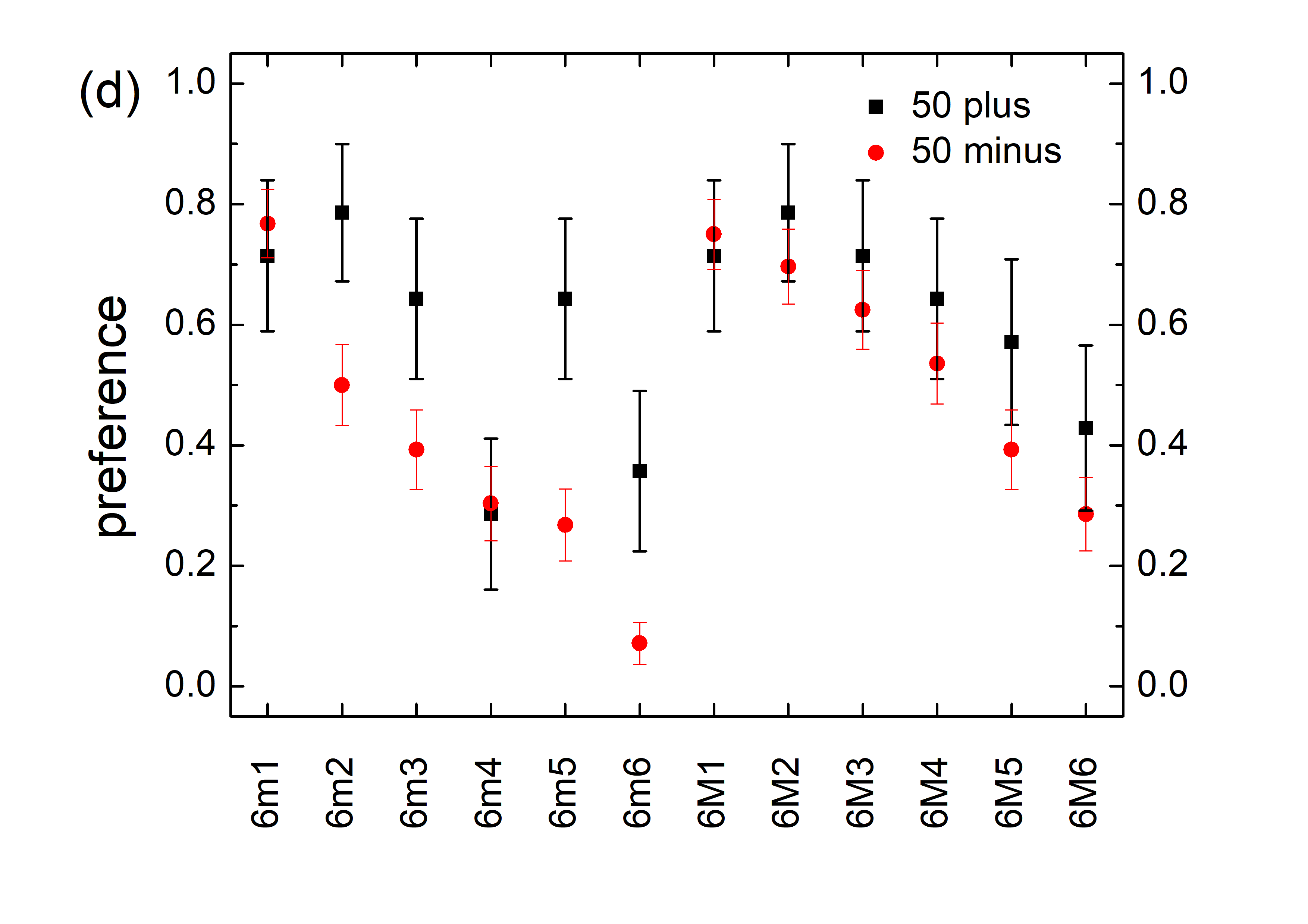}
\includegraphics[width=0.32\textwidth]{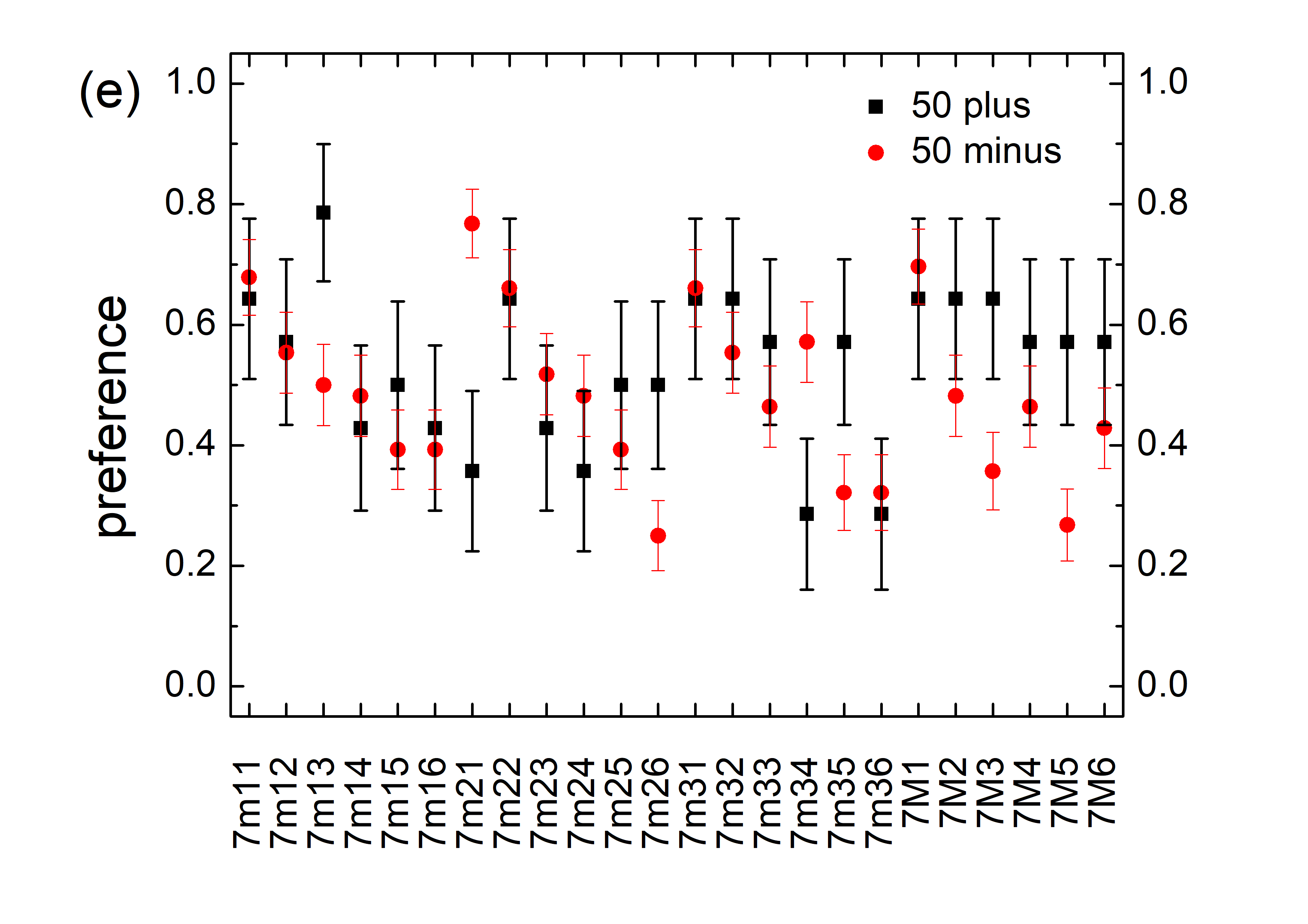}
\includegraphics[width=0.32\textwidth]{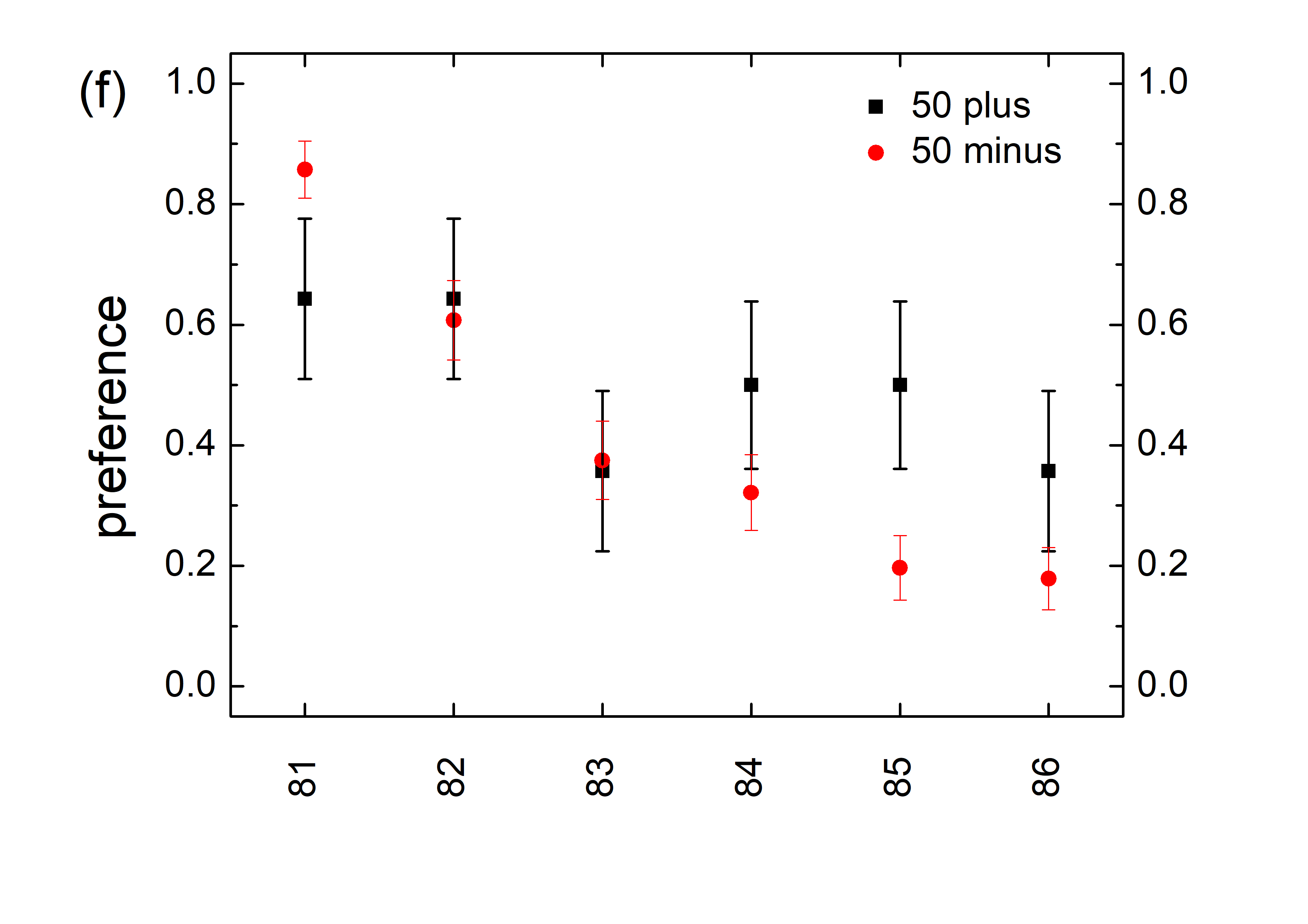}
\caption{Mean values and errors of responses or preference, $P$, for age groups \textit{above 50} (black squares, 14 persons) and \textit{below 50} (red circles, 56 persons): (a) seconds, (b) thirds, (c) fourths and fifths, (d) sixths, (e) sevenths, (f) octaves.}
\label{fig:PlusMinus50}
\end{figure*}

\begin{figure*}[htb]
\centering
\includegraphics[width=0.45\textwidth]{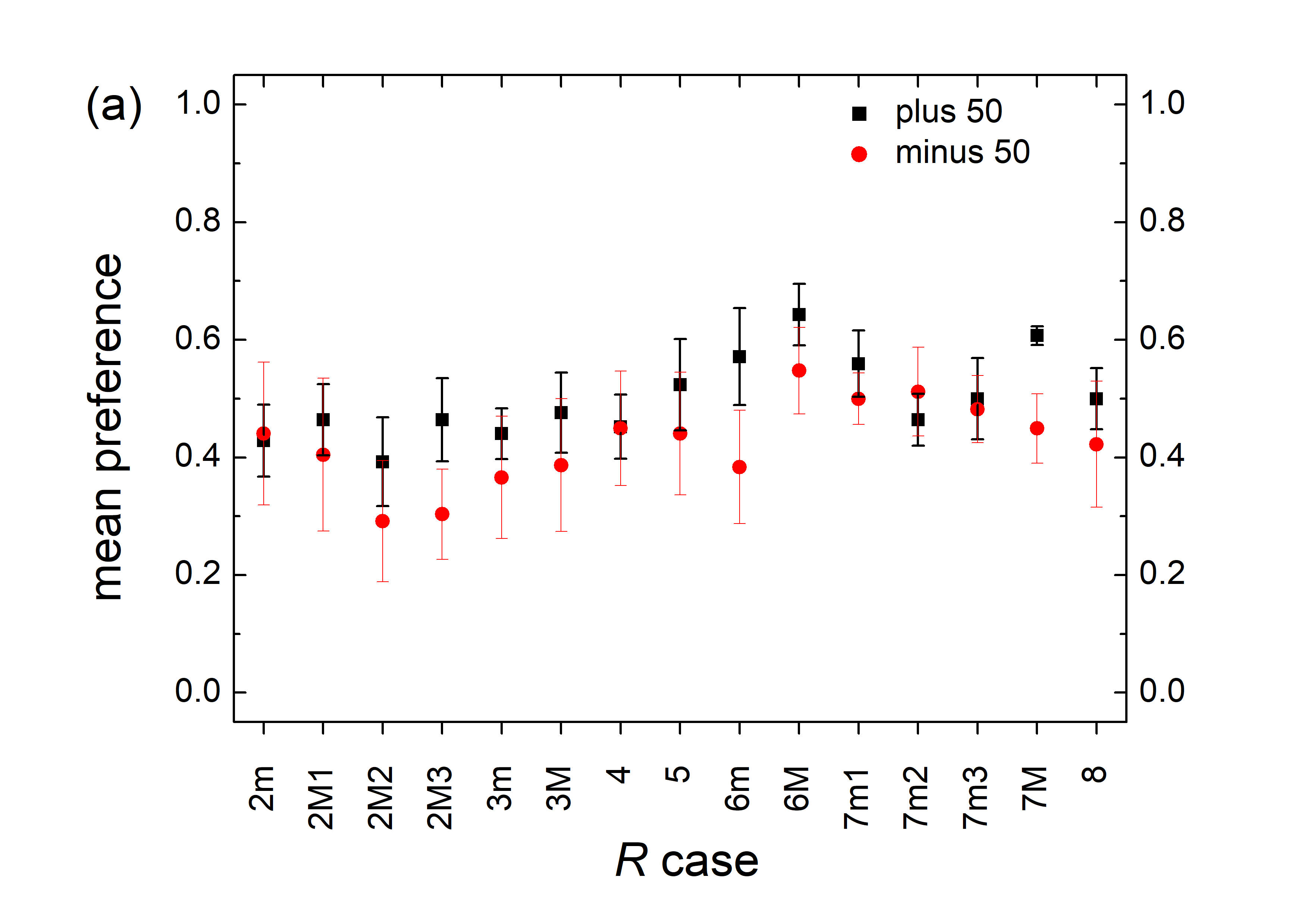} 
\includegraphics[width=0.45\textwidth]{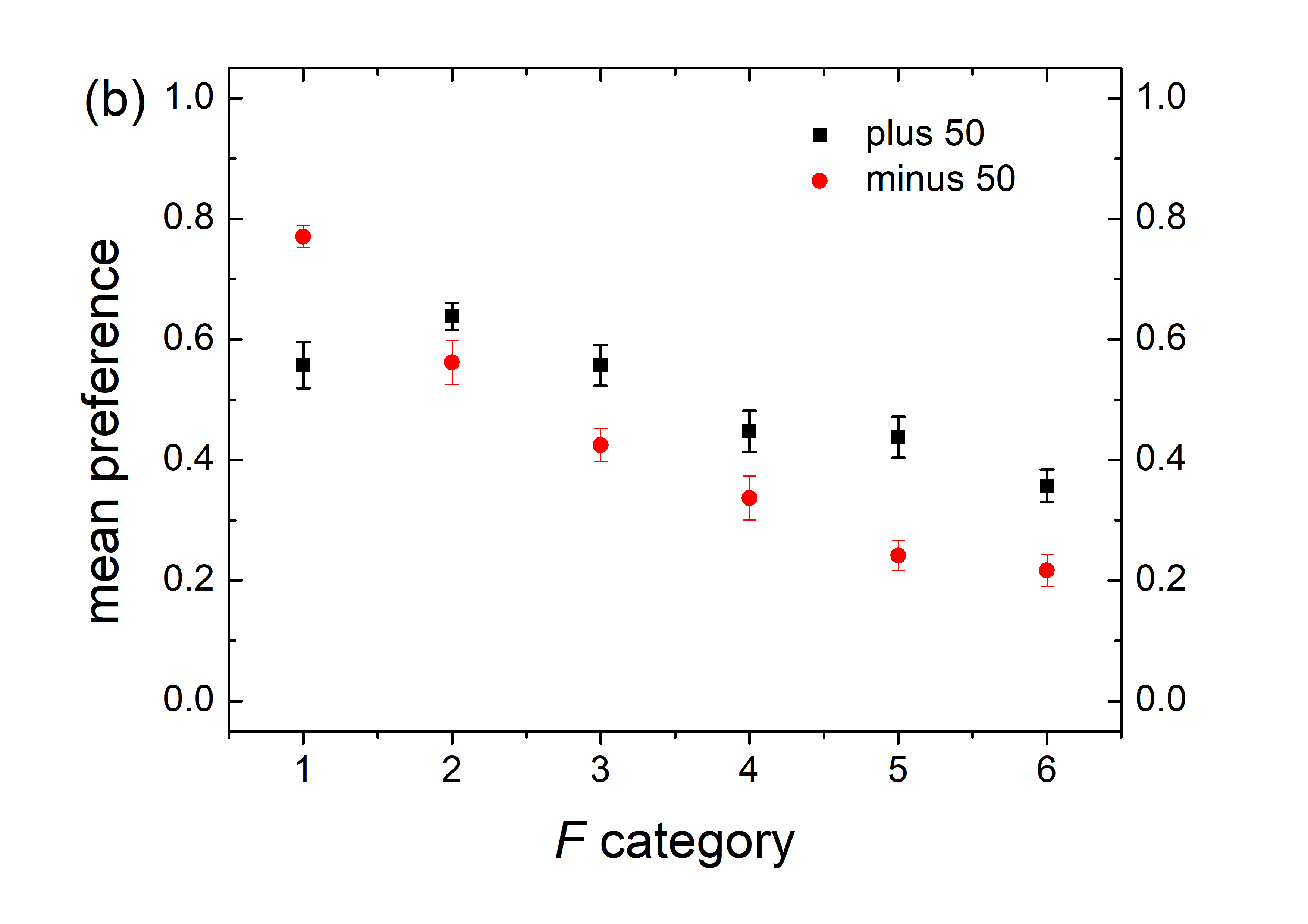}
\caption{Mean values and errors of $P$, for groups \textit{over 50} (black squares, 14 persons) and \textit{under 50} (red circles, 56 persons). (a) Mean over all $F$ categories for each $R$ case, $\overline{P}(R)$. (b) Mean over all $R$ cases for each $F$ category, $\overline{P}(F)$.}
\label{fig:meanvalues50RF}
\end{figure*}

Next, in Fig.~\ref{fig:PlusMinus50} we compare the preferences of subjects above 50 years of age (14 persons, black squares) with those below 50 years of age (56 persons, red circles).
In Fig.~\ref{fig:meanvalues50RF} we depict, for groups \textit{over 50} and \textit{under 50}: in (a) mean over all $F$ categories for each $R$ case, $\overline{P}(R)$ and in (b) mean over all $R$ cases for each $F$ category, $\overline{P}(F)$. 
The main observation from Fig.~\ref{fig:meanvalues50RF}(a) is that there some differences between the two age groups for the perception of frequency ratio $R$ cases; for some $R$ cases the mean values of one group are not within the mean errors of the other group. 
The main observation from
Fig.~\ref{fig:meanvalues50RF}(b) is that there  differences between the two age groups for the perception of mean frequency $F$ categories; for all $F$ categories the mean values of one group are not within the mean errors of the other group. 
This behavior is also confirmed by the MW tests shown in the Appendix, 
in Fig.~\ref{fig:MW-50} (for all $R$ cases) 
and in Fig.~\ref{fig:MW-50-F} (for all $F$ categories).
These MW tests confirm the inferences of Figs.~\ref{fig:meanvalues50RF} (a) and (b). It is impressive that in Fig.~\ref{fig:MW-50-F} the logarithmic vertical axis starts from 10$^{-10}$ in contrast to all other MW tests where is starts from  10$^{-4}$.
For the perception of mean frequency $F$ categories, the preferences of persons under 50 in Figs.~\ref{fig:meanvalues50RF}(b) cover a wider range.

Numerous studies have shown that aging affects auditory perception, including the perception of musical intervals and dyads. Older adults often experience reduced pitch discrimination and less distinct differentiation between consonant and dissonant dyads, even in the absence of clinically significant hearing loss. Bones and Plack~\cite{BonesPlack:2015} demonstrated that older listeners rate dissonant dyads as more pleasant and show reduced preference for consonance, a change attributed to diminished temporal neural coding in the auditory brainstem. Lentz et al.~\cite{Lentz:2022} found age-related declines in multiple psychoacoustic tasks—such as modulation detection and mistuning sensitivity—persist even after controlling for cognitive factors and hearing thresholds, suggesting central auditory aging plays a key role. Additionally, Tufts et al.~\cite{Tufts:2005} showed that sensorineural hearing loss, common in aging, leads to broadened auditory filters, resulting in a compressed perceptual range between consonant and dissonant dyads. These results imply that the reduced clarity in interval perception with age may arise from both peripheral (cochlear) and central (neural coding) factors.

Inference: It seems that age plays a role for the perception of $R$ and $F$. For the perception of mean frequency $F$ categories, the preferences of younger persons cover a wider range and there are also differences for the perception of frequency ratio $R$ cases and mean frequency $F$ categories.

\clearpage

\section{Conclusion}
\label{sec:conclusion}
We mainly studied two quantities, \textit{responses} [1 (like), 0 (dislike)], $\rho$, and  \textit{preferences}, $P$, defined as mean values of responses. Responses and preferences were statistically analyzed. Two important parameters were used: the ratio of frequencies, $R$, and the mean value of frequencies, $F$. We first studied the whole set of 70 persons and then we divided subjects into two groups (musicians - non-musicians, men - women, and age groups). Such divisions stem from the possibility that different perceptions of sound may exist due to a person's involvement with music for many years, due to gender or due to biological reasons such as differentiation of hearing in older ages. We utilized mean values and their errors, and compared responses and preferences with the Mann-Whitney rank sum test and the least squares method.  

\textit{For the general public:} Increased mean frequency $F$ makes simple-tone dyads less pleasant. For octaves, fifths, fourths, and sixths the feeling of consonance shows an almost ``linear'' fall, increasing $F$. Seconds and sevenths are more dispersed but still, increased $F$ increases the feeling of dissonance. Seconds seem to be the most unpleasant simple-tone dyads.

\textit{For groups musicians - non-musicians:} For octaves, the falls $P(F)$ show a similar slope between groups, but the overall preference level is higher for musicians. For fifths, a similar trend is observed, though less pronounced. Across all R cases, increasing pitch, consonance ratings tend to decrease for both groups. MW tests on preference scores for the combined group of fourths, fifths, and octaves did not show a statistically significant difference at the 0.05 level, though the result approaches this threshold. When analyzing individual $R$ cases, octaves yielded the lowest $p$-value among all intervals, indicating a possible trend, although not reaching statistical significance. In contrast, the MW test on responses (i.e., raw response data, without averaging) showed a statistically significant difference between musicians and non-musicians for octaves ($p \approx$ 0.0005). This result suggests that the perception of octaves differs reliably between the two groups. The averaged preference scores $P(R)$ indicate that musicians rate octaves (8), and possibly fourths (4), fifths (5), and major sixths (6M), higher than non-musicians. Both groups rated the intervals 2M2 and 2M3 lowest. Finally, $P(F)$ is a decreasing function both for musicians and non-musicians.

\textit{For the groups men - women:} As a general trend, for lower $F$ men show higher preferences, but this is reversed for higher $F$ where women show higher preferences. It seems that, men dislike high frequencies more than women. It also seems that the scores of men have higher variation than the scores of women.

\textit{For the age groups:} It seems that age plays an important role for the perception of $R$ and $F$. 
For the perception of mean frequency $F$ categories, the preferences of younger persons cover a wider range. There are also differences for the perception of frequency ratio $R$ cases (in some cases) and mean frequency $F$ categories (for all categories).

Relative to $R$, seconds accumulate most dislikes and are therefore found more dissonant. This is obvious from the mean preferences of the total set shown in Fig.~\ref{fig:meanvaluestotalsetRF} and from the relevant diagrams of group comparisons, i.e., Figs.~\ref{fig:meanvaluesMNMRF},~\ref{fig:meanvaluesMWRF},~\ref{fig:meanvalues35RF},~\ref{fig:meanvalues50RF}.
Among seconds, major seconds of type 2M2
($R=9/8$) and 2M3 ($R=8/7$) are generally perceived as more dissonant. From Fig.~\ref{fig:meanvaluestotalsetRF} and from Figs.~\ref{fig:meanvaluesMNMRF},~\ref{fig:meanvaluesMWRF},~\ref{fig:meanvalues35RF},~\ref{fig:meanvalues50RF}, we realize that 
as $R$ increases, simple-tone dyads are perceived as somewhat more acceptable. After seconds, there is a somehow upward trend in the listeners responses. As for the intervals widely considered consonant (fourths, fifths, major sixths, octaves) the general picture from the experiment is different. For the general set, the mean preferences for fourths, fifths and octaves are not greater than for sevenths; however, major sixths keep the maximum of mean preferences for the general set. Musicians find octaves more consonant than non-musicians as might be expected; also musicians almost find fourths, fifths and major sixths more consonant than non-musicians (cf. Fig.~ \ref{fig:meanvaluesMNMRF}).

Plomp and Levelt performed experiments with simple and composite tones, to study consonance relative to the frequency interval between the two tones~\cite{PlompLevelt:1965}. They showed differences between the case where dyads consisted of simple tones and the case where they consisted of complex sounds including harmonics.

Terhardt’s~\cite{Terhardt:1984} work referred to presenting dyads consisting of two tones—one fixed at La4 and the other varying gradually up to an octave (La5), while listeners rated these dyads for ``consonance’’ and ``roughness’’. Our results agree with the minimum at seconds shown by Terhardt~\cite{Terhardt:1984}, but do not confirm the monotonous increase up to the octave shown in his idealized curve.
In our results the increase is not monotonous for the general public (Fig.~\ref{fig:meanvaluestotalsetRF}(a)) and musicians show two local maxima at major sixth and octave, cf. Fig.~\ref{fig:meanvaluesMNMRF}(a).

Bowling and Purves included a diagram with composite musical intervals, listed by increasing degree of consonance~\cite{BowlingPurves:2015:PNAS}. This diagram was extracted from data, collected from 1898 to 2012, taken from various experiments with composite sounds: the intervals considered consonant are fourth, fifth and octave (the most consonant), third and sixth follow, while last are seventh and second. Hence, for composite sounds, including harmonics, it seems that there is a differentiation as to what seems consonant or dissonant.

It is worth noting that 2m1 has rather high preference. It is a minor second, $R=16/15$, $f_1 = 210$ Hz, $f_2 = 224 $ Hz, $F = 217$ Hz, $\Delta f = 14$ Hz. Clearly, a beat is formed, cf. Fig.~\ref{fig:100Hz200Hzand100Hz150Hzand500Hz600Hzand210Hz224Hz}(g)(h). We guess that this high preference stems from its low pitch. This extreme example fits our general conclusion that lower frequencies are more pleasant.

As for the mean frequency $F$, it is clear that higher frequencies collect more negative answers. A general trend is that increasing $F$, positive answers become less probable. This is a general observation for the total set, but it can also be seen for musicians - non-musicians, men and women and for age groups above/below 35 or above/below 50.
This lower preference to higher frequencies was also expressed immediately after the experiment, discussing with subjects.
From this decreasing of positive responses or preferences there is no exception to octaves or fifths and fourths, although the falls are not identical. 
One possible explanation for the decrease of likes with increasing frequency is that the human ear is most sensitive to sounds in the 2000–5000 Hz range, as indicated by the Fletcher–Munson curves~\cite{FletcherMunson:1933:Bell, FletcherMunson:1933:JASA}.
Hence, in this region, sounds will be somewhat piercing and will maximally stimulate our ear. 
In addition, in the event that a sound with low $F$ is followed by a sound with high $F$, our auditory system moves on the Fletcher-Munson curve suddenly from a high point to low point, so that it must respond to this change in a relatively short time. This effect could be strengthened by the fact that between two consecutive sounds there was a short time when no sound was heard, so absolute silence was succeeded by a sound with a high $F$ and therefore high perceived intensity.

In an experiment with single tones, with 8 male and 8 female University students, sounds with frequencies from 60 Hz to 5000 Hz were presented and subjects were asked to choose the most pleasant: sounds with high frequencies were rarely chosen~\cite{Vitz:1972}. 
In an experiment with 205 subjects, they had to choose a ``most pleasant'' sound delivered 
through an earphone by turning a control knob on a continuously variable audio oscillator ~\cite{Patchett:1979}. Most subjects chose a frequency in a relatively narrow 
($\approx$ 350 Hz) band centered around 400 Hz. The preferences did not appear influenced by sex or age~\cite{Patchett:1979}. 

Another possible reason for not preferring sounds with high frequencies might be that simple sounds in general are, due to the lack of harmonics, somewhat drier and more penetrating than complex sounds of corresponding frequencies. For example, a note with a frequency of about 3000 Hz on the piano corresponds to a note in its last, highest, octave.  This sound is not as piercing as its plain tone counterpart, but somewhat more rounded.

From the results of our experiments and from the discussion above it seems that consonance or dissonance is a complex multiparametric issue. It is certainly a topic that will not cease to concern people in general, whether they have something to do with music or simply because they find it interesting.

\noindent \textbf{Acknowledgments}
We thank Orestis Toufektsis, Institute of Composition, Theory of Music, History of Music, and Conducting, University of Music and Performing Arts, Graz, Austria, for critical reads of the manuscript and suggestions.

\noindent \textbf{Data availability:} The data that support the findings of this study were obtained from human participants under a protocol approved by the Ethics Committee of the National and Kapodistrian University of Athens requiring anonymization prior to sharing. The anonymized data are available from the corresponding author (Constantinos Simserides, csimseri@phys.uoa.gr) upon reasonable request. \\

\noindent \textbf{Author contributions:}
CS conceived the experiment. SK and CS conducted the experiment. SK and CS collaborated in the final arrangement of experimental details. 
CS and SK analyzed results. Both authors reviewed the manuscript. \\

\noindent \textbf{Additional information:} There are no competing interests.

\bibliographystyle{unsrt}  
\bibliography{references}  
%
%
%
%

\clearpage

\section*{Appendix}

\appendix

\setcounter{table}{0}
\setcounter{figure}{0}

\renewcommand{\thetable}{A\arabic{table}}
\renewcommand{\thefigure}{A\arabic{figure}}

\renewcommand{\theHtable}{A\arabic{table}}
\renewcommand{\theHfigure}{A\arabic{figure}}

\begin{figure}[h]
\centering
\includegraphics[width=0.33\textwidth]{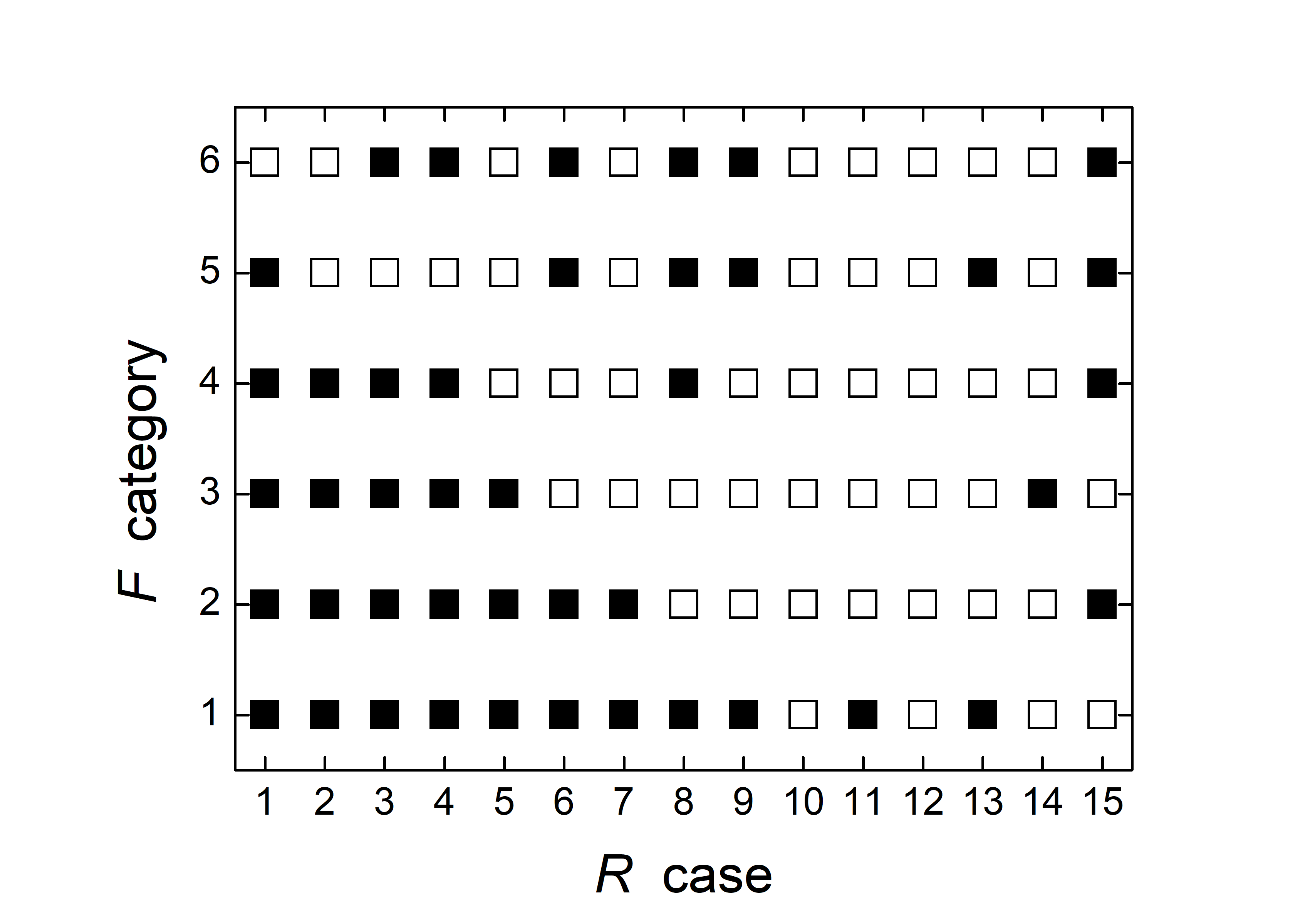} 
\caption{A random subject: Responses, $\rho$, for all simple-tone dyads, i.e., for each $F$ category and for each $R$ case. Yes = pleasant (full square) corresponds to $\rightarrow$ button. No = unpleasant (empty square)  corresponds to $\leftarrow$ button.}
\label{fig:SampleGraph}
\end{figure}

\begin{figure}[htb]
	\centering
	\includegraphics[width=0.42\textwidth]{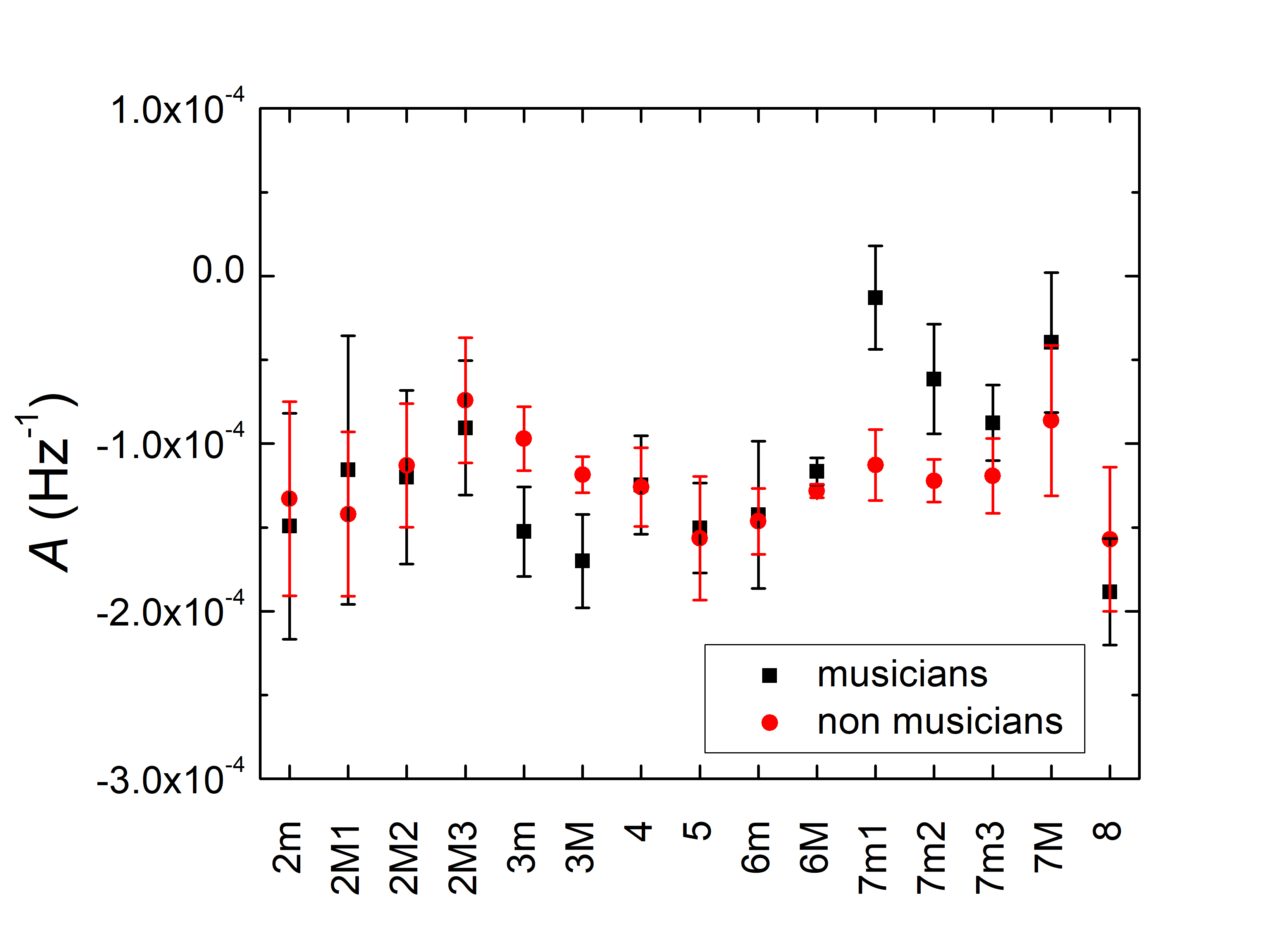}
	\includegraphics[width=0.42\textwidth]{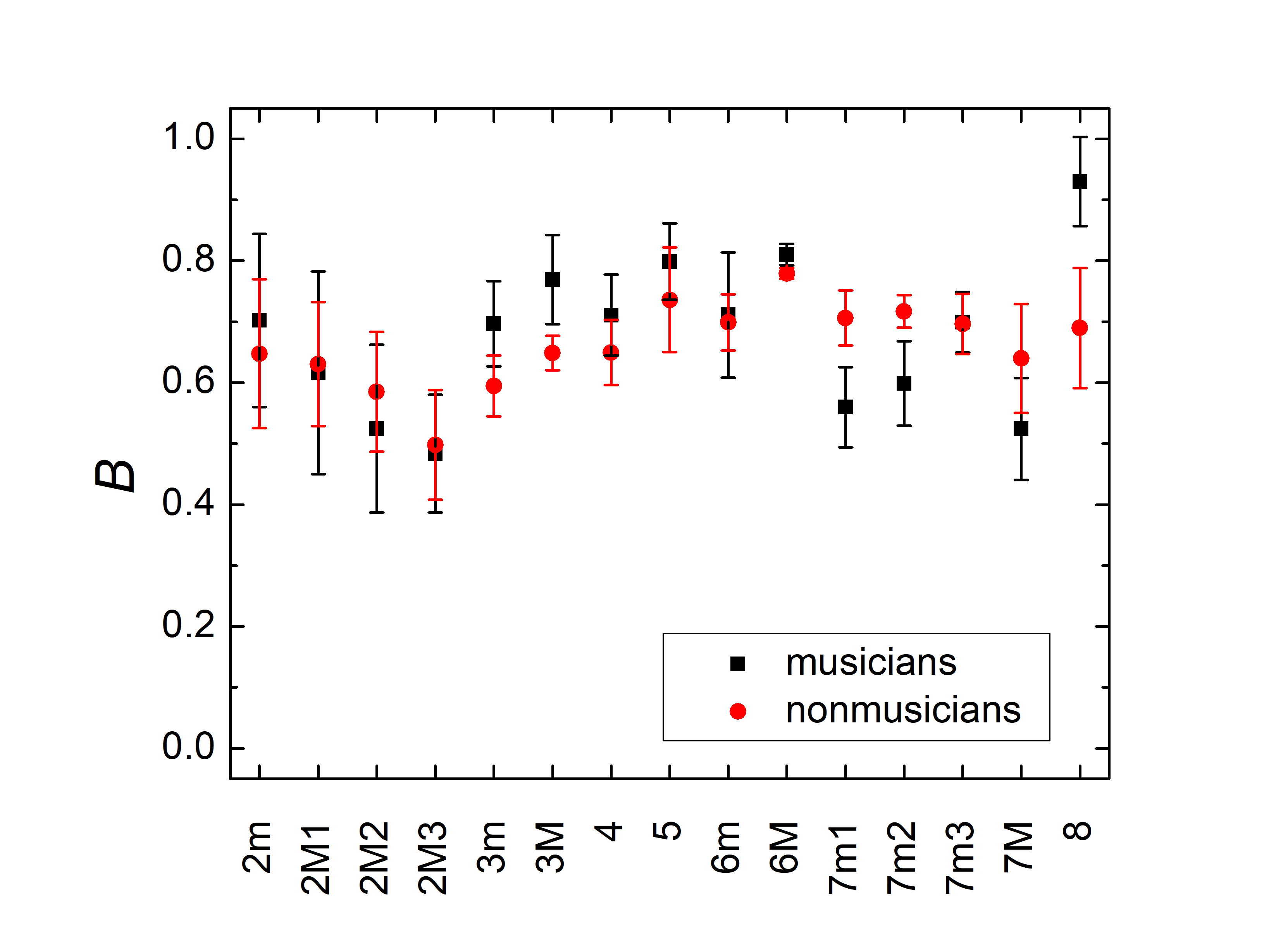} \\ 
	\includegraphics[width=0.42\textwidth]{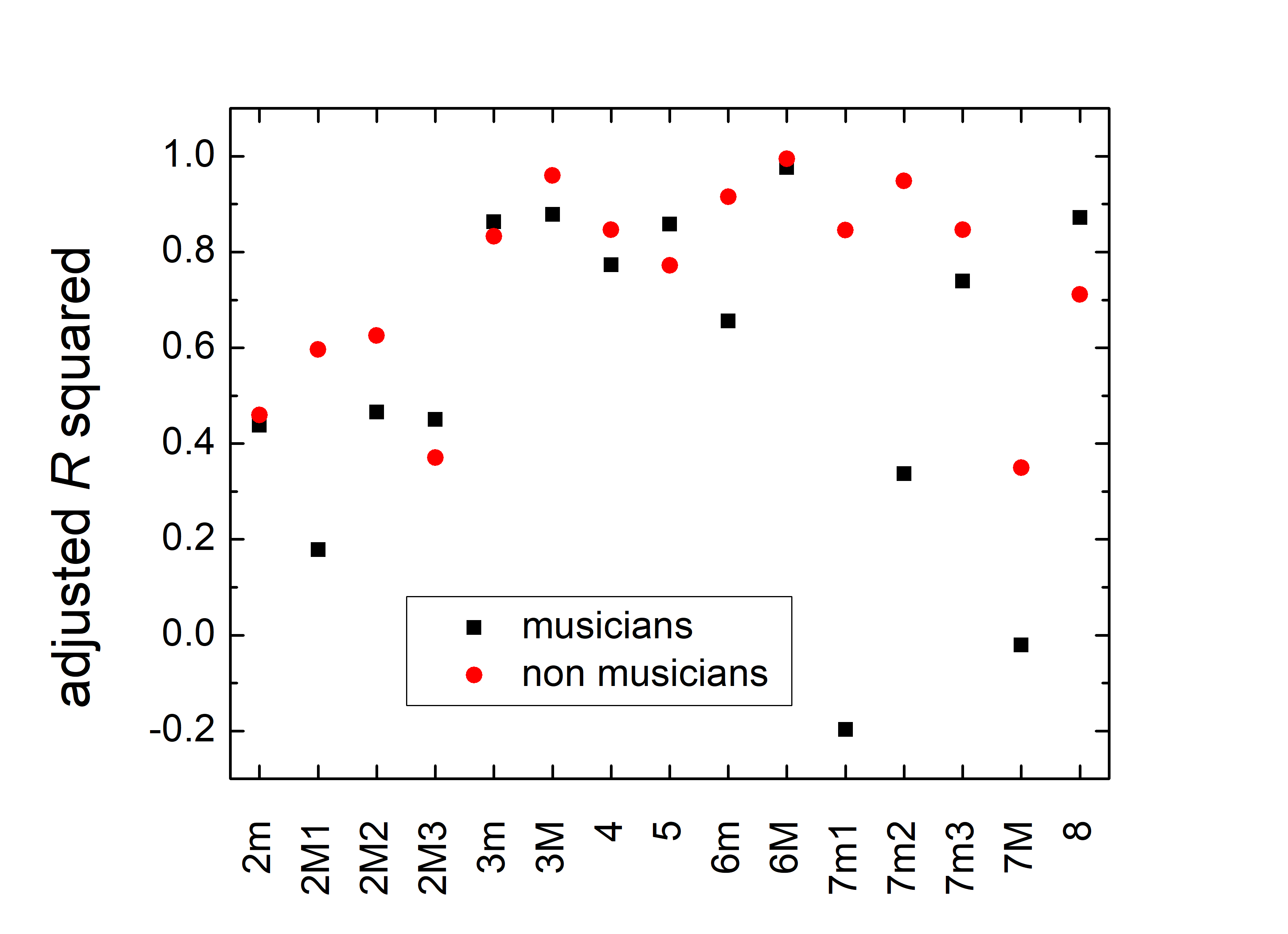} 
	\caption{Slopes, intercepts and adjusted $\mathcal{R}$ squared for linear fits $P=AF+B$, for the groups \textit{musicians} and \textit{non-musicians}.}
	\label{fig:ABR-MNM}
\end{figure}

\begin{figure}[h]
\centering
\includegraphics[width=0.65\textwidth]{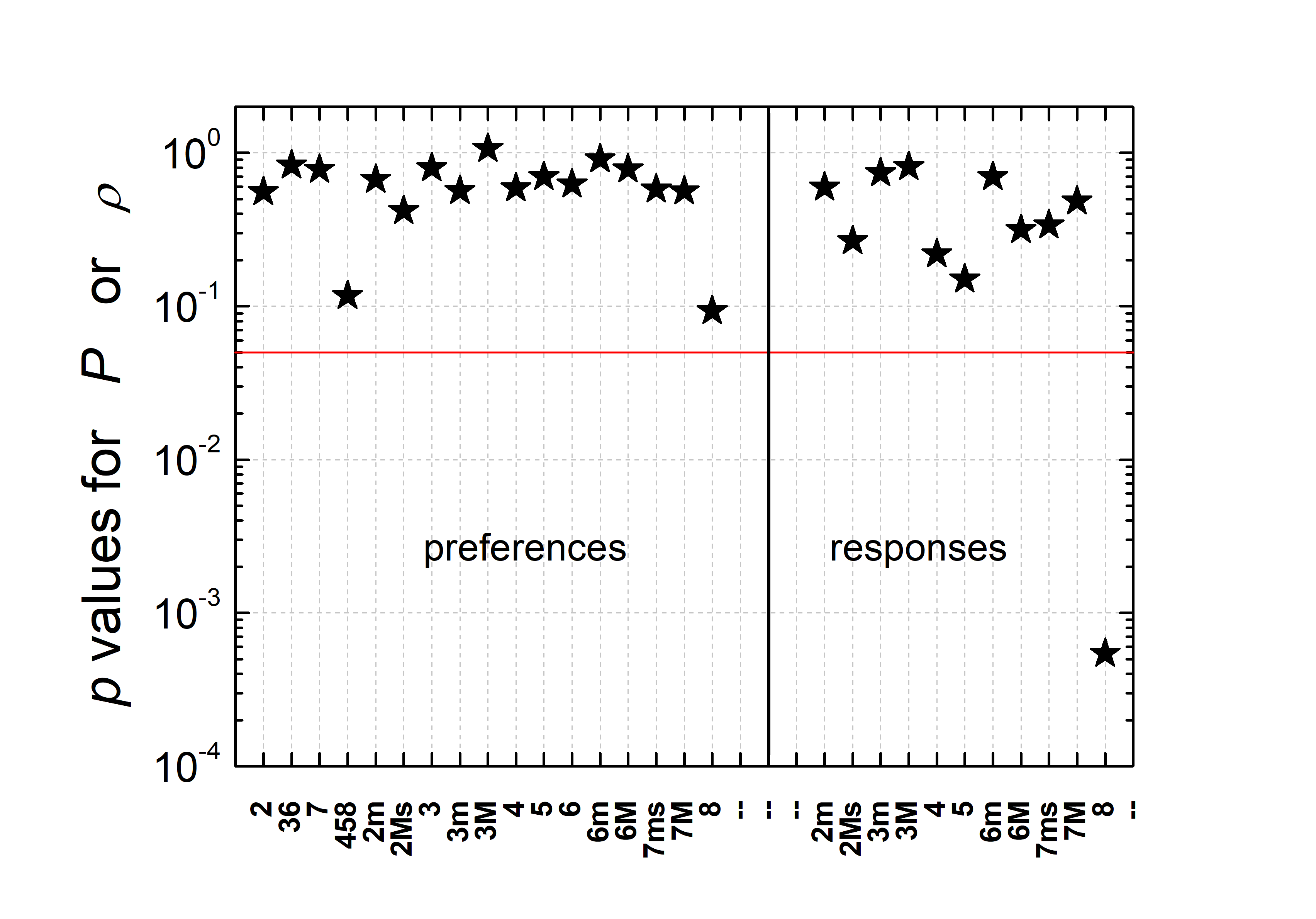}
\vspace{-0.2cm}
\caption{$R$ cases: $p$ values of Mann-Whitney tests on preferences ($P$) and responses ($\rho$), comparing \textit{musicians} and \textit{non-musicians}. The horizontal red line shows the  0.05 $p$-level. The vertical line divides preferences from responses.}
\label{fig:MW-MNM}
\end{figure}

\begin{figure}[h]
\centering
\includegraphics[width=0.65\textwidth]{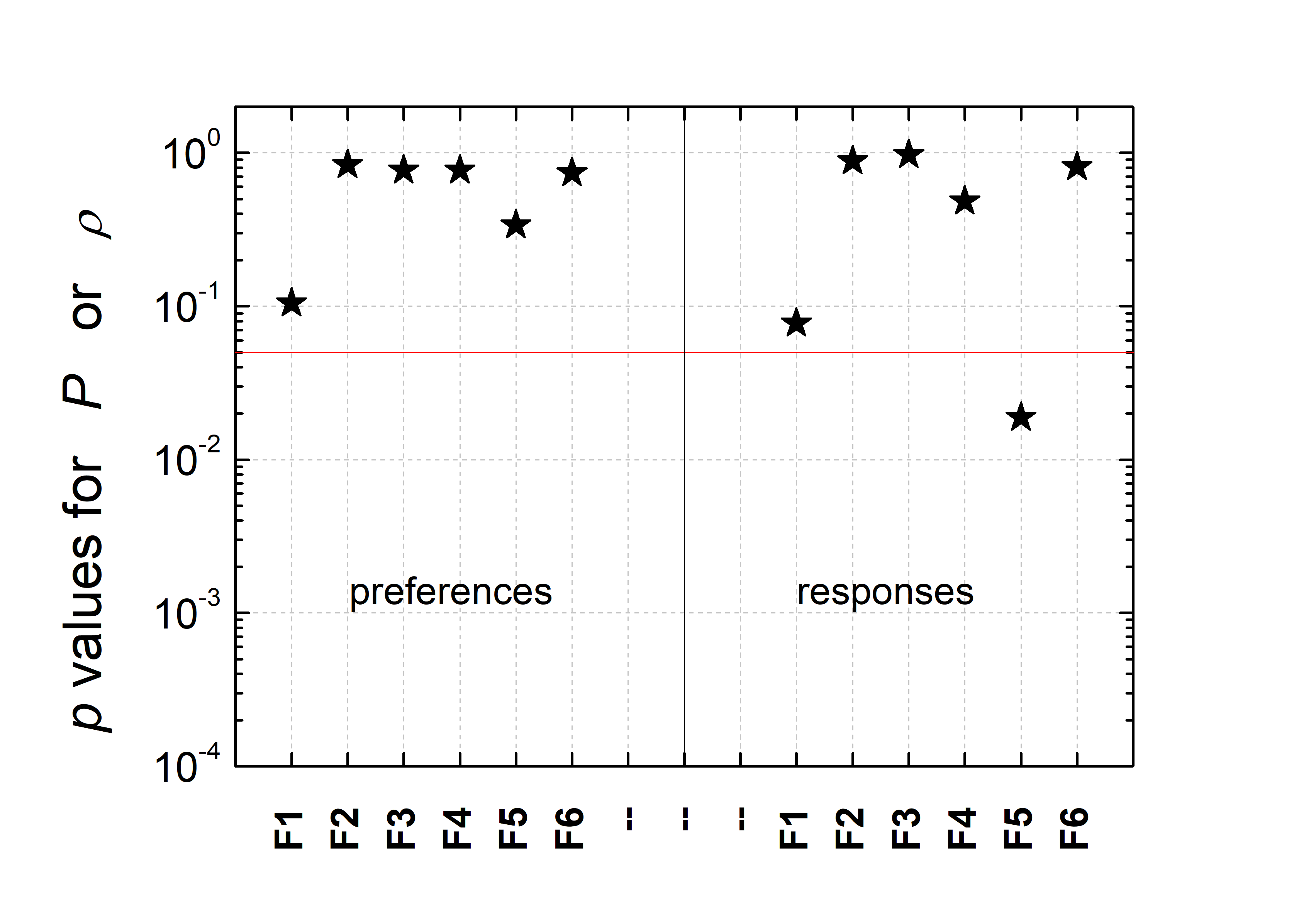}
\vspace{-0.2cm}
\caption{$F$ categories: $p$ values of Mann-Whitney tests on preferences ($P$) and responses ($\rho$), comparing \textit{musicians} and \textit{non-musicians}. The horizontal red line shows the  0.05 $p$-level. The vertical line divides preferences from responses.}
\label{fig:MW-MNM-F}
\end{figure}

\begin{figure}[h]
\centering
\includegraphics[width=0.65\textwidth]{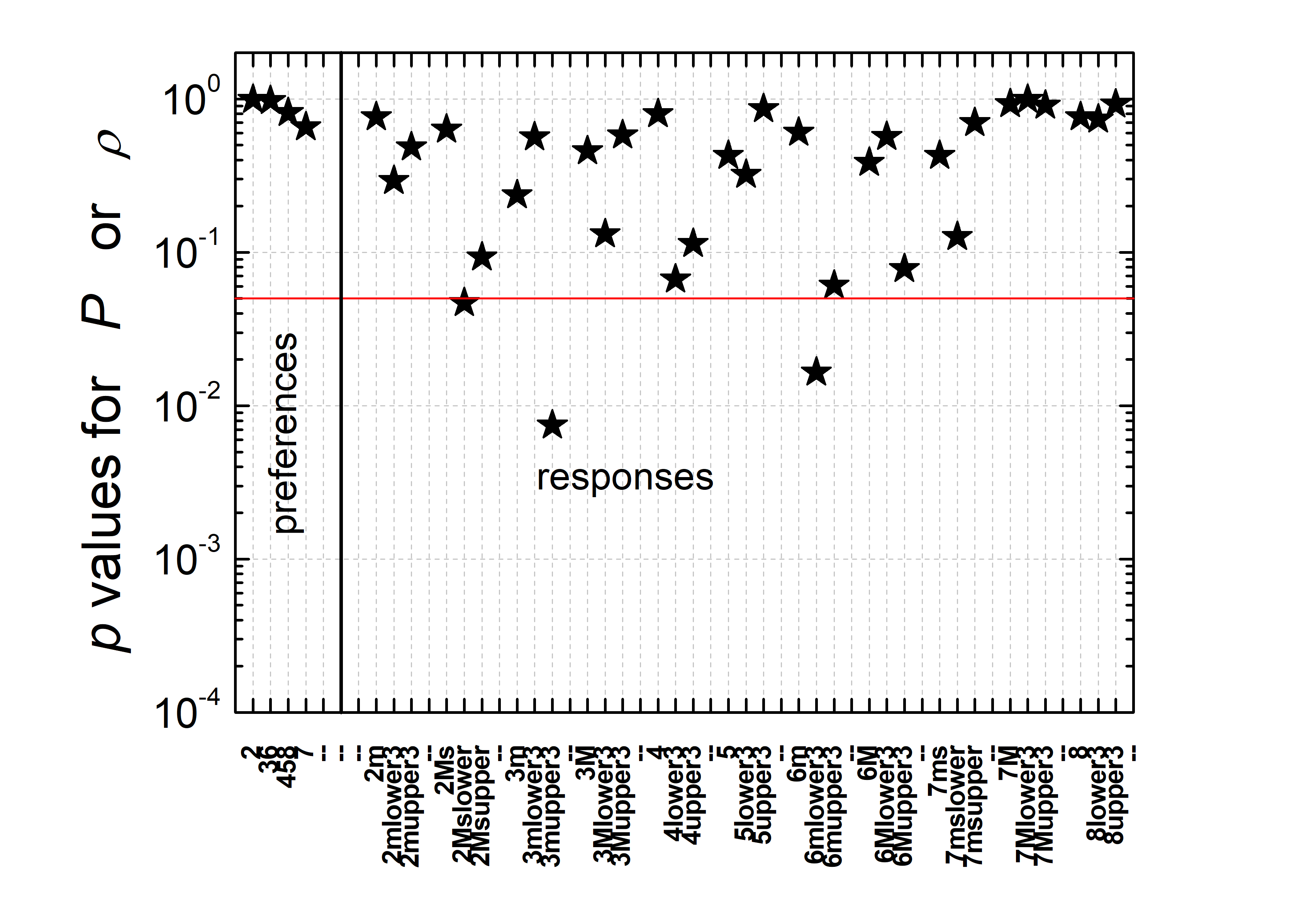}
\vspace{-0.2cm}
\caption{$R$ cases: $p$ values of the Mann-Whitney tests on preferences ($P$) and responses ($\rho$), comparing \textit{men} and \textit{women}. The horizontal red line shows the  0.05 $p$-level. The vertical line divides preferences from responses.}
\label{fig:MW-MW}
\end{figure}

\begin{figure}[h]
\centering
\includegraphics[width=0.65\textwidth]{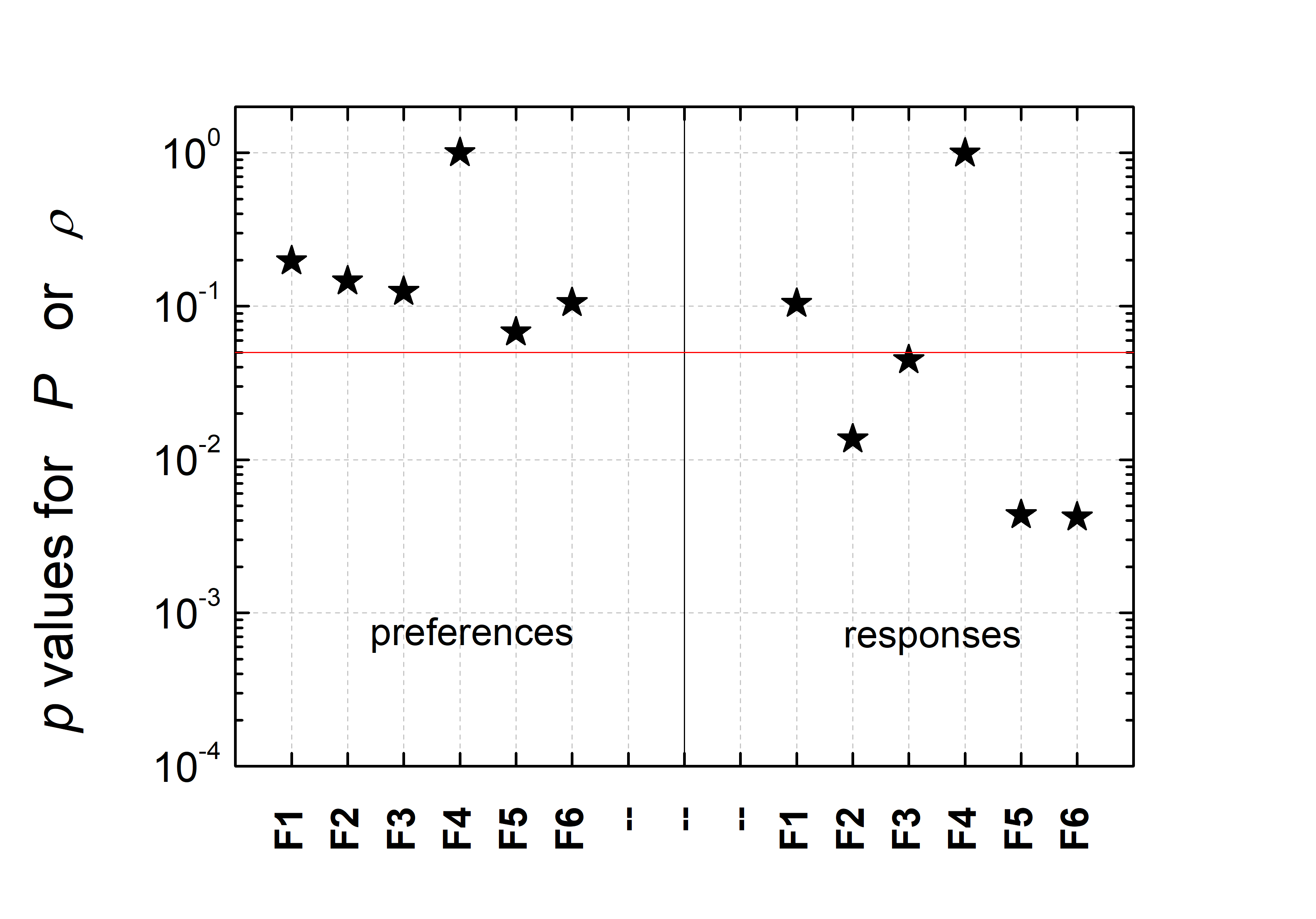}
\vspace{-0.2cm}
\caption{$F$ categories: $p$ values of the Mann-Whitney tests on preferences ($P$) and responses ($\rho$), comparing \textit{men} and \textit{women}. The horizontal red line shows the  0.05 $p$-level. The vertical line divides preferences from responses.}
\label{fig:MW-MW-F}
\end{figure}

\begin{figure}[t]
	\centering
	\includegraphics[width=0.65\textwidth]{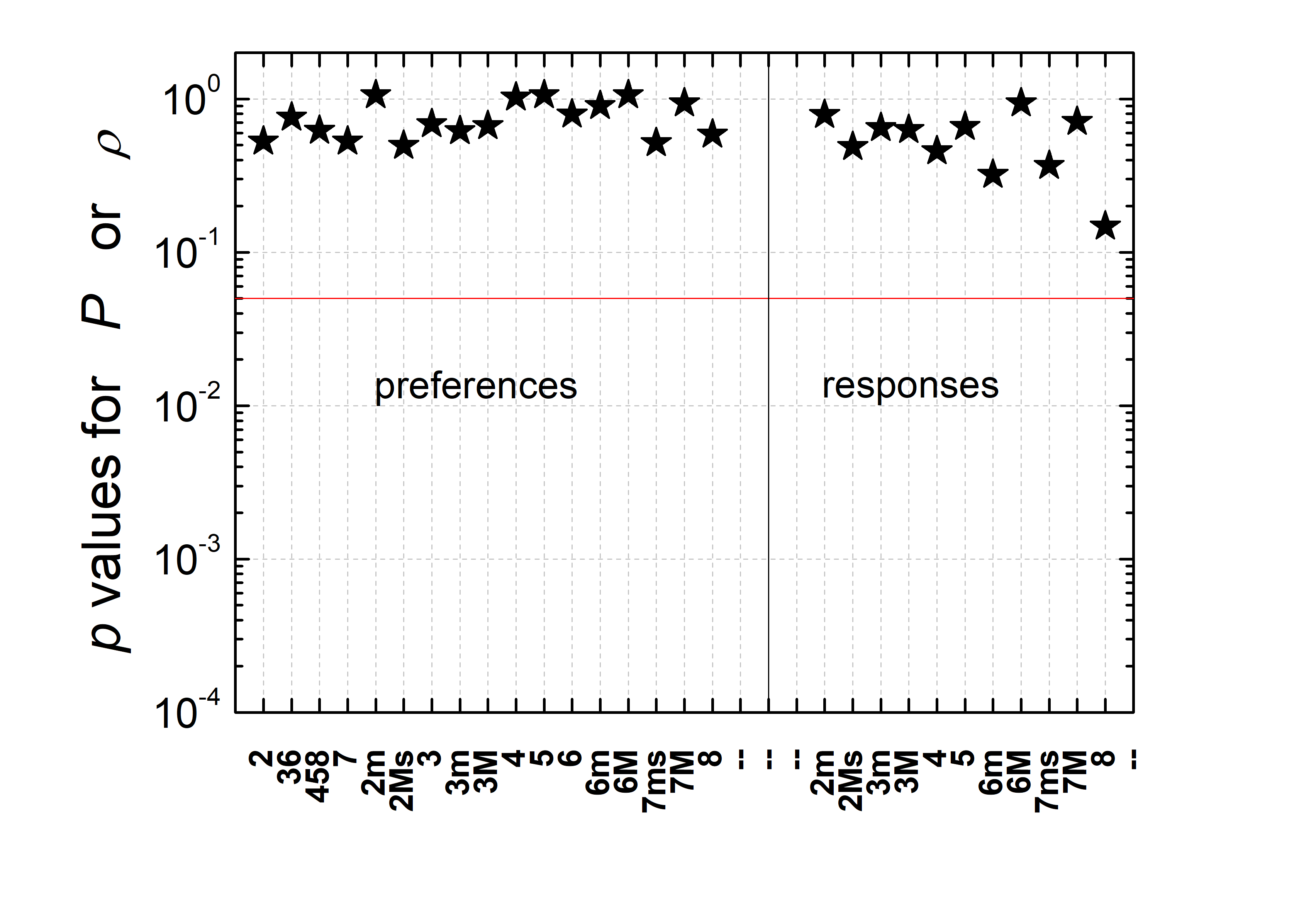}
	\caption{$p$ values of the Mann-Whitney tests on preferences ($P$) and responses ($\rho$), comparing the age groups \textit{above 35} and \textit{below 35}, relative to $R$ cases. The horizontal red line shows the 0.05 $p$-level. The vertical line divides preferences from responses.}
	\label{fig:MW-35}
\end{figure}

\begin{figure}[b]
	\centering
	\includegraphics[width=0.65\textwidth]{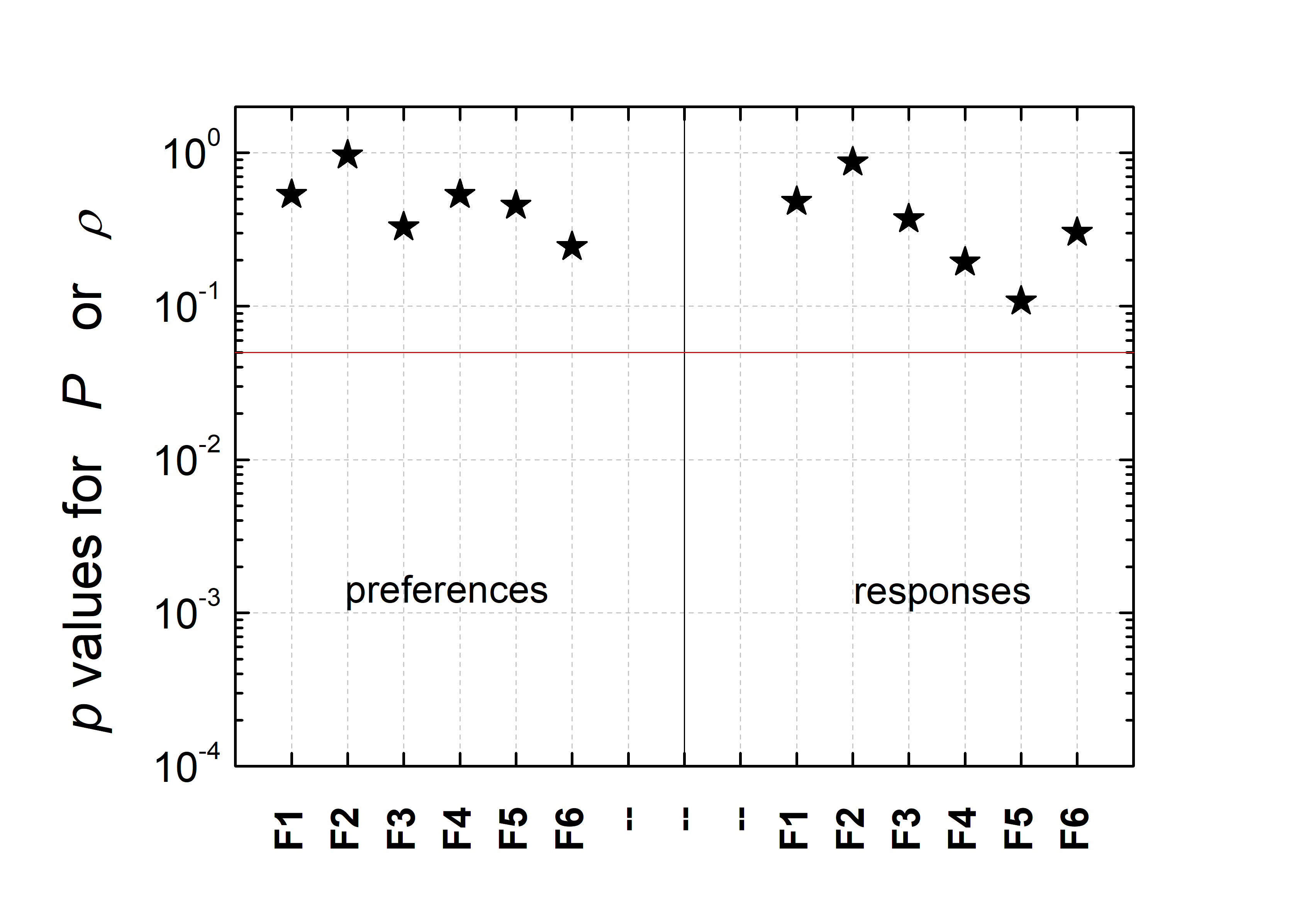}
	\caption{$p$ values of the Mann-Whitney tests on preferences ($P$) and responses ($\rho$), comparing the age groups \textit{above 35} and \textit{below 35}, relative to $F$ categories. The horizontal red line shows the  0.05 $p$-level. The vertical line divides preferences from responses.}
	\label{fig:MW-35-F}
\end{figure}

\begin{figure}[t]
	\centering
	\includegraphics[width=0.65\textwidth]{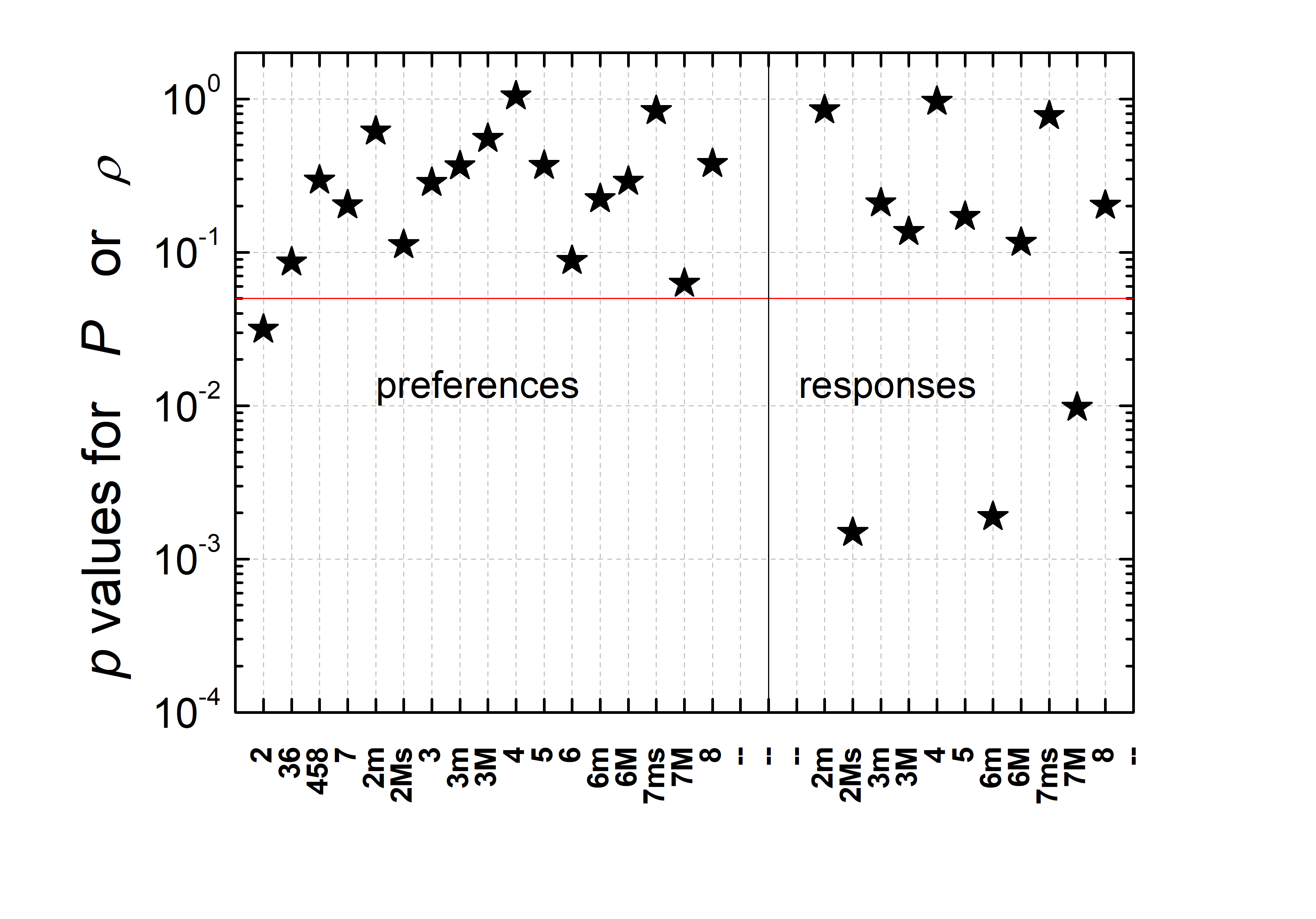}
	\caption{$p$ values of the Mann-Whitney tests on preferences ($P$) and responses ($\rho$), comparing the age groups \textit{above 50} and \textit{below 50}, relative to $R$ cases. The horizontal red line shows the 0.05 $p$-level. The vertical line divides preferences from responses.}
	\label{fig:MW-50}
\end{figure}

\begin{figure}[b]
	\centering
	\includegraphics[width=0.65\textwidth]{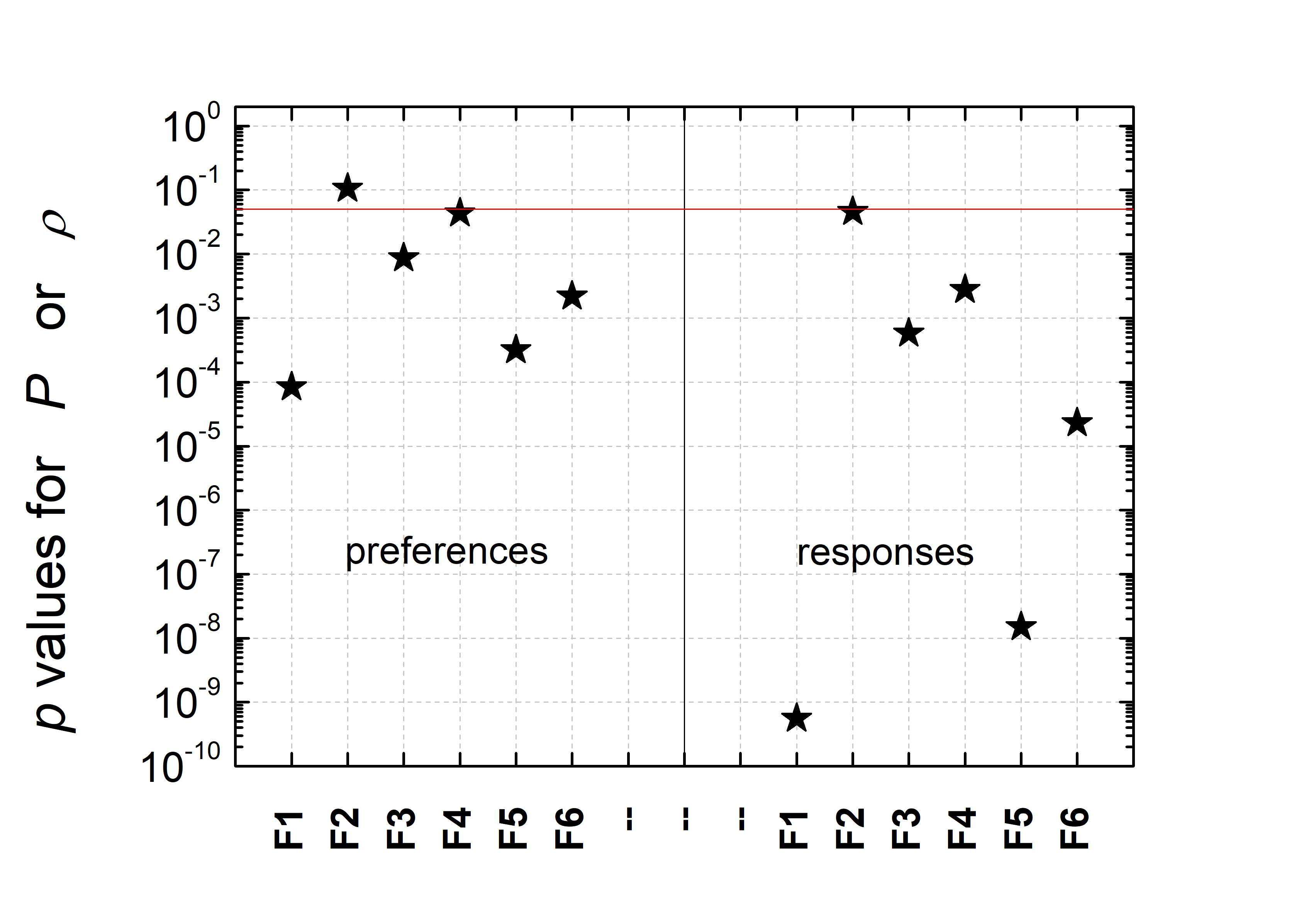}
	\caption{$p$ values of the Mann-Whitney tests on preferences ($P$) and responses ($\rho$), comparing the age groups \textit{above 50} and \textit{below 50}, relative to $F$ categories. The horizontal red line shows the  0.05 $p$-level. The vertical line divides preferences from responses.}
	\label{fig:MW-50-F}
\end{figure}

A presentation of responses, $\rho$, of a random subject is given in Fig.~\ref{fig:SampleGraph}. 
A complete list of simple-tone dyads used in this work is given in Table ~\ref{Table:ListDitonies}.
In Figure~\ref{fig:ABR-MNM} we present slopes, intercepts, and adjusted R squared of the linear fits $P=AF+B$, where $P$ is the preference and $F$ is the mean frequency of simple-tone dyads, for the groups musicians - non-musicians. The fits are not generally good: the relationships $P(F)$ are not straight lines. Still, we can make a crude estimation: all slopes are clearly negative (but sevenths have very small negative slopes). Hence, increasing pitch, the feeling of consonance decreases. We also notice that the most negative slopes are those for octaves and major thirds.
The $p$ values of the Mann-Whitney tests on preferences ($P$) and responses ($\rho$), comparing groups are given in Fig.~\ref{fig:MW-MNM} (musicians vs. non-musicians, $R$ cases), in Fig.~\ref{fig:MW-MNM-F} (musicians vs. non-musicians, $F$ categories), in Fig.~\ref{fig:MW-MW} (men vs. women, $R$ cases), in Fig.~\ref{fig:MW-MW-F} (men vs. women, $F$ categories), in Fig.~\ref{fig:MW-35} (age above vs. below 35, $R$ cases), in Fig.~\ref{fig:MW-35-F} (age above vs. below 35, $F$ categories), in Fig.~\ref{fig:MW-50} (age above vs. below 50, $R$ cases), in Fig.~\ref{fig:MW-50-F} (age above vs. below 50, $F$ categories).

\vspace{0.4cm}

\begin{table}[ht]
	\centering
	\footnotesize{
		\begin{tabular}{|c|c|c|c|c|} \hline
			interval  & codename & $R$  & $F$ (Hz)&$f_2$, $f_1$ (Hz) \\ \hline \hline 
			& 2m1 &         & 217 & 224,   210 \\
			& 2m2 &         & 514.6 & 531.2,   498 \\
			& 2m3 &  16/15  & 899 & 928,   870 \\
			2nd minor & 2m4 &  =      & $1725.\overline{6}$  & $1781.\overline{3}$,  1670 \\
			(1)   & 2m5 &  $1.0\overline{6}$ & 2433.5 & 2512,  2355 \\
			& 2m6 &        & $4081.\overline{6}$& 4213.3,  3950 \\ \hline
			& 2M11 &      & $242.\overline{7}$ & $255.\overline{5}$,   230 \\
			& 2M12 &      & $461.2\overline{7}$ & $485.\overline{5}$,   437 \\
			& 2M13 & 10/9 & $785.\overline{3}$ & $826.\overline{6}$,   744 \\
			2nd major & 2M14 & = & $1245.\overline{5}$ & $1311.\overline{1}$,  1180 \\
			(2)& 2M15 & $1.\overline{1}$    & $2596.\overline{6}$ & $2733.\overline{3}$,  2460 \\
			& 2M16 &      & $4074.\overline{4}$ & $4288.\overline{8}$,  3860 \\ \hline
			& 2M21 &     & 329.375 &  348.75,  310  \\
			& 2M22 &     & 828.75&  877.5,  780  \\
			& 2M23 & 9/8 &1168.75& 1237.5, 1100   \\
			2nd major & 2M24 & =   &3049.375& 3228.75, 2870  \\ 
			(2)& 2M25 &1.125&3453.125& 3656.25, 3250  \\
			& 2M26 &     &4356.25& 4612.5, 4100  \\ \hline
			& 2M31 &     & 417.857&  445.714,  390 \\
			& 2M32 &     & 797.142&  850.285, 744 \\
			& 2M33 & 8/7 &1052.142& 1122.285, 982 \\
			2nd major & 2M34 & =   &2089.285& 2228.571, 1950 \\
			(2)& 2M35 & $1.\overline{142857}$ &3375& 3600, 3150 \\ 
			& 2M36 &     &4146.428& 4422.857, 3870 \\ \hline
			& 3m1 &       & 220 &240, 200 \\
			& 3m2 &       & 550 &600, 500  \\
			& 3m3 & 6/5   &1650 &1800, 1500 \\
			3rd minor & 3m4 & =     & 2750 & 3000, 2500 \\
			(3)& 3m5 & 1.2 & 3300& 3600, 3000 \\
			& 3m6 &       & 4400 & 4800, 4000  \\ \hline
			& 3M1 &       & 292.5&325, 260 \\
			& 3M2 &       & 450  &500, 400 \\
			& 3M3 & 5/4   &1125  &1250, 1000 \\
			3rd major & 3M4 & =     &2475&2750, 2200 \\
			(4)& 3M5 &1.25   &3937.5&4375, 3500 \\
			& 3M6 &       &4275 &4750, 3800 \\ \hline
			& 41  &       &$256.\overline{6}$&$293.\overline{3}$, 220 \\
			& 42  &       &700 &800,  600 \\
			& 43  &  4/3  &910&1040, 780 \\
			4th pure  & 44  &$\approx$&$1283.\overline{3}$&$1466.\overline{6}$, 1100 \\
			(5)& 45  & $1.\overline{3}$ &$3033.\overline{3}$& $3466.\overline{6}$, 2600 \\
			& 46  &       &$4316.\overline{6}$&$4933.\overline{3}$, 3700 \\ \hline
	\end{tabular}}
\vspace{0.3cm}
	\caption{A complete list of simple-tone dyads used in this work. 1st column: name (number of semitones). 2nd column: codename. 3rd column: ratio of frequencies, $R = \frac{f_2}{f_1}$. 4th column: mean frequency, $F=\frac{f_1+f_2}{2}$. 5th column: couple of frequencies. \textit{Continues} ...}
	\label{Table:ListDitonies} 
\end{table}

\addtocounter{table}{-1}
\begin{table}[ht]
	\centering
	\footnotesize{
		\begin{tabular}{|c|c|c|c|c|} \hline
			interval  & codename & $R$  & $F$ (Hz)&$f_2$, $f_1$ (Hz) \\ \hline \hline 
			& 51  &       & 312.5 & 375, 250 \\
			& 52  &       & 812.5 & 975, 650 \\
			& 53  & 3/2   & 1277.5 & 1533, 1022 \\
			5th pure  & 54  & =     & 2312.5 & 2775, 1850 \\
			(7)& 55  & 1.5   & 2832.5 &3399, 2266 \\
			& 56  &       & 4082.5 &4899, 3266 \\ \hline
			& 6m1 &       & 299    & 368, 230 \\
			& 6m2 &       & 676    & 832, 520 \\
			& 6m3 & 8/5   & 1300   & 1600, 1000 \\
			6th minor & 6m4 & =     & 2405   & 2960, 1850 \\
			(8)& 6m5 & 1.6   & 2990   & 3680, 2300 \\
			& 6m6 &       & 3965   & 4880, 3050 \\ \hline 
			& 6M1 &      &$333.\overline{3}$&$416.\overline{6}$, 250 \\
			& 6M2 &      &$626.\overline{6}$&$783.\overline{3}$, 470 \\
			& 6M3 &  5/3 &1240& 1550, 930 \\
			6th major & 6M4 &  =   &$1866.\overline{6}$& $2333.\overline{3}$, 1400 \\
			(9)& 6M5 &$1.\overline{6}$ &2800& 3500, 2100 \\
			& 6M6 &      &$3866.\overline{6}$&  $4833.\overline{3}$, 2900 \\ \hline
			&7m11&      &481.25& 612.5, 350 \\
			&7m12&      &783.75& 997.5, 570 \\
			&7m13& 7/4  &1375  & 1750, 1000 \\
			7th minor &7m14&  =   &1856.25& 2362.5, 1350 \\
			(10)&7m15& 1.75 &2887.5& 3675, 2100 \\
			&7m16&      &3575& 4550, 2600 \\ \hline
			&7m21&      &$277.\overline{7}$& $355.\overline{5}$, 200 \\
			&7m22&      &$486.\overline{1}$& $622.\overline{2}$, 350 \\
			&7m23& 16/9 &1000  & 1280, 720 \\
			7th minor &7m24&  =   &$1944.\overline{4}$&$2488.\overline{8}$, 1400 \\
			(10)&7m25&$1.\overline{7}$ &3125&  4000, 2250 \\
			&7m26&      &$3472.\overline{2}$&$4444.\overline{4}$, 2500 \\ \hline
			&7m31&      & 588& 756, 420 \\
			&7m32&      &1204& 1548, 860 \\
			&7m33& 9/5  &1610& 2070, 1150  \\
			7th minor &7m34&  =   &1820& 2340, 1300 \\
			(10)&7m35& 1.8  &2660& 3420, 1900  \\
			&7m36&      &3780& 4860, 2700 \\ \hline
			&7M1  &      &388.125&  506.25, 270 \\
			&7M2  &      &848.125& 1106.25, 590  \\
			&7M3  & 15/8 &1243.4375& 1621.875, 865 \\
			7th major &7M4  & =    &1638.75& 2137.5, 1140 \\
			(11)&7M5  &1.875 &2716.875& 3543.75, 1890 \\
			&7M6  &      &3378.125& 4406.25, 2350 \\ \hline
			& 81  &      & 300&  400, 200 \\
			& 82  &      &1050& 1400, 700 \\
			& 83  &  2/1 &1500& 2000, 1000 \\
			8th pure  & 84  &  =   &2250& 3000, 1500 \\
			(12)& 85  &  2   &3000& 4000, 2000 \\
			& 86  &      &3750& 5000, 2500 \\ \hline
	\end{tabular}}
\vspace{0.3cm}
	\caption{\textit{Continued...} A complete list of simple-tone dyads used in this work. 1st column: name (number of semitones). 2nd column: codename. 3rd column: ratio of frequencies, $R = \frac{f_2}{f_1}$. 4th column: mean frequency, $F=\frac{f_1+f_2}{2}$. 5th column: couple of frequencies.}
\end{table}

\clearpage

\clearpage

\end{document}